\ifx\mnmacrosloaded\undefined 
%
%
%
%

\catcode `\@=11 

\def\@version{1.4}
\def\@verdate{22nd Feb 1994}

%
%
%
%


\newif\ifprod@font

\ifx\@typeface\undefined
  \def\@typeface{Comp. Modern}\prod@fontfalse
\else
  \prod@fonttrue 
\fi

\def\newfam{\alloc@8\fam\chardef\sixt@@n} 

\ifprod@font
\font\fiverm=mtr10 at 5pt
\font\fivebf=mtbx10 at 5pt
\font\fiveit=mtti10 at 5pt
\font\fivesl=mtsl10 at 5pt
\font\fivett=mttt10 at 5pt     \hyphenchar\fivett=-1
\font\fivecsc=mtcsc10 at 5pt
\font\fivesf=mtss10 at 5pt
\font\fivei=mtmi10 at 5pt      \skewchar\fivei='177
\font\fivemib=mtmib10 at 5pt   \skewchar\fivemib='177
\font\fivesy=mtsy10 at 5pt     \skewchar\fivesy='60
\font\fivesyb=mtbsy10 at 5pt   \skewchar\fivesyb='60

\font\sixrm=mtr10 at 6pt
\font\sixbf=mtbx10 at 6pt
\font\sixit=mtti10 at 6pt
\font\sixsl=mtsl10 at 6pt
\font\sixtt=mttt10 at 6pt      \hyphenchar\sixtt=-1
\font\sixcsc=mtcsc10 at 6pt
\font\sixsf=mtss10 at 6pt
\font\sixi=mtmi10 at 6pt       \skewchar\sixi='177
\font\sixmib=mtmib10 at 6pt    \skewchar\sixmib='177
\font\sixsy=mtsy10 at 6pt      \skewchar\sixsy='60
\font\sixsyb=mtbsy10 at 6pt    \skewchar\sixsyb='60

\font\sevenrm=mtr10 at 7pt
\font\sevenbf=mtbx10 at 7pt
\font\sevenit=mtti10 at 7pt
\font\sevensl=mtsl10 at 7pt
\font\seventt=mttt10 at 7pt     \hyphenchar\seventt=-1
\font\sevencsc=mtcsc10 at 7pt
\font\sevensf=mtss10 at 7pt
\font\seveni=mtmi10 at 7pt      \skewchar\seveni='177
\font\sevenmib=mtmib10 at 7pt   \skewchar\sevenmib='177
\font\sevensy=mtsy10 at 7pt     \skewchar\sevensy='60
\font\sevensyb=mtbsy10 at 7pt   \skewchar\sevensyb='60

\font\eightrm=mtr10 at 8pt
\font\eightbf=mtbx10 at 8pt
\font\eightit=mtti10 at 8pt
\font\eighti=mtmi10 at 8pt      \skewchar\eighti='177
\font\eightmib=mtmib10 at 8pt   \skewchar\eightmib='177
\font\eightsy=mtsy10 at 8pt     \skewchar\eightsy='60
\font\eightsyb=mtbsy10 at 8pt   \skewchar\eightsyb='60
\font\eightsl=mtsl10 at 8pt
\font\eighttt=mttt10 at 8pt     \hyphenchar\eighttt=-1
\font\eightcsc=mtcsc10 at 8pt
\font\eightsf=mtss10 at 8pt

\font\ninerm=mtr10 at 9pt
\font\ninebf=mtbx10 at 9pt
\font\nineit=mtti10 at 9pt
\font\ninei=mtmi10 at 9pt      \skewchar\ninei='177
\font\ninemib=mtmib10 at 9pt   \skewchar\ninemib='177
\font\ninesy=mtsy10 at 9pt     \skewchar\ninesy='60
\font\ninesyb=mtbsy10 at 9pt   \skewchar\ninesyb='60
\font\ninesl=mtsl10 at 9pt
\font\ninett=mttt10 at 9pt     \hyphenchar\ninett=-1
\font\ninecsc=mtcsc10 at 9pt
\font\ninesf=mtss10 at 9pt

\font\tenrm=mtr10
\font\tenbf=mtbx10
\font\tenit=mtti10
\font\teni=mtmi10		\skewchar\teni='177
\font\tenmib=mtmib10	\skewchar\tenmib='177
\font\tensy=mtsy10		\skewchar\tensy='60
\font\tensyb=mtbsy10	\skewchar\tensyb='60
\font\tenex=cmex10
\font\tensl=mtsl10
\font\tentt=mttt10		\hyphenchar\tentt=-1
\font\tencsc=mtcsc10
\font\tensf=mtss10

\font\elevenrm=mtr10 at 11pt
\font\elevenbf=mtbx10 at 11pt
\font\elevenit=mtti10 at 11pt
\font\eleveni=mtmi10 at 11pt      \skewchar\eleveni='177
\font\elevenmib=mtmib10 at 11pt   \skewchar\elevenmib='177
\font\elevensy=mtsy10 at 11pt     \skewchar\elevensy='60
\font\elevensyb=mtbsy10 at 11pt   \skewchar\elevensyb='60
\font\elevensl=mtsl10 at 11pt
\font\eleventt=mttt10 at 11pt     \hyphenchar\eleventt=-1
\font\elevencsc=mtcsc10 at 11pt
\font\elevensf=mtss10 at 11pt

\font\twelverm=mtr10 at 12pt
\font\twelvebf=mtbx10 at 12pt
\font\twelveit=mtti10 at 12pt
\font\twelvesl=mtsl10 at 12pt
\font\twelvett=mttt10 at 12pt     \hyphenchar\twelvett=-1
\font\twelvecsc=mtcsc10 at 12pt
\font\twelvesf=mtss10 at 12pt
\font\twelvei=mtmi10 at 12pt      \skewchar\twelvei='177
\font\twelvemib=mtmib10 at 12pt   \skewchar\twelvemib='177
\font\twelvesy=mtsy10 at 12pt     \skewchar\twelvesy='60
\font\twelvesyb=mtbsy10 at 12pt   \skewchar\twelvesyb='60

\font\fourteenrm=mtr10 at 14pt
\font\fourteenbf=mtbx10 at 14pt
\font\fourteenit=mtti10 at 14pt
\font\fourteeni=mtmi10 at 14pt      \skewchar\fourteeni='177
\font\fourteenmib=mtmib10 at 14pt   \skewchar\fourteenmib='177
\font\fourteensy=mtsy10 at 14pt     \skewchar\fourteensy='60
\font\fourteensyb=mtbsy10 at 14pt   \skewchar\fourteensyb='60
\font\fourteensl=mtsl10 at 14pt
\font\fourteentt=mttt10 at 14pt     \hyphenchar\fourteentt=-1
\font\fourteencsc=mtcsc10 at 14pt
\font\fourteensf=mtss10 at 14pt

\font\seventeenrm=mtr10 at 17pt
\font\seventeenbf=mtbx10 at 17pt
\font\seventeenit=mtti10 at 17pt
\font\seventeeni=mtmi10 at 17pt      \skewchar\seventeeni='177
\font\seventeenmib=mtmib10 at 17pt   \skewchar\seventeenmib='177
\font\seventeensy=mtsy10 at 17pt     \skewchar\seventeensy='60
\font\seventeensyb=mtbsy10 at 17pt   \skewchar\seventeensyb='60
\font\seventeensl=mtsl10 at 17pt
\font\seventeentt=mttt10 at 17pt     \hyphenchar\seventeentt=-1
\font\seventeencsc=mtcsc10 at 17pt
\font\seventeensf=mtss10 at 17pt


\newfam\xmfam
\newfam\ymfam

\font\fivexm=mtxm10 at 5pt
\font\sixxm=mtxm10 at 6pt
\font\sevenxm=mtxm10 at 7pt
\font\eightxm=mtxm10 at 8pt
\font\ninexm=mtxm10 at 9pt
\font\tenxm=mtxm10
\font\elevenxm=mtxm10 at 11pt
\font\twelvexm=mtxm10 at 12pt
\font\fourteenxm=mtxm10 at 14pt
\font\seventeenxm=mtxm10 at 17pt

\font\fiveym=mtym10 at 5pt
\font\sixym=mtym10 at 6pt
\font\sevenym=mtym10 at 7pt
\font\eightym=mtym10 at 8pt
\font\nineym=mtym10 at 9pt
\font\tenym=mtym10
\font\elevenym=mtym10 at 11pt
\font\twelveym=mtym10 at 12pt
\font\fourteenym=mtym10 at 14pt
\font\seventeenym=mtym10 at 17pt
\else
\font\fiverm=cmr5
\font\fivei=cmmi5             \skewchar\fivei='177
\font\fivemib=cmmib10 at 5pt  \skewchar\fivemib='177
\font\fivesy=cmsy5            \skewchar\fivesy='60
\font\fivesyb=cmbsy10 at 5pt  \skewchar\fivesyb='60
\font\fivebf=cmbx5

\font\sixrm=cmr6
\font\sixi=cmmi6             \skewchar\sixi='177
\font\sixmib=cmmib10 at 6pt  \skewchar\sixmib='177
\font\sixsy=cmsy6            \skewchar\sixsy='60
\font\sixsyb=cmbsy10 at 6pt  \skewchar\sixsyb='60
\font\sixbf=cmbx6

\font\sevenrm=cmr7
\font\seveni=cmmi7             \skewchar\seveni='177
\font\sevenmib=cmmib10 at 7pt  \skewchar\sevenmib='177
\font\sevensy=cmsy7            \skewchar\sevensy='60
\font\sevensyb=cmbsy10 at 7pt  \skewchar\sevensyb='60
\font\sevenbf=cmbx7

\font\eightrm=cmr8
\font\eightbf=cmbx8
\font\eightit=cmti8
\font\eighti=cmmi8			\skewchar\eighti='177
\font\eightmib=cmmib10 at 8pt	\skewchar\eightmib='177
\font\eightsy=cmsy8			\skewchar\eightsy='60
\font\eightsyb=cmbsy10 at 8pt	\skewchar\eightsyb='60
\font\eightsl=cmsl8
\font\eighttt=cmtt8			\hyphenchar\eighttt=-1
\font\eightcsc=cmcsc10 at 8pt
\font\eightsf=cmss8

\font\ninerm=cmr9
\font\ninebf=cmbx9
\font\nineit=cmti9
\font\ninei=cmmi9			\skewchar\ninei='177
\font\ninemib=cmmib10 at 9pt	\skewchar\ninemib='177
\font\ninesy=cmsy9			\skewchar\ninesy='60
\font\ninesyb=cmbsy10 at 9pt	\skewchar\ninesyb='60
\font\ninesl=cmsl9
\font\ninett=cmtt9			\hyphenchar\ninett=-1
\font\ninecsc=cmcsc10 at 9pt
\font\ninesf=cmss9

\font\tenrm=cmr10
\font\tenbf=cmbx10
\font\tenit=cmti10
\font\teni=cmmi10		\skewchar\teni='177
\font\tenmib=cmmib10	\skewchar\tenmib='177
\font\tensy=cmsy10		\skewchar\tensy='60
\font\tensyb=cmbsy10	\skewchar\tensyb='60
\font\tenex=cmex10
\font\tensl=cmsl10
\font\tentt=cmtt10		\hyphenchar\tentt=-1
\font\tencsc=cmcsc10
\font\tensf=cmss10

\font\elevenrm=cmr10 scaled \magstephalf
\font\elevenbf=cmbx10 scaled \magstephalf
\font\elevenit=cmti10 scaled \magstephalf
\font\eleveni=cmmi10 scaled \magstephalf	\skewchar\eleveni='177
\font\elevenmib=cmmib10 scaled \magstephalf	\skewchar\elevenmib='177
\font\elevensy=cmsy10 scaled \magstephalf	\skewchar\elevensy='60
\font\elevensyb=cmbsy10 scaled \magstephalf	\skewchar\elevensyb='60
\font\elevensl=cmsl10 scaled \magstephalf
\font\eleventt=cmtt10 scaled \magstephalf	\hyphenchar\eleventt=-1
\font\elevencsc=cmcsc10 scaled \magstephalf
\font\elevensf=cmss10 scaled \magstephalf

\font\twelverm=cmr10 scaled \magstep1
\font\twelvebf=cmbx10 scaled \magstep1
\font\twelvei=cmmi10 scaled \magstep1      \skewchar\twelvei='177
\font\twelvemib=cmmib10 scaled \magstep1   \skewchar\twelvemib='177
\font\twelvesy=cmsy10 scaled \magstep1     \skewchar\twelvesy='60
\font\twelvesyb=cmbsy10 scaled \magstep1   \skewchar\twelvesyb='60

\font\fourteenrm=cmr10 scaled \magstep2
\font\fourteenbf=cmbx10 scaled \magstep2
\font\fourteenit=cmti10 scaled \magstep2
\font\fourteeni=cmmi10 scaled \magstep2		\skewchar\fourteeni='177
\font\fourteenmib=cmmib10 scaled \magstep2	\skewchar\fourteenmib='177
\font\fourteensy=cmsy10 scaled \magstep2	\skewchar\fourteensy='60
\font\fourteensyb=cmbsy10 scaled \magstep2	\skewchar\fourteensyb='60
\font\fourteensl=cmsl10 scaled \magstep2
\font\fourteentt=cmtt10 scaled \magstep2	\hyphenchar\fourteentt=-1
\font\fourteencsc=cmcsc10 scaled \magstep2
\font\fourteensf=cmss10 scaled \magstep2

\font\seventeenrm=cmr10 scaled \magstep3
\font\seventeenbf=cmbx10 scaled \magstep3
\font\seventeenit=cmti10 scaled \magstep3
\font\seventeeni=cmmi10 scaled \magstep3	\skewchar\seventeeni='177
\font\seventeenmib=cmmib10 scaled \magstep3	\skewchar\seventeenmib='177
\font\seventeensy=cmsy10 scaled \magstep3	\skewchar\seventeensy='60
\font\seventeensyb=cmbsy10 scaled \magstep3	\skewchar\seventeensyb='60
\font\seventeensl=cmsl10 scaled \magstep3
\font\seventeentt=cmtt10 scaled \magstep3	\hyphenchar\seventeentt=-1
\font\seventeencsc=cmcsc10 scaled \magstep3
\font\seventeensf=cmss10 scaled \magstep3
\fi

\def\hexnumber#1{\ifcase#1 0\or1\or2\or3\or4\or5\or6\or7\or8\or9\or
  A\or B\or C\or D\or E\or F\fi}

\ifprod@font
  \edef\@xm{\hexnumber\xmfam}
  \edef\@ym{\hexnumber\ymfam}
\fi

\def\makestrut{%
  \setbox\strutbox=\hbox{%
    \vrule height.7\baselineskip depth.3\baselineskip width \z@}%
}

\def\baselinestretch{1}
\newskip\tmp@bls

\def\b@ls#1{
  \tmp@bls=#1\relax
  \baselineskip=#1\relax\makestrut
  \normalbaselineskip=\baselinestretch\tmp@bls
  \normalbaselines
}

\def\nostb@ls#1{
  \normalbaselineskip=#1\relax
  \normalbaselines
  \makestrut
}

%

\newfam\mibfam 
\newfam\sybfam 
\newfam\scfam  
\newfam\sffam  

\def\mit{\fam\@ne}

\def\cal{\fam\tw@}

\def\em{\ifdim\fontdimen1\font>\z@ \rm\else\it\fi}

\textfont3=\tenex
\scriptfont3=\tenex
\scriptscriptfont3=\tenex

\setbox0=\hbox{\tenex B} \p@renwd=\wd0 

\def\eightpoint{
  \def\rm{\fam0\eightrm}%
  \textfont0=\eightrm \scriptfont0=\sixrm \scriptscriptfont0=\fiverm%
  \textfont1=\eighti  \scriptfont1=\sixi  \scriptscriptfont1=\fivei%
  \textfont2=\eightsy \scriptfont2=\sixsy \scriptscriptfont2=\fivesy%
  \textfont\itfam=\eightit\def\it{\fam\itfam\eightit}%
  \ifprod@font
    \scriptfont\itfam=\sixit
      \scriptscriptfont\itfam=\fiveit
  \else
    \scriptfont\itfam=\eightit
      \scriptscriptfont\itfam=\eightit
  \fi
  \textfont\bffam=\eightbf%
    \scriptfont\bffam=\sixbf%
      \scriptscriptfont\bffam=\fivebf%
  \def\bf{\fam\bffam\eightbf}%
  \textfont\slfam=\eightsl\def\sl{\fam\slfam\eightsl}%
  \ifprod@font
    \scriptfont\slfam=\sixsl
      \scriptscriptfont\slfam=\fivesl
  \else
    \scriptfont\slfam=\eightsl
      \scriptscriptfont\slfam=\eightsl
  \fi
  \textfont\ttfam=\eighttt\def\tt{\fam\ttfam\eighttt}%
  \ifprod@font
    \scriptfont\ttfam=\sixtt
      \scriptscriptfont\ttfam=\fivett
  \else
    \scriptfont\ttfam=\eighttt
      \scriptscriptfont\ttfam=\eighttt
  \fi
  \textfont\scfam=\eightcsc\def\sc{\fam\scfam\eightcsc}%
  \ifprod@font
    \scriptfont\scfam=\sixcsc
      \scriptscriptfont\scfam=\fivecsc
  \else
    \scriptfont\scfam=\eightcsc
      \scriptscriptfont\scfam=\eightcsc
  \fi
  \textfont\sffam=\eightsf\def\sf{\fam\sffam\eightsf}%
  \ifprod@font
    \scriptfont\sffam=\sixsf
      \scriptscriptfont\sffam=\fivesf
  \else
    \scriptfont\sffam=\eightsf
      \scriptscriptfont\sffam=\eightsf
  \fi
  \textfont\mibfam=\eightmib
    \scriptfont\mibfam=\sixmib
      \scriptscriptfont\mibfam=\fivemib
  \textfont\sybfam=\eightsyb
    \scriptfont\sybfam=\sixsyb
      \scriptscriptfont\sybfam=\fivesyb
  \ifprod@font
    \textfont\xmfam=\eightxm
      \scriptfont\xmfam=\sixxm
        \scriptscriptfont\xmfam=\fivexm
    \textfont\ymfam=\eightym
      \scriptfont\ymfam=\sixym
        \scriptscriptfont\ymfam=\fiveym
  \fi
  \def\oldstyle{\fam\@ne\eighti}%
  \def\boldstyle{\fam\mibfam\eightmib}%
  \b@ls{10pt}\rm%
}

\def\ninepoint{
  \def\rm{\fam0\ninerm}%
  \textfont0=\ninerm \scriptfont0=\sixrm \scriptscriptfont0=\fiverm%
  \textfont1=\ninei  \scriptfont1=\sixi  \scriptscriptfont1=\fivei%
  \textfont2=\ninesy \scriptfont2=\sixsy \scriptscriptfont2=\fivesy%
  \textfont\itfam=\nineit\def\it{\fam\itfam\nineit}%
  \ifprod@font
    \scriptfont\itfam=\sixit
      \scriptscriptfont\itfam=\fiveit
  \else
    \scriptfont\itfam=\nineit
      \scriptscriptfont\itfam=\nineit
  \fi
  \textfont\bffam=\ninebf%
    \scriptfont\bffam=\sixbf%
      \scriptscriptfont\bffam=\fivebf%
  \def\bf{\fam\bffam\ninebf}%
  \textfont\slfam=\ninesl\def\sl{\fam\slfam\ninesl}%
  \ifprod@font
    \scriptfont\slfam=\sixsl
      \scriptscriptfont\slfam=\fivesl
  \else
    \scriptfont\slfam=\ninesl
      \scriptscriptfont\slfam=\ninesl
  \fi
  \textfont\ttfam=\ninett\def\tt{\fam\ttfam\ninett}%
  \ifprod@font
    \scriptfont\ttfam=\sixtt
      \scriptscriptfont\ttfam=\fivett
  \else
    \scriptfont\ttfam=\ninett
      \scriptscriptfont\ttfam=\ninett
  \fi
  \textfont\scfam=\ninecsc\def\sc{\fam\scfam\ninecsc}%
  \ifprod@font
    \scriptfont\scfam=\sixcsc
      \scriptscriptfont\scfam=\fivecsc
  \else
    \scriptfont\scfam=\ninecsc
      \scriptscriptfont\scfam=\ninecsc
  \fi
  \textfont\sffam=\ninesf\def\sf{\fam\sffam\ninesf}%
  \ifprod@font
    \scriptfont\sffam=\sixsf
      \scriptscriptfont\sffam=\fivesf
  \else
    \scriptfont\sffam=\ninesf
      \scriptscriptfont\sffam=\ninesf
  \fi
  \textfont\mibfam=\ninemib
    \scriptfont\mibfam=\sixmib
      \scriptscriptfont\mibfam=\fivemib
  \textfont\sybfam=\ninesyb
    \scriptfont\sybfam=\sixsyb
      \scriptscriptfont\sybfam=\fivesyb
  \ifprod@font
    \textfont\xmfam=\ninexm
      \scriptfont\xmfam=\sixxm
        \scriptscriptfont\xmfam=\fivexm
    \textfont\ymfam=\nineym
      \scriptfont\ymfam=\sixym
        \scriptscriptfont\ymfam=\fiveym
  \fi
  \def\oldstyle{\fam\@ne\ninei}%
  \def\boldstyle{\fam\mibfam\ninemib}%
  \b@ls{\TextLeading plus \Feathering}\rm%
}

\def\tenpoint{
  \def\rm{\fam0\tenrm}%
  \textfont0=\tenrm \scriptfont0=\sevenrm \scriptscriptfont0=\fiverm%
  \textfont1=\teni  \scriptfont1=\seveni  \scriptscriptfont1=\fivei%
  \textfont2=\tensy \scriptfont2=\sevensy \scriptscriptfont2=\fivesy%
  \textfont\itfam=\tenit\def\it{\fam\itfam\tenit}%
  \ifprod@font
    \scriptfont\itfam=\sevenit
      \scriptscriptfont\itfam=\fiveit
  \else
    \scriptfont\itfam=\tenit
      \scriptscriptfont\itfam=\tenit
  \fi
  \textfont\bffam=\tenbf%
    \scriptfont\bffam=\sevenbf%
      \scriptscriptfont\bffam=\fivebf%
  \def\bf{\fam\bffam\tenbf}%
  \textfont\slfam=\tensl\def\sl{\fam\slfam\tensl}%
  \ifprod@font
    \scriptfont\slfam=\sevensl
      \scriptscriptfont\slfam=\fivesl
  \else
    \scriptfont\slfam=\tensl
      \scriptscriptfont\slfam=\tensl
  \fi
  \textfont\ttfam=\tentt\def\tt{\fam\ttfam\tentt}%
  \ifprod@font
    \scriptfont\ttfam=\seventt
      \scriptscriptfont\ttfam=\fivett
  \else
    \scriptfont\ttfam=\tentt
      \scriptscriptfont\ttfam=\tentt
  \fi
  \textfont\scfam=\tencsc\def\sc{\fam\scfam\tencsc}%
  \ifprod@font
    \scriptfont\scfam=\sevencsc
      \scriptscriptfont\scfam=\fivecsc
  \else
    \scriptfont\scfam=\tencsc
      \scriptscriptfont\scfam=\tencsc
  \fi
  \textfont\sffam=\tensf\def\sf{\fam\sffam\tensf}%
  \ifprod@font
    \scriptfont\sffam=\sevensf
      \scriptscriptfont\sffam=\fivesf
  \else
    \scriptfont\sffam=\tensf
      \scriptscriptfont\sffam=\tensf
  \fi
  \textfont\mibfam=\tenmib
    \scriptfont\mibfam=\sevenmib
      \scriptscriptfont\mibfam=\fivemib
  \textfont\sybfam=\tensyb
    \scriptfont\sybfam=\sevensyb
      \scriptscriptfont\sybfam=\fivesyb
  \ifprod@font
    \textfont\xmfam=\tenxm
      \scriptfont\xmfam=\sevenxm
        \scriptscriptfont\xmfam=\fivexm
    \textfont\ymfam=\tenym
      \scriptfont\ymfam=\sevenym
        \scriptscriptfont\ymfam=\fiveym
  \fi
  \def\oldstyle{\fam\@ne\teni}%
  \def\boldstyle{\fam\mibfam\tenmib}%
  \b@ls{11pt}\rm%
}

\def\elevenpoint{
  \def\rm{\fam0\elevenrm}%
  \textfont0=\elevenrm \scriptfont0=\eightrm \scriptscriptfont0=\sixrm%
  \textfont1=\eleveni  \scriptfont1=\eighti  \scriptscriptfont1=\sixi%
  \textfont2=\elevensy \scriptfont2=\eightsy \scriptscriptfont2=\sixsy%
  \textfont\itfam=\elevenit\def\it{\fam\itfam\elevenit}%
  \ifprod@font
    \scriptfont\itfam=\eightit
      \scriptscriptfont\itfam=\sixit
  \else
    \scriptfont\itfam=\elevenit
      \scriptscriptfont\itfam=\elevenit
  \fi
  \textfont\bffam=\elevenbf%
    \scriptfont\bffam=\eightbf%
      \scriptscriptfont\bffam=\sixbf%
  \def\bf{\fam\bffam\elevenbf}%
  \textfont\slfam=\elevensl\def\sl{\fam\slfam\elevensl}%
  \ifprod@font
    \scriptfont\slfam=\eightsl
      \scriptscriptfont\slfam=\sixsl
  \else
    \scriptfont\slfam=\elevensl
      \scriptscriptfont\slfam=\elevensl
  \fi
  \textfont\ttfam=\eleventt\def\tt{\fam\ttfam\eleventt}%
  \ifprod@font
    \scriptfont\ttfam=\eighttt
      \scriptscriptfont\ttfam=\sixtt
  \else
    \scriptfont\ttfam=\eleventt
      \scriptscriptfont\ttfam=\eleventt
  \fi
  \textfont\scfam=\elevencsc\def\sc{\fam\scfam\elevencsc}%
  \ifprod@font
    \scriptfont\scfam=\eightcsc
      \scriptscriptfont\scfam=\sixcsc
  \else
    \scriptfont\scfam=\elevencsc
      \scriptscriptfont\scfam=\elevencsc
  \fi
  \textfont\sffam=\elevensf\def\sf{\fam\sffam\elevensf}%
  \ifprod@font
    \scriptfont\sffam=\eightsf
      \scriptscriptfont\sffam=\sixsf
  \else
    \scriptfont\sffam=\elevensf
      \scriptscriptfont\sffam=\elevensf
  \fi
  \textfont\mibfam=\elevenmib
    \scriptfont\mibfam=\eightmib
      \scriptscriptfont\mibfam=\sixmib
  \textfont\sybfam=\elevensyb
    \scriptfont\sybfam=\eightsyb
      \scriptscriptfont\sybfam=\sixsyb
  \ifprod@font
    \textfont\xmfam=\elevenxm
      \scriptfont\xmfam=\eightxm
       \scriptscriptfont\xmfam=\sixxm
    \textfont\ymfam=\elevenym
      \scriptfont\ymfam=\eightym
        \scriptscriptfont\ymfam=\sixym
   \fi
  \def\oldstyle{\fam\@ne\eleveni}%
  \def\boldstyle{\fam\mibfam\elevenmib}%
  \b@ls{13pt}\rm%
}

\def\fourteenpoint{
  \def\rm{\fam0\fourteenrm}%
  \textfont0\fourteenrm  \scriptfont0\tenrm  \scriptscriptfont0\sevenrm%
  \textfont1\fourteeni   \scriptfont1\teni   \scriptscriptfont1\seveni%
  \textfont2\fourteensy  \scriptfont2\tensy  \scriptscriptfont2\sevensy%
  \textfont\itfam=\fourteenit\def\it{\fam\itfam\fourteenit}%
  \ifprod@font
    \scriptfont\itfam=\tenit
      \scriptscriptfont\itfam=\sevenit
  \else
    \scriptfont\itfam=\fourteenit
      \scriptscriptfont\itfam=\fourteenit
  \fi
  \textfont\bffam=\fourteenbf%
    \scriptfont\bffam=\tenbf%
      \scriptscriptfont\bffam=\sevenbf%
  \def\bf{\fam\bffam\fourteenbf}%
  \textfont\slfam=\fourteensl\def\sl{\fam\slfam\fourteensl}%
  \ifprod@font
    \scriptfont\slfam=\tensl
      \scriptscriptfont\slfam=\sevensl
  \else
    \scriptfont\slfam=\fourteensl
      \scriptscriptfont\slfam=\fourteensl
  \fi
  \textfont\ttfam=\fourteentt\def\tt{\fam\ttfam\fourteentt}%
  \ifprod@font
    \scriptfont\ttfam=\tentt
      \scriptscriptfont\ttfam=\seventt
  \else
    \scriptfont\ttfam=\fourteentt
      \scriptscriptfont\ttfam=\fourteentt
  \fi
  \textfont\scfam=\fourteencsc\def\sc{\fam\scfam\fourteencsc}%
  \ifprod@font
    \scriptfont\scfam=\tencsc
      \scriptscriptfont\scfam=\sevencsc
  \else
    \scriptfont\scfam=\fourteencsc
      \scriptscriptfont\scfam=\fourteencsc
  \fi
  \textfont\sffam=\fourteensf\def\sf{\fam\sffam\fourteensf}%
  \ifprod@font
    \scriptfont\sffam=\tensf
      \scriptscriptfont\sffam=\sevensf
  \else
    \scriptfont\sffam=\fourteensf
      \scriptscriptfont\sffam=\fourteensf
  \fi
  \textfont\mibfam=\fourteenmib
    \scriptfont\mibfam=\tenmib
      \scriptscriptfont\mibfam=\sevenmib
  \textfont\sybfam=\fourteensyb
    \scriptfont\sybfam=\tensyb
      \scriptscriptfont\sybfam=\sevensyb
  \ifprod@font
    \textfont\xmfam=\fourteenxm
      \scriptfont\xmfam=\tenxm
        \scriptscriptfont\xmfam=\sevenxm
   \textfont\ymfam=\fourteenym
      \scriptfont\ymfam=\tenym
        \scriptscriptfont\ymfam=\sevenym
  \fi
  \def\oldstyle{\fam\@ne\fourteeni}%
  \def\boldstyle{\fam\mibfam\fourteenmib}%
  \b@ls{17pt}\rm%
}

\def\seventeenpoint{
  \def\rm{\fam0\seventeenrm}%
  \textfont0\seventeenrm  \scriptfont0\twelverm  \scriptscriptfont0\tenrm%
  \textfont1\seventeeni   \scriptfont1\twelvei   \scriptscriptfont1\teni%
  \textfont2\seventeensy  \scriptfont2\twelvesy  \scriptscriptfont2\tensy%
  \textfont\itfam=\seventeenit\def\it{\fam\itfam\seventeenit}%
  \ifprod@font
    \scriptfont\itfam=\twelveit
      \scriptscriptfont\itfam=\tenit
  \else
    \scriptfont\itfam=\seventeenit
      \scriptscriptfont\itfam=\seventeenit
  \fi
  \textfont\bffam=\seventeenbf%
    \scriptfont\bffam=\twelvebf%
      \scriptscriptfont\bffam=\tenbf%
  \def\bf{\fam\bffam\seventeenbf}%
  \textfont\slfam=\seventeensl\def\sl{\fam\slfam\seventeensl}%
  \ifprod@font
    \scriptfont\slfam=\twelvesl
      \scriptscriptfont\slfam=\tensl
  \else
    \scriptfont\slfam=\seventeensl
      \scriptscriptfont\slfam=\seventeensl
  \fi
  \textfont\ttfam=\seventeentt\def\tt{\fam\ttfam\seventeentt}%
  \ifprod@font
    \scriptfont\ttfam=\twelvett
      \scriptscriptfont\ttfam=\tentt
  \else
    \scriptfont\ttfam=\seventeentt
      \scriptscriptfont\ttfam=\seventeentt
  \fi
  \textfont\scfam=\seventeencsc\def\sc{\fam\scfam\seventeencsc}%
  \ifprod@font
    \scriptfont\scfam=\twelvecsc
      \scriptscriptfont\scfam=\tencsc
  \else
    \scriptfont\scfam=\seventeencsc
      \scriptscriptfont\scfam=\seventeencsc
  \fi
  \textfont\sffam=\seventeensf\def\sf{\fam\sffam\seventeensf}%
  \ifprod@font
    \scriptfont\sffam=\twelvesf
      \scriptscriptfont\sffam=\tensf
  \else
    \scriptfont\sffam=\seventeensf
      \scriptscriptfont\sffam=\seventeensf
  \fi
  \textfont\mibfam=\seventeenmib
    \scriptfont\mibfam=\twelvemib
      \scriptscriptfont\mibfam=\tenmib
  \textfont\sybfam=\seventeensyb
    \scriptfont\sybfam=\twelvesyb
      \scriptscriptfont\sybfam=\tensyb
  \ifprod@font
    \textfont\xmfam=\seventeenxm
      \scriptfont\xmfam=\twelvexm
        \scriptscriptfont\xmfam=\tenxm
    \textfont\ymfam=\seventeenym
      \scriptfont\ymfam=\twelveym
        \scriptscriptfont\ymfam=\tenym
  \fi
  \def\oldstyle{\fam\@ne\seventeeni}%
  \def\boldstyle{\fam\mibfam\seventeenmib}%
  \b@ls{20pt}\rm%
}

\lineskip=1pt      \normallineskip=\lineskip
\lineskiplimit=\z@ \normallineskiplimit=\lineskiplimit



\def\la{\mathrel{\mathchoice {\vcenter{\offinterlineskip\halign{\hfil
$\displaystyle##$\hfil\cr<\cr\sim\cr}}}
{\vcenter{\offinterlineskip\halign{\hfil$\textstyle##$\hfil\cr
<\cr\sim\cr}}}
{\vcenter{\offinterlineskip\halign{\hfil$\scriptstyle##$\hfil\cr
<\cr\sim\cr}}}
{\vcenter{\offinterlineskip\halign{\hfil$\scriptscriptstyle##$\hfil\cr
<\cr\sim\cr}}}}}

\def\ga{\mathrel{\mathchoice {\vcenter{\offinterlineskip\halign{\hfil
$\displaystyle##$\hfil\cr>\cr\sim\cr}}}
{\vcenter{\offinterlineskip\halign{\hfil$\textstyle##$\hfil\cr
>\cr\sim\cr}}}
{\vcenter{\offinterlineskip\halign{\hfil$\scriptstyle##$\hfil\cr
>\cr\sim\cr}}}
{\vcenter{\offinterlineskip\halign{\hfil$\scriptscriptstyle##$\hfil\cr
>\cr\sim\cr}}}}}

\def\getsto{\mathrel{\mathchoice {\vcenter{\offinterlineskip
\halign{\hfil
$\displaystyle##$\hfil\cr\gets\cr\to\cr}}}
{\vcenter{\offinterlineskip\halign{\hfil$\textstyle##$\hfil\cr\gets
\cr\to\cr}}}
{\vcenter{\offinterlineskip\halign{\hfil$\scriptstyle##$\hfil\cr\gets
\cr\to\cr}}}
{\vcenter{\offinterlineskip\halign{\hfil$\scriptscriptstyle##$\hfil\cr
\gets\cr\to\cr}}}}}

\def\lid{\mathrel{\mathchoice {\vcenter{\offinterlineskip\halign{\hfil
$\displaystyle##$\hfil\cr<\cr\noalign{\vskip1.2pt}=\cr}}}
{\vcenter{\offinterlineskip\halign{\hfil$\textstyle##$\hfil\cr<\cr
\noalign{\vskip1.2pt}=\cr}}}
{\vcenter{\offinterlineskip\halign{\hfil$\scriptstyle##$\hfil\cr<\cr
\noalign{\vskip1pt}=\cr}}}
{\vcenter{\offinterlineskip\halign{\hfil$\scriptscriptstyle##$\hfil\cr
<\cr
\noalign{\vskip0.9pt}=\cr}}}}}

\def\gid{\mathrel{\mathchoice {\vcenter{\offinterlineskip\halign{\hfil
$\displaystyle##$\hfil\cr>\cr\noalign{\vskip1.2pt}=\cr}}}
{\vcenter{\offinterlineskip\halign{\hfil$\textstyle##$\hfil\cr>\cr
\noalign{\vskip1.2pt}=\cr}}}
{\vcenter{\offinterlineskip\halign{\hfil$\scriptstyle##$\hfil\cr>\cr
\noalign{\vskip1pt}=\cr}}}
{\vcenter{\offinterlineskip\halign{\hfil$\scriptscriptstyle##$\hfil\cr
>\cr
\noalign{\vskip0.9pt}=\cr}}}}}

\def\grole{\mathrel{\mathchoice {\vcenter{\offinterlineskip\halign{\hfil
$\displaystyle##$\hfil\cr>\cr\noalign{\vskip-1.5pt}<\cr}}}
{\vcenter{\offinterlineskip\halign{\hfil$\textstyle##$\hfil\cr
>\cr\noalign{\vskip-1.5pt}<\cr}}}
{\vcenter{\offinterlineskip\halign{\hfil$\scriptstyle##$\hfil\cr
>\cr\noalign{\vskip-1pt}<\cr}}}
{\vcenter{\offinterlineskip\halign{\hfil$\scriptscriptstyle##$\hfil\cr
>\cr\noalign{\vskip-0.5pt}<\cr}}}}}

\def\leogr{\mathrel{\mathchoice {\vcenter{\offinterlineskip\halign{\hfil
$\displaystyle##$\hfil\cr<\cr\noalign{\vskip-1.5pt}>\cr}}}
{\vcenter{\offinterlineskip\halign{\hfil$\textstyle##$\hfil\cr
<\cr\noalign{\vskip-1.5pt}>\cr}}}
{\vcenter{\offinterlineskip\halign{\hfil$\scriptstyle##$\hfil\cr
<\cr\noalign{\vskip-1pt}>\cr}}}
{\vcenter{\offinterlineskip\halign{\hfil$\scriptscriptstyle##$\hfil\cr
<\cr\noalign{\vskip-0.5pt}>\cr}}}}}

\def\loa{\mathrel{\mathchoice {\vcenter{\offinterlineskip\halign{\hfil
$\displaystyle##$\hfil\cr<\cr\approx\cr}}}
{\vcenter{\offinterlineskip\halign{\hfil$\textstyle##$\hfil\cr
<\cr\approx\cr}}}
{\vcenter{\offinterlineskip\halign{\hfil$\scriptstyle##$\hfil\cr
<\cr\approx\cr}}}
{\vcenter{\offinterlineskip\halign{\hfil$\scriptscriptstyle##$\hfil\cr
<\cr\approx\cr}}}}}

\def\goa{\mathrel{\mathchoice {\vcenter{\offinterlineskip\halign{\hfil
$\displaystyle##$\hfil\cr>\cr\approx\cr}}}
{\vcenter{\offinterlineskip\halign{\hfil$\textstyle##$\hfil\cr
>\cr\approx\cr}}}
{\vcenter{\offinterlineskip\halign{\hfil$\scriptstyle##$\hfil\cr
>\cr\approx\cr}}}
{\vcenter{\offinterlineskip\halign{\hfil$\scriptscriptstyle##$\hfil\cr
>\cr\approx\cr}}}}}

\def\diameter{{\ifmmode\mathchoice
{\ooalign{\hfil\hbox{$\displaystyle/$}\hfil\crcr
{\hbox{$\displaystyle\mathchar"20D$}}}}
{\ooalign{\hfil\hbox{$\textstyle/$}\hfil\crcr
{\hbox{$\textstyle\mathchar"20D$}}}}
{\ooalign{\hfil\hbox{$\scriptstyle/$}\hfil\crcr
{\hbox{$\scriptstyle\mathchar"20D$}}}}
{\ooalign{\hfil\hbox{$\scriptscriptstyle/$}\hfil\crcr
{\hbox{$\scriptscriptstyle\mathchar"20D$}}}}
\else{\ooalign{\hfil/\hfil\crcr\mathhexbox20D}}%
\fi}}

\def\sq{\ifmmode\squareforqed\else{\unskip\nobreak\hfil
\penalty50\hskip1em\null\nobreak\hfil\squareforqed
\parfillskip=0pt\finalhyphendemerits=0\endgraf}\fi}
\def\squareforqed{\hbox{\rlap{$\sqcap$}$\sqcup$}}


\def\bbbc{{\mathchoice {\setbox0=\hbox{$\displaystyle\rm C$}\hbox{\hbox
to0pt{\kern0.4\wd0\vrule height0.9\ht0\hss}\box0}}
{\setbox0=\hbox{$\textstyle\rm C$}\hbox{\hbox
to0pt{\kern0.4\wd0\vrule height0.9\ht0\hss}\box0}}
{\setbox0=\hbox{$\scriptstyle\rm C$}\hbox{\hbox
to0pt{\kern0.4\wd0\vrule height0.9\ht0\hss}\box0}}
{\setbox0=\hbox{$\scriptscriptstyle\rm C$}\hbox{\hbox
to0pt{\kern0.4\wd0\vrule height0.9\ht0\hss}\box0}}}}
\def\bbbq{{\mathchoice {\setbox0=\hbox{$\displaystyle\rm
Q$}\hbox{\raise
0.15\ht0\hbox to0pt{\kern0.4\wd0\vrule height0.8\ht0\hss}\box0}}
{\setbox0=\hbox{$\textstyle\rm Q$}\hbox{\raise
0.15\ht0\hbox to0pt{\kern0.4\wd0\vrule height0.8\ht0\hss}\box0}}
{\setbox0=\hbox{$\scriptstyle\rm Q$}\hbox{\raise
0.15\ht0\hbox to0pt{\kern0.4\wd0\vrule height0.7\ht0\hss}\box0}}
{\setbox0=\hbox{$\scriptscriptstyle\rm Q$}\hbox{\raise
0.15\ht0\hbox to0pt{\kern0.4\wd0\vrule height0.7\ht0\hss}\box0}}}}
\def\bbbt{{\mathchoice {\setbox0=\hbox{$\displaystyle\rm
T$}\hbox{\hbox to0pt{\kern0.3\wd0\vrule height0.9\ht0\hss}\box0}}
{\setbox0=\hbox{$\textstyle\rm T$}\hbox{\hbox
to0pt{\kern0.3\wd0\vrule height0.9\ht0\hss}\box0}}
{\setbox0=\hbox{$\scriptstyle\rm T$}\hbox{\hbox
to0pt{\kern0.3\wd0\vrule height0.9\ht0\hss}\box0}}
{\setbox0=\hbox{$\scriptscriptstyle\rm T$}\hbox{\hbox
to0pt{\kern0.3\wd0\vrule height0.9\ht0\hss}\box0}}}}
\def\bbbs{{\mathchoice
{\setbox0=\hbox{$\displaystyle     \rm S$}\hbox{\raise0.5\ht0\hbox
to0pt{\kern0.35\wd0\vrule height0.45\ht0\hss}\hbox
to0pt{\kern0.55\wd0\vrule height0.5\ht0\hss}\box0}}
{\setbox0=\hbox{$\textstyle        \rm S$}\hbox{\raise0.5\ht0\hbox
to0pt{\kern0.35\wd0\vrule height0.45\ht0\hss}\hbox
to0pt{\kern0.55\wd0\vrule height0.5\ht0\hss}\box0}}
{\setbox0=\hbox{$\scriptstyle      \rm S$}\hbox{\raise0.5\ht0\hbox
to0pt{\kern0.35\wd0\vrule height0.45\ht0\hss}\raise0.05\ht0\hbox
to0pt{\kern0.5\wd0\vrule height0.45\ht0\hss}\box0}}
{\setbox0=\hbox{$\scriptscriptstyle\rm S$}\hbox{\raise0.5\ht0\hbox
to0pt{\kern0.4\wd0\vrule height0.45\ht0\hss}\raise0.05\ht0\hbox
to0pt{\kern0.55\wd0\vrule height0.45\ht0\hss}\box0}}}}
\def\bbbz{{\mathchoice {\hbox{$\sf\textstyle Z\kern-0.4em Z$}}
{\hbox{$\sf\textstyle Z\kern-0.4em Z$}}
{\hbox{$\sf\scriptstyle Z\kern-0.3em Z$}}
{\hbox{$\sf\scriptscriptstyle Z\kern-0.2em Z$}}}}


\ifprod@font
  \mathchardef\la="3\@xm2E
  \mathchardef\getsto="3\@xm1C
  \mathchardef\lid="3\@xm35
  \mathchardef\grole="3\@xm3F
  \mathchardef\loa="3\@xm2F
  \mathchardef\ga="3\@xm26
  \mathchardef\gid="3\@xm3D
  \mathchardef\leogr="3\@xm37
  \mathchardef\goa="3\@xm27
  \mathchardef\sq="0\@xm03
%
%
\def\diameter{{%
  \ifmmode
    \mathchoice
    {\ooalign{\hfil\hbox{$\displaystyle/$}\hfil\crcr
    {\lower.2ex\hbox{$\displaystyle\mathchar"20D$}}}}%
    {\ooalign{\hfil\hbox{$\textstyle/$}\hfil\crcr
    {\lower.2ex\hbox{$\textstyle\mathchar"20D$}}}}%
    {\ooalign{\hfil\hbox{$\scriptstyle/$}\hfil\crcr
    {\lower.1ex\hbox{$\scriptstyle\mathchar"20D$}}}}%
    {\ooalign{\hfil\hbox{$\scriptscriptstyle/$}\hfil\crcr
    {\lower.1ex\hbox{$\scriptscriptstyle\mathchar"20D$}}}}%
  \else
    {\ooalign{\hfil/\hfil\crcr\lower.2ex\hbox{\mathhexbox20D}}}%
  \fi
}}
%
%

\def\bbbc{{\Bbb{C}}}
\def\bbbq{{\Bbb{Q}}}
\def\bbbt{{\Bbb{T}}}
\def\bbbs{{\Bbb{S}}}
\def\bbbz{{\Bbb{Z}}}
\fi


\ifprod@font
\mathchardef\boxdot="2\@xm00
\mathchardef\boxplus="2\@xm01
\mathchardef\boxtimes="2\@xm02
\mathchardef\square="0\@xm03
\mathchardef\blacksquare="0\@xm04
\mathchardef\centerdot="2\@xm05
\mathchardef\lozenge="0\@xm06
\mathchardef\blacklozenge="0\@xm07
\mathchardef\circlearrowright="3\@xm08
\mathchardef\circlearrowleft="3\@xm09
\mathchardef\rightleftharpoons="3\@xm0A
\mathchardef\leftrightharpoons="3\@xm0B
\mathchardef\boxminus="2\@xm0C
\mathchardef\Vdash="3\@xm0D
\mathchardef\Vvdash="3\@xm0E
\mathchardef\vDash="3\@xm0F
\mathchardef\twoheadrightarrow="3\@xm10
\mathchardef\twoheadleftarrow="3\@xm11
\mathchardef\leftleftarrows="3\@xm12
\mathchardef\rightrightarrows="3\@xm13
\mathchardef\upuparrows="3\@xm14
\mathchardef\downdownarrows="3\@xm15
\mathchardef\upharpoonright="3\@xm16

\mathchardef\downharpoonright="3\@xm17
\mathchardef\upharpoonleft="3\@xm18
\mathchardef\downharpoonleft="3\@xm19
\mathchardef\rightarrowtail="3\@xm1A
\mathchardef\leftarrowtail="3\@xm1B
\mathchardef\leftrightarrows="3\@xm1C
\mathchardef\rightleftarrows="3\@xm1D
\mathchardef\Lsh="3\@xm1E
\mathchardef\Rsh="3\@xm1F
\mathchardef\rightsquigarrow="3\@xm20
\mathchardef\leftrightsquigarrow="3\@xm21
\mathchardef\looparrowleft="3\@xm22
\mathchardef\looparrowright="3\@xm23
\mathchardef\circeq="3\@xm24
\mathchardef\succsim="3\@xm25
\mathchardef\gtrsim="3\@xm26
\mathchardef\gtrapprox="3\@xm27
\mathchardef\multimap="3\@xm28
\mathchardef\therefore="3\@xm29
\mathchardef\because="3\@xm2A
\mathchardef\doteqdot="3\@xm2B

\mathchardef\triangleq="3\@xm2C
\mathchardef\precsim="3\@xm2D
\mathchardef\lesssim="3\@xm2E
\mathchardef\lessapprox="3\@xm2F
\mathchardef\eqslantless="3\@xm30
\mathchardef\eqslantgtr="3\@xm31
\mathchardef\curlyeqprec="3\@xm32
\mathchardef\curlyeqsucc="3\@xm33
\mathchardef\preccurlyeq="3\@xm34
\mathchardef\leqq="3\@xm35
\mathchardef\leqslant="3\@xm36
\mathchardef\lessgtr="3\@xm37
\mathchardef\backprime="0\@xm38
\mathchardef\risingdotseq="3\@xm3A
\mathchardef\fallingdotseq="3\@xm3B
\mathchardef\succcurlyeq="3\@xm3C
\mathchardef\geqq="3\@xm3D
\mathchardef\geqslant="3\@xm3E
\mathchardef\gtrless="3\@xm3F
\mathchardef\sqsubset="3\@xm40
\mathchardef\sqsupset="3\@xm41
\mathchardef\vartriangleright="3\@xm42
\mathchardef\vartriangleleft="3\@xm43
\mathchardef\trianglerighteq="3\@xm44
\mathchardef\trianglelefteq="3\@xm45
\mathchardef\bigstar="0\@xm46
\mathchardef\between="3\@xm47
\mathchardef\blacktriangledown="0\@xm48
\mathchardef\blacktriangleright="3\@xm49
\mathchardef\blacktriangleleft="3\@xm4A
\mathchardef\vartriangle="0\@xm4D
\mathchardef\blacktriangle="0\@xm4E
\mathchardef\triangledown="0\@xm4F
\mathchardef\eqcirc="3\@xm50
\mathchardef\lesseqgtr="3\@xm51
\mathchardef\gtreqless="3\@xm52
\mathchardef\lesseqqgtr="3\@xm53
\mathchardef\gtreqqless="3\@xm54
\mathchardef\Rrightarrow="3\@xm56
\mathchardef\Lleftarrow="3\@xm57
\mathchardef\veebar="2\@xm59
\mathchardef\barwedge="2\@xm5A
\mathchardef\doublebarwedge="2\@xm5B
\mathchardef\angle="0\@xm5C
\mathchardef\measuredangle="0\@xm5D
\mathchardef\sphericalangle="0\@xm5E
\mathchardef\varpropto="3\@xm5F
\mathchardef\smallsmile="3\@xm60
\mathchardef\smallfrown="3\@xm61
\mathchardef\Subset="3\@xm62
\mathchardef\Supset="3\@xm63
\mathchardef\Cup="2\@xm64

\mathchardef\Cap="2\@xm65

\mathchardef\curlywedge="2\@xm66
\mathchardef\curlyvee="2\@xm67
\mathchardef\leftthreetimes="2\@xm68
\mathchardef\rightthreetimes="2\@xm69
\mathchardef\subseteqq="3\@xm6A
\mathchardef\supseteqq="3\@xm6B
\mathchardef\bumpeq="3\@xm6C
\mathchardef\Bumpeq="3\@xm6D
\mathchardef\lll="3\@xm6E

\mathchardef\ggg="3\@xm6F

\mathchardef\circledS="0\@xm73
\mathchardef\pitchfork="3\@xm74
\mathchardef\dotplus="2\@xm75
\mathchardef\backsim="3\@xm76
\mathchardef\backsimeq="3\@xm77
\mathchardef\complement="0\@xm7B
\mathchardef\intercal="2\@xm7C
\mathchardef\circledcirc="2\@xm7D
\mathchardef\circledast="2\@xm7E
\mathchardef\circleddash="2\@xm7F
\def\ulcorner{\delimiter"4\@xm70\@xm70 }
\def\urcorner{\delimiter"5\@xm71\@xm71 }
\def\llcorner{\delimiter"4\@xm78\@xm78 }
\def\lrcorner{\delimiter"5\@xm79\@xm79 }
\def\yen{\mathhexbox\@xm55 }
\def\checkmark{\mathhexbox\@xm58 }
\def\circledR{\mathhexbox\@xm72 }
\def\maltese{\mathhexbox\@xm7A }
\mathchardef\lvertneqq="3\@ym00
\mathchardef\gvertneqq="3\@ym01
\mathchardef\nleq="3\@ym02
\mathchardef\ngeq="3\@ym03
\mathchardef\nless="3\@ym04
\mathchardef\ngtr="3\@ym05
\mathchardef\nprec="3\@ym06
\mathchardef\nsucc="3\@ym07
\mathchardef\lneqq="3\@ym08
\mathchardef\gneqq="3\@ym09
\mathchardef\nleqslant="3\@ym0A
\mathchardef\ngeqslant="3\@ym0B
\mathchardef\lneq="3\@ym0C
\mathchardef\gneq="3\@ym0D
\mathchardef\npreceq="3\@ym0E
\mathchardef\nsucceq="3\@ym0F
\mathchardef\precnsim="3\@ym10
\mathchardef\succnsim="3\@ym11
\mathchardef\lnsim="3\@ym12
\mathchardef\gnsim="3\@ym13
\mathchardef\nleqq="3\@ym14
\mathchardef\ngeqq="3\@ym15
\mathchardef\precneqq="3\@ym16
\mathchardef\succneqq="3\@ym17
\mathchardef\precnapprox="3\@ym18
\mathchardef\succnapprox="3\@ym19
\mathchardef\lnapprox="3\@ym1A
\mathchardef\gnapprox="3\@ym1B
\mathchardef\nsim="3\@ym1C
\mathchardef\ncong="3\@ym1D

\mathchardef\varsubsetneq="3\@ym20
\mathchardef\varsupsetneq="3\@ym21
\mathchardef\nsubseteqq="3\@ym22
\mathchardef\nsupseteqq="3\@ym23
\mathchardef\subsetneqq="3\@ym24
\mathchardef\supsetneqq="3\@ym25
\mathchardef\varsubsetneqq="3\@ym26
\mathchardef\varsupsetneqq="3\@ym27
\mathchardef\subsetneq="3\@ym28
\mathchardef\supsetneq="3\@ym29
\mathchardef\nsubseteq="3\@ym2A
\mathchardef\nsupseteq="3\@ym2B
\mathchardef\nparallel="3\@ym2C
\mathchardef\nmid="3\@ym2D
\mathchardef\nshortmid="3\@ym2E
\mathchardef\nshortparallel="3\@ym2F
\mathchardef\nvdash="3\@ym30
\mathchardef\nVdash="3\@ym31
\mathchardef\nvDash="3\@ym32
\mathchardef\nVDash="3\@ym33
\mathchardef\ntrianglerighteq="3\@ym34
\mathchardef\ntrianglelefteq="3\@ym35
\mathchardef\ntriangleleft="3\@ym36
\mathchardef\ntriangleright="3\@ym37
\mathchardef\nleftarrow="3\@ym38
\mathchardef\nrightarrow="3\@ym39
\mathchardef\nLeftarrow="3\@ym3A
\mathchardef\nRightarrow="3\@ym3B
\mathchardef\nLeftrightarrow="3\@ym3C
\mathchardef\nleftrightarrow="3\@ym3D
\mathchardef\divideontimes="2\@ym3E
\mathchardef\varnothing="0\@ym3F
\mathchardef\nexists="0\@ym40
\mathchardef\mho="0\@ym66
\mathchardef\eth="0\@ym67
\mathchardef\eqsim="3\@ym68
\mathchardef\beth="0\@ym69
\mathchardef\gimel="0\@ym6A
\mathchardef\daleth="0\@ym6B
\mathchardef\lessdot="3\@ym6C
\mathchardef\gtrdot="3\@ym6D
\mathchardef\ltimes="2\@ym6E
\mathchardef\rtimes="2\@ym6F
\mathchardef\shortmid="3\@ym70
\mathchardef\shortparallel="3\@ym71
\mathchardef\smallsetminus="2\@ym72
\mathchardef\thicksim="3\@ym73
\mathchardef\thickapprox="3\@ym74
\mathchardef\approxeq="3\@ym75
\mathchardef\succapprox="3\@ym76
\mathchardef\precapprox="3\@ym77
\mathchardef\curvearrowleft="3\@ym78
\mathchardef\curvearrowright="3\@ym79
\mathchardef\digamma="0\@ym7A
\mathchardef\varkappa="0\@ym7B
\mathchardef\hslash="0\@ym7D
\mathchardef\hbar="0\@ym7E
\mathchardef\backepsilon="3\@ym7F


\def\Bbb{\ifmmode\let\next\Bbb@\else
\def\next{\errmessage{Use \string\Bbb\space only in math mode}}\fi\next}
\def\Bbb@#1{{\Bbb@@{#1}}}
\def\Bbb@@#1{\fam\ymfam#1}
\fi


\def\Nulle{0} 
\def\Afe{1}   
\def\Hae{2}   
\def\Hbe{3}   
\def\Hce{4}   
\def\Hde{5}   


\newcount\LastMac       \LastMac=\Nulle

\newskip\half      \half=5.5pt plus 1.5pt minus 2.25pt
\newskip\one       \one=11pt plus 3pt minus 5.5pt
\newskip\onehalf   \onehalf=16.5pt plus 5.5pt minus 8.25pt
\newskip\two       \two=22pt plus 5.5pt minus 11pt

\def\Half{\addvspace{\half}}
\def\One{\addvspace{\one}}
\def\OneHalf{\addvspace{\onehalf}}
\def\Two{\addvspace{\two}}


\def\Raggedright{
  \rightskip=\z@ plus \hsize\relax
}

\def\Fullout{
  \rightskip=\z@\relax
}

\def\Hang#1#2{
  \hangindent=#1%
  \hangafter=#2\relax
}


\newif\ifsp@page
\def\pagestyle#1{\csname ps@#1\endcsname}
\def\thispagestyle#1{\global\sp@pagetrue\gdef\sp@type{#1}}

\def\ps@titlepage{%
  \def\@oddhead{\eightpoint\noindent \the\CatchLine
    \ifprod@font\else\qquad Printed\ \today\fi \hfil}%
  \let\@evenhead=\@oddhead
}

\def\ps@headings{%
  \def\@oddhead{\elevenpoint\it\noindent
    \hfill\the\RightHeader\hskip1.5em\rm\folio}%
  \def\@evenhead{\elevenpoint\noindent
    \folio\hskip1.5em\it\the\LeftHeader\hfill}%
}

\def\ps@plate{%
  \def\@oddhead{\eightpoint\noindent\plt@cap\hfil}%
  \def\@evenhead{\eightpoint\noindent\plt@cap\hfil}%
}



\def\title#1{
  \bgroup
    \vbox to 8pt{\vss}%
    \seventeenpoint
    \Raggedright
    \noindent \strut{\bf #1}\par
  \egroup
}

\def\author#1{
  \bgroup
    \ifnum\LastMac=\Afe \OneHalf\else \vskip 21pt\fi
    \fourteenpoint
    \Raggedright
    \noindent \strut #1\par
    \vskip 3pt%
  \egroup
}

\def\affiliation#1{
  \bgroup
    \vskip -4pt%
    \eightpoint
    \Raggedright
    \noindent \strut {\it #1}\par
  \egroup
  \LastMac=\Afe\relax
}

\def\acceptedline#1{
  \bgroup
    \Two
    \eightpoint
    \Raggedright
    \noindent \strut #1\par
  \egroup
}

\long\def\abstract#1{%
  \bgroup
    \vskip 20pt%
    \everypar{\Hang{11pc}{0}}%
    \noindent{\ninebf ABSTRACT}\par
    \tenpoint
    \Fullout
    \noindent #1\par
  \egroup
}

\long\def\keywords#1{
  \bgroup
    \Half
    \everypar{\Hang{11pc}{0}}%
    \tenpoint
    \Fullout
    \noindent\hbox{\bf Key words:}\ #1\par
  \egroup
}


\def\maketitle{%
  \EndOpening
  \ifsinglecol \else \MakePage\fi
}


\def\pageoffset#1#2{\hoffset=#1\relax\voffset=#2\relax}


\def\Autonumber{
  \global\AutoNumbertrue  
}

\newif\ifAutoNumber \AutoNumberfalse
\newcount\Sec        
\newcount\SecSec
\newcount\SecSecSec

\Sec=\z@

\def\:{\let\@sptoken= } \:  
\def\:{\@xifnch} \expandafter\def\: {\futurelet\@tempc\@ifnch}

\def\@ifnextchar#1#2#3{%
  \let\@tempMACe #1%
  \def\@tempMACa{#2}%
  \def\@tempMACb{#3}%
  \futurelet \@tempMACc\@ifnch%
}

\def\@ifnch{%
\ifx \@tempMACc \@sptoken%
  \let\@tempMACd\@xifnch%
\else%
  \ifx \@tempMACc \@tempMACe%
    \let\@tempMACd\@tempMACa%
  \else%
    \let\@tempMACd\@tempMACb%
  \fi%
\fi%
\@tempMACd%
}

\def\@ifstar#1#2{\@ifnextchar *{\def\@tempMACa*{#1}\@tempMACa}{#2}}

\newskip\@tempskipb

\def\addvspace#1{%
  \ifvmode\else \endgraf\fi%
  \ifdim\lastskip=\z@%
    \vskip #1\relax%
  \else%
    \@tempskipb#1\relax\@xaddvskip%
  \fi%
}

\def\@xaddvskip{%
  \ifdim\lastskip<\@tempskipb%
    \vskip-\lastskip%
    \vskip\@tempskipb\relax%
  \else%
    \ifdim\@tempskipb<\z@%
      \ifdim\lastskip<\z@ \else%
        \advance\@tempskipb\lastskip%
        \vskip-\lastskip\vskip\@tempskipb%
      \fi%
    \fi%
  \fi%
}

\newskip\@tmpSKIP

\def\addpen#1{%
  \ifvmode
    \if@nobreak
    \else
      \ifdim\lastskip=\z@
        \penalty#1\relax
      \else
        \@tmpSKIP=\lastskip
        \vskip -\lastskip
        \penalty#1\vskip\@tmpSKIP
      \fi
    \fi
  \fi
}

\newcount\@clubpen   \@clubpen=\clubpenalty
\newif\if@nobreak    \@nobreakfalse

\def\@noafterindent{%
  \global\@nobreaktrue
  \everypar{\if@nobreak
              \global\@nobreakfalse
              \clubpenalty \@M
              {\setbox\z@\lastbox}%
              \LastMac=\Nulle\relax%
            \else
              \clubpenalty \@clubpen
              \everypar{}%
            \fi}
}

\newcount\gds@cbrk   \gds@cbrk=-300

\def\@nohdbrk{\interlinepenalty \@M\relax}

\let\@par=\par
\def\@restorepar{\def\par{\@par}}

\newif\if@endpe   \@endpefalse
 
\def\@doendpe{\@endpetrue \@nobreakfalse \LastMac=\Nulle\relax%
     \def\par{\@restorepar\everypar{}\par\@endpefalse}%
              \everypar{\setbox\z@\lastbox\everypar{}\@endpefalse}%
}

\def\section{\@ifstar{\@ssection}{\@section}}

\def\@section#1{
  \if@nobreak
    \everypar{}%
    \ifnum\LastMac=\Hae \addvspace{\half}\fi
  \else
    \addpen{\gds@cbrk}%
    \addvspace{\two}%
  \fi
  \bgroup
    \ninepoint\bf
    \Raggedright
    \ifAutoNumber
      \global\advance\Sec \@ne
      \noindent\@nohdbrk\number\Sec\hskip 1pc \uppercase{#1}\par
      \global\SecSec=\z@
    \else
      \noindent\@nohdbrk\uppercase{#1}\par
    \fi
  \egroup
  \nobreak
  \vskip\half
  \nobreak
  \@noafterindent
  \LastMac=\Hae\relax
}

\def\@ssection#1{
  \if@nobreak
    \everypar{}%
    \ifnum\LastMac=\Hae \addvspace{\half}\fi
  \else
    \addpen{\gds@cbrk}%
    \addvspace{\two}%
  \fi
  \bgroup
    \ninepoint\bf
    \Raggedright
    \noindent\@nohdbrk\uppercase{#1}\par
  \egroup
  \nobreak
  \vskip\half
  \nobreak
  \@noafterindent
  \LastMac=\Hae\relax
}

\def\subsection#1{
  \if@nobreak
    \everypar{}%
    \ifnum\LastMac=\Hae \addvspace{1pt plus 1pt minus .5pt}\fi
  \else
    \addpen{\gds@cbrk}%
    \addvspace{\onehalf}%
  \fi
  \bgroup
    \ninepoint\bf
    \Raggedright
    \ifAutoNumber
      \global\advance\SecSec \@ne
      \noindent\@nohdbrk\number\Sec.\number\SecSec \hskip 1pc\relax #1\par
      \global\SecSecSec=\z@
    \else
      \noindent\@nohdbrk #1\par
    \fi
  \egroup
  \nobreak
  \vskip\half
  \nobreak
  \@noafterindent
  \LastMac=\Hbe\relax
}

\def\subsubsection#1{
  \if@nobreak
    \everypar{}%
    \ifnum\LastMac=\Hbe \addvspace{1pt plus 1pt minus .5pt}\fi
  \else
    \addpen{\gds@cbrk}%
    \addvspace{\onehalf}%
  \fi
  \bgroup
    \ninepoint\it
    \Raggedright
    \ifAutoNumber
      \global\advance\SecSecSec \@ne
      \noindent\@nohdbrk\number\Sec.\number\SecSec.\number\SecSecSec
        \hskip 1pc\relax #1\par
    \else
      \noindent\@nohdbrk #1\par
    \fi
  \egroup
  \nobreak
  \vskip\half
  \nobreak
  \@noafterindent
  \LastMac=\Hce\relax
}

\def\paragraph#1{
  \if@nobreak
    \everypar{}%
  \else
    \addpen{\gds@cbrk}%
    \addvspace{\one}%
  \fi%
  \bgroup%
    \ninepoint\it
    \noindent #1\ \nobreak%
  \egroup
  \LastMac=\Hde\relax
  \ignorespaces
}




\def\beginlist{%
  \par\if@nobreak \else\addvspace{\half}\fi%
  \bgroup%
    \ninepoint
    \let\item=\list@item%
}

\def\list@item{%
  \par\noindent\hskip 1em\relax%
  \ignorespaces%
}

\def\endlist{\par\egroup\addvspace{\half}\@doendpe}


\def\beginrefs{%
  \par
  \bgroup
    \eightpoint
    \Raggedright
    \let\bibitem=\bib@item
}

\def\bib@item{%
  \par\parindent=1.5em\Hang{1.5em}{1}%
  \everypar={\Hang{1.5em}{1}\ignorespaces}%
  \noindent\ignorespaces
}

\def\endrefs{\par\egroup\@doendpe}


\newtoks\CatchLine

\def\@journal{Mon.\ Not.\ R.\ Astron.\ Soc.\ }  
\def\@pubyear{1994}        
\def\@pagerange{000--000}  
\def\@volume{000}          
\def\@microfiche{}         %

\def\pubyear#1{\gdef\@pubyear{#1}\@makecatchline}
\def\pagerange#1{\gdef\@pagerange{#1}\@makecatchline}
\def\volume#1{\gdef\@volume{#1}\@makecatchline}
\def\microfiche#1{\gdef\@microfiche{and Microfiche\ #1}\@makecatchline}

\def\@makecatchline{%
  \global\CatchLine{%
    {\rm \@journal {\bf \@volume},\ \@pagerange\ (\@pubyear)\ \@microfiche}}%
}

\@makecatchline 

\newtoks\LeftHeader
\def\shortauthor#1{
  \global\LeftHeader{#1}%
}

\newtoks\RightHeader
\def\shorttitle#1{
  \global\RightHeader{#1}%
}

\def\PageHead{
  \begingroup
    \ifsp@page
      \csname ps@\sp@type\endcsname
      \global\sp@pagefalse
    \fi
    \ifodd\pageno
      \let\the@head=\@oddhead
    \else
      \let\the@head=\@evenhead
    \fi
    \vbox to \z@{\vskip-22.5\p@%
      \hbox to \PageWidth{\vbox to8.5\p@{}%
        \the@head
      }%
    \vss}%
  \endgroup
  \nointerlineskip
}

\def\today{%
  \number\day\space
  \ifcase\month\or January\or February\or March\or April\or May\or June\or
    July\or August\or September\or October\or November\or December\fi
  \space\number\year%
}

\def\PageFoot{} 

\def\authorcomment#1{%
  \gdef\PageFoot{%
    \nointerlineskip%
    \vbox to 22pt{\vfil%
      \hbox to \PageWidth{\elevenpoint\noindent \hfil #1 \hfil}}%
  }%
}


\newif\ifplate@page
\newbox\plt@box

\def\beginplatepage{%
  \let\plate=\plate@head
  \let\caption=\fig@caption
  \global\setbox\plt@box=\vbox\bgroup
  \TEMPDIMEN=\PageWidth 
  \hsize=\PageWidth\relax
}

\def\endplatepage{\par\egroup\global\plate@pagetrue}
\def\plate@head#1{\gdef\plt@cap{#1}}


\def\letters{%
  \gdef\folio{\ifnum\pageno<\z@ L\romannumeral-\pageno
    \else L\number\pageno \fi}%
}


\everydisplay{\displaysetup}

\newif\ifeqno
\newif\ifleqno

\def\displaysetup#1$${%
 \displaytest#1\eqno\eqno\displaytest
}

\def\displaytest#1\eqno#2\eqno#3\displaytest{%
 \if!#3!\ldisplaytest#1\leqno\leqno\ldisplaytest
 \else\eqnotrue\leqnofalse\def\eqn{#2}\def\eq{#1}\fi
 \generaldisplay$$}

\def\ldisplaytest#1\leqno#2\leqno#3\ldisplaytest{%
 \def\eq{#1}%
 \if!#3!\eqnofalse\else\eqnotrue\leqnotrue
  \def\eqn{#2}\fi}

\def\generaldisplay{%
\ifeqno \ifleqno 
   \hbox to \hsize{\noindent
     $\displaystyle\eq$\hfil$\displaystyle\eqn$}
  \else
    \hbox to \hsize{\noindent
     $\displaystyle\eq$\hfil$\displaystyle\eqn$}
  \fi
 \else
 \hbox to \hsize{\vbox{\noindent
  $\displaystyle\eq$\hfil}}
 \fi
}


\def\@notice{%
  \par\Two%
  \noindent{\b@ls{11pt}\ninerm This paper has been produced using the
    Blackwell Scientific Publications \TeX\ macros.\par}%
}

\outer\def\bye{\@notice\par\vfill\supereject\end}


\def\start@mess{%
  Monthly notices of the RAS journal style (\@typeface)\space
    v\@version,\space \@verdate.%
}

\everyjob{\Warn{\start@mess}}



\newif\if@debug \@debugfalse  

\def\Print#1{\if@debug\immediate\write16{#1}\else \fi}
\def\Warn#1{\immediate\write16{#1}}
\def\wlog#1{}

\newcount\Iteration 

\def\Single{0} \def\Double{1}                 
\def\Figure{0} \def\Table{1}                  

\def\InStack{0}  
\def\InZoneA{1}
\def\InZoneB{2}
\def\InZoneC{3}

\newcount\TEMPCOUNT 
\newdimen\TEMPDIMEN 
\newbox\TEMPBOX     
\newbox\VOIDBOX     

\newcount\LengthOfStack 
\newcount\MaxItems      
\newcount\StackPointer
\newcount\Point         
\newcount\NextFigure    
\newcount\NextTable     
\newcount\NextItem      

\newcount\StatusStack   
\newcount\NumStack      
\newcount\TypeStack     
\newcount\SpanStack     
\newcount\BoxStack      

\newcount\ItemSTATUS    
\newcount\ItemNUMBER    
\newcount\ItemTYPE      
\newcount\ItemSPAN      
\newbox\ItemBOX         
\newdimen\ItemSIZE      

\newdimen\PageHeight    
\newdimen\TextLeading   
\newdimen\Feathering    
\newcount\LinesPerPage  
\newdimen\ColumnWidth   
\newdimen\ColumnGap     
\newdimen\PageWidth     
\newdimen\BodgeHeight   
\newcount\Leading       

\newdimen\ZoneBSize  
\newdimen\TextSize   
\newbox\ZoneABOX     
\newbox\ZoneBBOX     
\newbox\ZoneCBOX     

\newif\ifFirstSingleItem
\newif\ifFirstZoneA
\newif\ifMakePageInComplete
\newif\ifMoreFigures \MoreFiguresfalse 
\newif\ifMoreTables  \MoreTablesfalse  

\newif\ifFigInZoneB 
\newif\ifFigInZoneC 
\newif\ifTabInZoneB 
\newif\ifTabInZoneC

\newif\ifZoneAFullPage

\newbox\MidBOX    
\newbox\LeftBOX
\newbox\RightBOX
\newbox\PageBOX   

\newif\ifLeftCOL  
\LeftCOLtrue

\newdimen\ZoneBAdjust

\newcount\ItemFits
\def\Yes{1}
\def\No{2}


\MaxItems=15
\NextFigure=\z@        
\NextTable=\@ne

\BodgeHeight=6pt
\TextLeading=11pt    
\Leading=11
\Feathering=\z@      
\LinesPerPage=61     
\topskip=\TextLeading
\ColumnWidth=20pc    
\ColumnGap=2pc       

\newskip\ItemSepamount  
\ItemSepamount=\TextLeading plus \TextLeading minus 4pt

\parskip=\z@ plus .1pt
\parindent=18pt
\widowpenalty=\z@
\clubpenalty=10000
\tolerance=1500
\hbadness=1500
\abovedisplayskip=6pt plus 2pt minus 2pt
\belowdisplayskip=6pt plus 2pt minus 2pt
\abovedisplayshortskip=6pt plus 2pt minus 2pt
\belowdisplayshortskip=6pt plus 2pt minus 2pt

\ninepoint 


\PageHeight=682pt

\PageWidth=2\ColumnWidth
\advance\PageWidth by \ColumnGap

\pagestyle{headings}




\newcount\DUMMY \StatusStack=\allocationnumber
\newcount\DUMMY \newcount\DUMMY \newcount\DUMMY 
\newcount\DUMMY \newcount\DUMMY \newcount\DUMMY 
\newcount\DUMMY \newcount\DUMMY \newcount\DUMMY
\newcount\DUMMY \newcount\DUMMY \newcount\DUMMY 
\newcount\DUMMY \newcount\DUMMY \newcount\DUMMY

\newcount\DUMMY \NumStack=\allocationnumber
\newcount\DUMMY \newcount\DUMMY \newcount\DUMMY 
\newcount\DUMMY \newcount\DUMMY \newcount\DUMMY 
\newcount\DUMMY \newcount\DUMMY \newcount\DUMMY 
\newcount\DUMMY \newcount\DUMMY \newcount\DUMMY 
\newcount\DUMMY \newcount\DUMMY \newcount\DUMMY

\newcount\DUMMY \TypeStack=\allocationnumber
\newcount\DUMMY \newcount\DUMMY \newcount\DUMMY 
\newcount\DUMMY \newcount\DUMMY \newcount\DUMMY 
\newcount\DUMMY \newcount\DUMMY \newcount\DUMMY 
\newcount\DUMMY \newcount\DUMMY \newcount\DUMMY 
\newcount\DUMMY \newcount\DUMMY \newcount\DUMMY

\newcount\DUMMY \SpanStack=\allocationnumber
\newcount\DUMMY \newcount\DUMMY \newcount\DUMMY 
\newcount\DUMMY \newcount\DUMMY \newcount\DUMMY 
\newcount\DUMMY \newcount\DUMMY \newcount\DUMMY 
\newcount\DUMMY \newcount\DUMMY \newcount\DUMMY 
\newcount\DUMMY \newcount\DUMMY \newcount\DUMMY

\newbox\DUMMY   \BoxStack=\allocationnumber
\newbox\DUMMY   \newbox\DUMMY \newbox\DUMMY 
\newbox\DUMMY   \newbox\DUMMY \newbox\DUMMY 
\newbox\DUMMY   \newbox\DUMMY \newbox\DUMMY 
\newbox\DUMMY   \newbox\DUMMY \newbox\DUMMY 
\newbox\DUMMY   \newbox\DUMMY \newbox\DUMMY

\def\wlog{\immediate\write\m@ne}


\def\GetItemAll#1{%
 \GetItemSTATUS{#1}
 \GetItemNUMBER{#1}
 \GetItemTYPE{#1}
 \GetItemSPAN{#1}
 \GetItemBOX{#1}
}

\def\GetItemSTATUS#1{%
 \Point=\StatusStack
 \advance\Point by #1
 \global\ItemSTATUS=\count\Point
}

\def\GetItemNUMBER#1{%
 \Point=\NumStack
 \advance\Point by #1
 \global\ItemNUMBER=\count\Point
}

\def\GetItemTYPE#1{%
 \Point=\TypeStack
 \advance\Point by #1
 \global\ItemTYPE=\count\Point
}

\def\GetItemSPAN#1{%
 \Point\SpanStack
 \advance\Point by #1
 \global\ItemSPAN=\count\Point
}

\def\GetItemBOX#1{%
 \Point=\BoxStack
 \advance\Point by #1
 \global\setbox\ItemBOX=\vbox{\copy\Point}
 \global\ItemSIZE=\ht\ItemBOX
 \global\advance\ItemSIZE by \dp\ItemBOX
 \TEMPCOUNT=\ItemSIZE
 \divide\TEMPCOUNT by \Leading
 \divide\TEMPCOUNT by 65536
 \advance\TEMPCOUNT \@ne
 \ItemSIZE=\TEMPCOUNT pt
 \global\multiply\ItemSIZE by \Leading
}


\def\JoinStack{%
 \ifnum\LengthOfStack=\MaxItems 
  \Warn{WARNING: Stack is full...some items will be lost!}
 \else
  \Point=\StatusStack
  \advance\Point by \LengthOfStack
  \global\count\Point=\ItemSTATUS
  \Point=\NumStack
  \advance\Point by \LengthOfStack
  \global\count\Point=\ItemNUMBER
  \Point=\TypeStack
  \advance\Point by \LengthOfStack
  \global\count\Point=\ItemTYPE
  \Point\SpanStack
  \advance\Point by \LengthOfStack
  \global\count\Point=\ItemSPAN
  \Point=\BoxStack
  \advance\Point by \LengthOfStack
  \global\setbox\Point=\vbox{\copy\ItemBOX}
  \global\advance\LengthOfStack \@ne
  \ifnum\ItemTYPE=\Figure 
   \global\MoreFigurestrue
  \else
   \global\MoreTablestrue
  \fi
 \fi
}


\def\LeaveStack#1{%
 {\Iteration=#1
 \loop
 \ifnum\Iteration<\LengthOfStack
  \advance\Iteration \@ne
  \GetItemSTATUS{\Iteration}
   \advance\Point by \m@ne
   \global\count\Point=\ItemSTATUS
  \GetItemNUMBER{\Iteration}
   \advance\Point by \m@ne
   \global\count\Point=\ItemNUMBER
  \GetItemTYPE{\Iteration}
   \advance\Point by \m@ne
   \global\count\Point=\ItemTYPE
  \GetItemSPAN{\Iteration}
   \advance\Point by \m@ne
   \global\count\Point=\ItemSPAN
  \GetItemBOX{\Iteration}
   \advance\Point by \m@ne
   \global\setbox\Point=\vbox{\copy\ItemBOX}
 \repeat}
 \global\advance\LengthOfStack by \m@ne
}


\newif\ifStackNotClean

\def\CleanStack{%
 \StackNotCleantrue
 {\Iteration=\z@
  \loop
   \ifStackNotClean
    \GetItemSTATUS{\Iteration}
    \ifnum\ItemSTATUS=\InStack
     \advance\Iteration \@ne
     \else
      \LeaveStack{\Iteration}
    \fi
   \ifnum\LengthOfStack<\Iteration
    \StackNotCleanfalse
   \fi
 \repeat}
}


\def\FindItem#1#2{%
 \global\StackPointer=\m@ne 
 {\Iteration=\z@
  \loop
  \ifnum\Iteration<\LengthOfStack
   \GetItemSTATUS{\Iteration}
   \ifnum\ItemSTATUS=\InStack
    \GetItemTYPE{\Iteration}
    \ifnum\ItemTYPE=#1
     \GetItemNUMBER{\Iteration}
     \ifnum\ItemNUMBER=#2
      \global\StackPointer=\Iteration
      \Iteration=\LengthOfStack 
     \fi
    \fi
   \fi
  \advance\Iteration \@ne
 \repeat}
}


\def\FindNext{%
 \global\StackPointer=\m@ne 
 {\Iteration=\z@
  \loop
  \ifnum\Iteration<\LengthOfStack
   \GetItemSTATUS{\Iteration}
   \ifnum\ItemSTATUS=\InStack
    \GetItemTYPE{\Iteration}
   \ifnum\ItemTYPE=\Figure
    \ifMoreFigures
      \global\NextItem=\Figure
      \global\StackPointer=\Iteration
      \Iteration=\LengthOfStack 
    \fi
   \fi
   \ifnum\ItemTYPE=\Table
    \ifMoreTables
      \global\NextItem=\Table
      \global\StackPointer=\Iteration
      \Iteration=\LengthOfStack 
    \fi
   \fi
  \fi
  \advance\Iteration \@ne
 \repeat}
}


\def\ChangeStatus#1#2{%
 \Point=\StatusStack
 \advance\Point by #1
 \global\count\Point=#2
}



\def\Zone{\InZoneA}

\ZoneBAdjust=\z@

\def\MakePage{
 \global\ZoneBSize=\PageHeight
 \global\TextSize=\ZoneBSize
 \global\ZoneAFullPagefalse
 \global\topskip=\TextLeading
 \MakePageInCompletetrue
 \MoreFigurestrue
 \MoreTablestrue
 \FigInZoneBfalse
 \FigInZoneCfalse
 \TabInZoneBfalse
 \TabInZoneCfalse
 \global\FirstSingleItemtrue
 \global\FirstZoneAtrue
 \global\setbox\ZoneABOX=\box\VOIDBOX
 \global\setbox\ZoneBBOX=\box\VOIDBOX
 \global\setbox\ZoneCBOX=\box\VOIDBOX
 \loop
  \ifMakePageInComplete
 \FindNext
 \ifnum\StackPointer=\m@ne
  \NextItem=\m@ne
  \MoreFiguresfalse
  \MoreTablesfalse
 \fi
 \ifnum\NextItem=\Figure
   \FindItem{\Figure}{\NextFigure}
   \ifnum\StackPointer=\m@ne \global\MoreFiguresfalse
   \else
    \GetItemSPAN{\StackPointer}
    \ifnum\ItemSPAN=\Single \def\Zone{\InZoneB}\relax
     \ifFigInZoneC \global\MoreFiguresfalse\fi
    \else
     \def\Zone{\InZoneA}
     \ifFigInZoneB \def\Zone{\InZoneC}\fi
    \fi
   \fi
   \ifMoreFigures\Print{}\FigureItems\fi
 \fi
\ifnum\NextItem=\Table
   \FindItem{\Table}{\NextTable}
   \ifnum\StackPointer=\m@ne \global\MoreTablesfalse
   \else
    \GetItemSPAN{\StackPointer}
    \ifnum\ItemSPAN=\Single\relax
     \ifTabInZoneC \global\MoreTablesfalse\fi
    \else
     \def\Zone{\InZoneA}
     \ifTabInZoneB \def\Zone{\InZoneC}\fi
    \fi
   \fi
   \ifMoreTables\Print{}\TableItems\fi
 \fi
   \MakePageInCompletefalse 
   \ifMoreFigures\MakePageInCompletetrue\fi
   \ifMoreTables\MakePageInCompletetrue\fi
 \repeat
 \ifZoneAFullPage
  \global\TextSize=\z@
  \global\ZoneBSize=\z@
  \global\vsize=\z@\relax
  \global\topskip=\z@\relax
  \vbox to \z@{\vss}
  \eject
 \else
 \global\advance\ZoneBSize by -\ZoneBAdjust
 \global\vsize=\ZoneBSize
 \global\hsize=\ColumnWidth
 \global\ZoneBAdjust=\z@
 \ifdim\TextSize<23pt
 \Warn{}
 \Warn{* Making column fall short: TextSize=\the\TextSize *}
 \vskip-\lastskip\eject\fi
 \fi
}

\def\MakeRightCol{
 \global\TextSize=\ZoneBSize
 \MakePageInCompletetrue
 \MoreFigurestrue
 \MoreTablestrue
 \global\FirstSingleItemtrue
 \global\setbox\ZoneBBOX=\box\VOIDBOX
 \def\Zone{\InZoneB}
 \loop
  \ifMakePageInComplete
 \FindNext
 \ifnum\StackPointer=\m@ne
  \NextItem=\m@ne
  \MoreFiguresfalse
  \MoreTablesfalse
 \fi
 \ifnum\NextItem=\Figure
   \FindItem{\Figure}{\NextFigure}
   \ifnum\StackPointer=\m@ne \MoreFiguresfalse
   \else
    \GetItemSPAN{\StackPointer}
    \ifnum\ItemSPAN=\Double\relax
     \MoreFiguresfalse\fi
   \fi
   \ifMoreFigures\Print{}\FigureItems\fi
 \fi
 \ifnum\NextItem=\Table
   \FindItem{\Table}{\NextTable}
   \ifnum\StackPointer=\m@ne \MoreTablesfalse
   \else
    \GetItemSPAN{\StackPointer}
    \ifnum\ItemSPAN=\Double\relax
     \MoreTablesfalse\fi
   \fi
   \ifMoreTables\Print{}\TableItems\fi
 \fi
   \MakePageInCompletefalse 
   \ifMoreFigures\MakePageInCompletetrue\fi
   \ifMoreTables\MakePageInCompletetrue\fi
 \repeat
 \ifZoneAFullPage
  \global\TextSize=\z@
  \global\ZoneBSize=\z@
  \global\vsize=\z@\relax
  \global\topskip=\z@\relax
  \vbox to \z@{\vss}
  \eject
 \else
 \global\vsize=\ZoneBSize
 \global\hsize=\ColumnWidth
 \ifdim\TextSize<23pt
 \Warn{}
 \Warn{* Making column fall short: TextSize=\the\TextSize *}
 \vskip-\lastskip\eject\fi
\fi
}

\def\FigureItems{
 \Print{Considering...}
 \ShowItem{\StackPointer}
 \GetItemBOX{\StackPointer} 
 \GetItemSPAN{\StackPointer}
  \CheckFitInZone 
  \ifnum\ItemFits=\Yes
   \ifnum\ItemSPAN=\Single
     \ChangeStatus{\StackPointer}{\InZoneB} 
     \global\FigInZoneBtrue
     \ifFirstSingleItem
      \hbox{}\vskip-\BodgeHeight
     \global\advance\ItemSIZE by \TextLeading
     \fi
     \unvbox\ItemBOX\ItemSep
     \global\FirstSingleItemfalse
     \global\advance\TextSize by -\ItemSIZE
     \global\advance\TextSize by -\TextLeading
   \else
    \ifFirstZoneA
     \global\advance\ItemSIZE by \TextLeading
     \global\FirstZoneAfalse\fi
    \global\advance\TextSize by -\ItemSIZE
    \global\advance\TextSize by -\TextLeading
    \global\advance\ZoneBSize by -\ItemSIZE
    \global\advance\ZoneBSize by -\TextLeading
    \ifFigInZoneB\relax
     \else
     \ifdim\TextSize<3\TextLeading
     \global\ZoneAFullPagetrue
     \fi
    \fi
    \ChangeStatus{\StackPointer}{\Zone}
    \ifnum\Zone=\InZoneC \global\FigInZoneCtrue\fi
  \fi
   \Print{TextSize=\the\TextSize}
   \Print{ZoneBSize=\the\ZoneBSize}
  \global\advance\NextFigure \@ne
   \Print{This figure has been placed.}
  \else
   \Print{No space available for this figure...holding over.}
   \Print{}
   \global\MoreFiguresfalse
  \fi
}

\def\TableItems{
 \Print{Considering...}
 \ShowItem{\StackPointer}
 \GetItemBOX{\StackPointer} 
 \GetItemSPAN{\StackPointer}
  \CheckFitInZone 
  \ifnum\ItemFits=\Yes
   \ifnum\ItemSPAN=\Single
    \ChangeStatus{\StackPointer}{\InZoneB}
     \global\TabInZoneBtrue
     \ifFirstSingleItem
      \hbox{}\vskip-\BodgeHeight
     \global\advance\ItemSIZE by \TextLeading
     \fi
     \unvbox\ItemBOX\ItemSep
     \global\FirstSingleItemfalse
     \global\advance\TextSize by -\ItemSIZE
     \global\advance\TextSize by -\TextLeading
   \else
    \ifFirstZoneA
    \global\advance\ItemSIZE by \TextLeading
    \global\FirstZoneAfalse\fi
    \global\advance\TextSize by -\ItemSIZE
    \global\advance\TextSize by -\TextLeading
    \global\advance\ZoneBSize by -\ItemSIZE
    \global\advance\ZoneBSize by -\TextLeading
    \ifFigInZoneB\relax
     \else
     \ifdim\TextSize<3\TextLeading
     \global\ZoneAFullPagetrue
     \fi
    \fi
    \ChangeStatus{\StackPointer}{\Zone}
    \ifnum\Zone=\InZoneC \global\TabInZoneCtrue\fi
   \fi
  \global\advance\NextTable \@ne
   \Print{This table has been placed.}
  \else
  \Print{No space available for this table...holding over.}
   \Print{}
   \global\MoreTablesfalse
  \fi
}


\def\CheckFitInZone{%
{\advance\TextSize by -\ItemSIZE
 \advance\TextSize by -\TextLeading
 \ifFirstSingleItem
  \advance\TextSize by \TextLeading
 \fi
 \ifnum\Zone=\InZoneA\relax
  \else \advance\TextSize by -\ZoneBAdjust
 \fi
 \ifdim\TextSize<3\TextLeading \global\ItemFits=\No
 \else \global\ItemFits=\Yes\fi}
}

\def\BeginOpening{%
  \thispagestyle{titlepage}%
  \global\setbox\ItemBOX=\vbox\bgroup%
    \hsize=\PageWidth%
    \hrule height \z@
    \ifsinglecol\vskip 6pt\fi 
}

\let\begintopmatter=\BeginOpening  

\def\EndOpening{%
  \One
  \egroup
  \ifsinglecol
    \box\ItemBOX%
    \vskip\TextLeading plus 2\TextLeading
    \@noafterindent
  \else
    \ItemNUMBER=\z@%
    \ItemTYPE=\Figure
    \ItemSPAN=\Double
    \ItemSTATUS=\InStack
    \JoinStack
  \fi
}


\newif\if@here  \@herefalse

\def\no@float{\global\@heretrue}
\let\nofloat=\relax 

\def\beginfigure{%
  \@ifstar{\global\@dfloattrue \@bfigure}{\global\@dfloatfalse \@bfigure}%
}

\def\@bfigure#1{%
  \par
  \if@dfloat
    \ItemSPAN=\Double
    \TEMPDIMEN=\PageWidth
  \else
    \ItemSPAN=\Single
    \TEMPDIMEN=\ColumnWidth
  \fi
  \ifsinglecol
    \TEMPDIMEN=\PageWidth
  \else
    \ItemSTATUS=\InStack
    \ItemNUMBER=#1%
    \ItemTYPE=\Figure
  \fi
  \bgroup
    \hsize=\TEMPDIMEN
    \global\setbox\ItemBOX=\vbox\bgroup
      \eightpoint\nostb@ls{10pt}%
      \let\caption=\fig@caption
      \ifsinglecol \let\nofloat=\no@float\fi
}

\def\fig@caption#1{%
  \vskip 5.5pt plus 6pt%
  \bgroup 
    \eightpoint\nostb@ls{10pt}%
    \setbox\TEMPBOX=\hbox{#1}%
    \ifdim\wd\TEMPBOX>\TEMPDIMEN
      \noindent \unhbox\TEMPBOX\par
    \else
      \hbox to \hsize{\hfil\unhbox\TEMPBOX\hfil}%
    \fi
  \egroup
}

\def\endfigure{%
  \par\egroup 
  \egroup
  \ifsinglecol
    \if@here \midinsert\global\@herefalse\else \topinsert\fi
      \unvbox\ItemBOX
    \endinsert
  \else
    \JoinStack
    \Print{Processing source for figure \the\ItemNUMBER}%
  \fi
}


\newbox\tab@cap@box
\def\tab@caption#1{\global\setbox\tab@cap@box=\hbox{#1\par}}

\newtoks\tab@txt@toks
\long\def\tab@txt#1{\global\tab@txt@toks={#1}\global\table@txttrue}

\newif\iftable@txt  \table@txtfalse
\newif\if@dfloat    \@dfloatfalse

\def\begintable{%
  \@ifstar{\global\@dfloattrue \@btable}{\global\@dfloatfalse \@btable}%
}

\def\@btable#1{%
  \par
  \if@dfloat
    \ItemSPAN=\Double
    \TEMPDIMEN=\PageWidth
  \else
    \ItemSPAN=\Single
    \TEMPDIMEN=\ColumnWidth
  \fi
  \ifsinglecol
    \TEMPDIMEN=\PageWidth
  \else
    \ItemSTATUS=\InStack
    \ItemNUMBER=#1%
    \ItemTYPE=\Table
  \fi
  \bgroup
    \eightpoint\nostb@ls{10pt}%
    \global\setbox\ItemBOX=\vbox\bgroup
      \let\caption=\tab@caption
      \let\tabletext=\tab@txt
      \ifsinglecol \let\nofloat=\no@float\fi
}

\def\endtable{%
  \par\egroup 
  \egroup
  \setbox\TEMPBOX=\hbox to \TEMPDIMEN{%
    \hss
    \vbox{%
      \hsize=\wd\ItemBOX
      \ifvoid\tab@cap@box
      \else
        \noindent\unhbox\tab@cap@box
        \vskip 5.5pt plus 6pt%
      \fi
      \box\ItemBOX
      \iftable@txt
        \vskip 10pt%
        \eightpoint\nostb@ls{10pt}%
        \noindent\the\tab@txt@toks
        \global\table@txtfalse
      \fi
    }%
    \hss
  }%
  \ifsinglecol
    \if@here \midinsert\global\@herefalse\else \topinsert\fi
      \box\TEMPBOX
    \endinsert
  \else
    \global\setbox\ItemBOX=\box\TEMPBOX
    \JoinStack
    \Print{Processing source for table \the\ItemNUMBER}%
  \fi
}

\def\UnloadZoneA{%
\FirstZoneAtrue
 \Iteration=\z@
  \loop
   \ifnum\Iteration<\LengthOfStack
    \GetItemSTATUS{\Iteration}
    \ifnum\ItemSTATUS=\InZoneA
     \GetItemBOX{\Iteration}
     \ifFirstZoneA \vbox to \BodgeHeight{\vfil}%
     \FirstZoneAfalse\fi
     \unvbox\ItemBOX\ItemSep
     \LeaveStack{\Iteration}
     \else
     \advance\Iteration \@ne
   \fi
 \repeat
}

\def\UnloadZoneC{%
\Iteration=\z@
  \loop
   \ifnum\Iteration<\LengthOfStack
    \GetItemSTATUS{\Iteration}
    \ifnum\ItemSTATUS=\InZoneC
     \GetItemBOX{\Iteration}
     \ItemSep\unvbox\ItemBOX
     \LeaveStack{\Iteration}
     \else
     \advance\Iteration \@ne
   \fi
 \repeat
}


\def\ShowItem#1{
  {\GetItemAll{#1}
  \Print{\the#1:
  {TYPE=\ifnum\ItemTYPE=\Figure Figure\else Table\fi}
  {NUMBER=\the\ItemNUMBER}
  {SPAN=\ifnum\ItemSPAN=\Single Single\else Double\fi}
  {SIZE=\the\ItemSIZE}}}
}

\def\ShowStack{%
 \Print{}
 \Print{LengthOfStack = \the\LengthOfStack}
 \ifnum\LengthOfStack=\z@ \Print{Stack is empty}\fi
 \Iteration=\z@
 \loop
 \ifnum\Iteration<\LengthOfStack
  \ShowItem{\Iteration}
  \advance\Iteration \@ne
 \repeat
}

\def\B#1#2{%
\hbox{\vrule\kern-0.4pt\vbox to #2{%
\hrule width #1\vfill\hrule}\kern-0.4pt\vrule}
}


\newif\ifsinglecol   \singlecolfalse

\def\onecolumn{%
  \global\output={\singlecoloutput}%
  \global\hsize=\PageWidth
  \global\vsize=\PageHeight
  \global\ColumnWidth=\hsize
  \global\TextLeading=12pt
  \global\Leading=12
  \global\singlecoltrue
  \global\let\onecolumn=\relax
  \global\let\footnote=\sing@footnote
  \global\let\vfootnote=\sing@vfootnote
  \ninepoint 
  \message{(Single column)}%
}

\def\singlecoloutput{%
  \shipout\vbox{\PageHead\pagebody\PageFoot}%
  \advancepageno
  \ifplate@page
    \shipout\vbox{%
      \sp@pagetrue
      \def\sp@type{plate}%
      \global\plate@pagefalse
      \PageHead\vbox to \PageHeight{\unvbox\plt@box\vfil}\PageFoot%
    }%
    \message{[plate]}%
    \advancepageno
  \fi
  \ifnum\outputpenalty>-\@MM \else\dosupereject\fi%
}

\def\ItemSep{\vskip\ItemSepamount\relax}

\def\ItemSepbreak{\par\ifdim\lastskip<\ItemSepamount
  \removelastskip\penalty-200\ItemSep\fi%
}


\let\@@endinsert=\endinsert 

\def\endinsert{\egroup 
  \if@mid \dimen@\ht\z@ \advance\dimen@\dp\z@ \advance\dimen@12\p@
    \advance\dimen@\pagetotal \advance\dimen@-\pageshrink
    \ifdim\dimen@>\pagegoal\@midfalse\p@gefalse\fi\fi
  \if@mid \ItemSep\box\z@\ItemSepbreak
  \else\insert\topins{\penalty100 
    \splittopskip\z@skip
    \splitmaxdepth\maxdimen \floatingpenalty\z@
    \ifp@ge \dimen@\dp\z@
    \vbox to\vsize{\unvbox\z@\kern-\dimen@}
    \else \box\z@\nobreak\ItemSep\fi}\fi\endgroup%
}


\def\gobbleone#1{}
\def\gobbletwo#1#2{}
\let\footnote=\gobbletwo 
\let\vfootnote=\gobbleone

\def\sing@footnote#1{\let\@sf\empty 
  \ifhmode\edef\@sf{\spacefactor\the\spacefactor}\/\fi
  \hbox{$^{\hbox{\eightpoint #1}}$}\@sf\sing@vfootnote{#1}%
}

\def\sing@vfootnote#1{\insert\footins\bgroup\eightpoint\b@ls{9pt}%
  \interlinepenalty\interfootnotelinepenalty
  \splittopskip\ht\strutbox 
  \splitmaxdepth\dp\strutbox \floatingpenalty\@MM
  \leftskip\z@skip \rightskip\z@skip \spaceskip\z@skip \xspaceskip\z@skip
  \noindent $^{\scriptstyle\hbox{#1}}$\hskip 4pt%
    \footstrut\futurelet\next\fo@t%
}

\def\footnoterule{\kern-3\p@ \hrule height \z@ \kern 3\p@}

\skip\footins=19.5pt plus 12pt minus 1pt
\count\footins=1000
\dimen\footins=\maxdimen


\def\landscape{%
  \global\TEMPDIMEN=\PageWidth
  \global\PageWidth=\PageHeight
  \global\PageHeight=\TEMPDIMEN
  \global\let\landscape=\relax
  \onecolumn
  \message{(landscape)}%
  \raggedbottom
}


\output{%
  \ifLeftCOL
    \global\setbox\LeftBOX=\vbox to \ZoneBSize{\box255\unvbox\ZoneBBOX}%
    \global\LeftCOLfalse
    \MakeRightCol
  \else
    \setbox\RightBOX=\vbox to \ZoneBSize{\box255\unvbox\ZoneBBOX}%
    \setbox\MidBOX=\hbox{\box\LeftBOX\hskip\ColumnGap\box\RightBOX}%
    \setbox\PageBOX=\vbox to \PageHeight{%
      \UnloadZoneA\box\MidBOX\UnloadZoneC}%
    \shipout\vbox{\PageHead\box\PageBOX\PageFoot}%
    \advancepageno
    \ifplate@page
      \shipout\vbox{%
        \sp@pagetrue
        \def\sp@type{plate}%
        \global\plate@pagefalse
        \PageHead\vbox to \PageHeight{\unvbox\plt@box\vfil}\PageFoot%
      }%
      \message{[plate]}%
      \advancepageno
    \fi
    \global\LeftCOLtrue
    \CleanStack
    \MakePage
  \fi
}


\Warn{\start@mess}

\def\mnmacrosloaded{} 

\catcode `\@=12 


\fi

\input psfig.sty



\def\eg{{e.g.~}}
\def\ie{{i.e.~}}
\def\etal{et~al.~}
\def\eV{e\kern-.15em V}                 
\def\keV{ke\kern-.15em V}                

\def\tbrem{$T_{\rm brem}$}
\def\aox{$\alpha_{\rm ox}$}
\def\aos{$\alpha_{\rm os}$}
\def\ax{$\alpha_{\rm x}$}
\def\aopt{$\alpha_{\rm opt}$}
\def\lhard{$L_{\rm 2keV}$}
\def\lsoft{$L_{\rm 0.2keV}$}

\def\luv{$L_{2500}$}

\def\pcorr{$p_{\rm corr}$}

\def\nh{$N_{\rm H}$}
\def\nhgal{$N_{\rm HGal}$}
\def\nhmod{$N_{\rm Hmod}$}
\def\nhint{$N_{\rm Hint}$}
\def\axmod{$\alpha_{\rm xmod}$}
\def\ratio{$R_{0.2keV}$}


\pageoffset{-2.5pc}{0pc}

%
  
%
%
%

\Autonumber  


\pagerange{000--000}    
\pubyear{0000}
\volume{000}

\begintopmatter  

\title{Optical and X-ray properties of the RIXOS AGN: I - The continua}
\author{E. M. Puchnarewicz,$^1$ K.~O.~Mason,$^1$ E.~Romero-Colmenero,$^1$
F.~J.~Carrera,$^1$  G.~Hasinger,$^2$ R.~M$^c$Mahon,$^3$ J.~P.~D.~Mittaz,$^1$
M.~J.~Page$^1$ and R.~Carballo$^{4,5}$}

\affiliation{$^1$ Mullard Space Science  Laboratory, University College London,
Holmbury St. Mary, Dorking, Surrey RH5 6NT, UK.}
\affiliation{$^2$ Astrophysikalisches Institut Potsdam, An der Sternwarte 16,
Potsdam, Germany.}
\affiliation{$^3$ Institute of Astronomy, Madingley Road, Cambridge CB3 0HA,
UK.}
\affiliation{$^4$ Instituto de Fisica de Cantabria, Avda. de los 
Castros s/n 39005, Santander, Spain.}
\affiliation{$^5$ Dept. Fisica Moderna, Universidad de Cantabria, Avda. de los
Castros s/n 39005, Santander, Spain.}

\shortauthor{E. M. Puchnarewicz et al.}
\shorttitle{Properties of RIXOS AGN: I - continua}



\abstract {We present measurements of the optical and X-ray continua  of 108
AGN (Seyfert 1s and quasars) from the Rosat International X-ray/Optical Survey
(RIXOS). The sample covers a wide range in redshift ($0<z<3.3$), in X-ray
spectral slope\break (--1.5$<$\ax$<$2.6) and in optical-to-X-ray ratio
(0.4$<$\aox$<$1.5). A correlation is found between \ax\ and \aox; similar
correlations have recently been reported in other X-ray and optical samples. We
also identify previously unreported relationships between  the optical slope
(\aopt) and \ax\ (particularly at high redshifts) and between \aopt\ and \aox.
These trends show that the overall optical-to-X-ray continuum changes from
convex to concave as \ax\ hardens,  demonstrating a strong behavioural link
between the optical/UV big blue bump (BBB) and the soft X-ray excess, which is
consistent with them being part of the same spectral component.

By constructing models of the optical-to-X-ray continuum,  we demonstrate that
the observed correlations are consistent with an intrinsic spectrum which is
absorbed through different amounts of cold gas and dust. The intrinsic spectrum
is the sum of an optical-to-soft X-ray `big bump' component and an \ax=1 power
law; the column density of the cold gas ranges from 0 to $\sim4\times10^{21}$
cm$^{-2}$ and the  dust-to-gas ratio is assumed to be Galactic. The `big bump'
may be represented by a \tbrem$\sim10^6$K thermal bremsstrahlung or an
accretion disk with a surrounding hot corona. The scatter in the data can
accommodate a wide range in big bump temperature (or black hole mass) and
strength. A source for the absorbing gas may be the dusty, molecular torus
which lies beyond the broad line-emitting regions, although with a much lower
column density than observed in Seyfert 2 galaxies. Alternatively, it may be
the bulge of a spiral  host galaxy or an elliptical host galaxy.}

\keywords {Quasars: general -- Galaxies: Active -- Galaxies: Seyfert -- 
X-rays: general.}

\maketitle  

\section{Introduction}

Three major continuum components are observed in the optical-to-X-ray spectra 
of non-blazar AGN; the big blue bump (BBB), which rises through the optical and
UV (\eg Edelson \&\ Malkan 1986; Elvis \etal 1994),   the `canonical' hard
X-ray power-law, which dominates above $\sim$2~\keV\ (\eg Mushotzky 1984;
Turner \& Pounds 1989; Comastri \etal\ 1992) and the soft X-ray ($E<$1.0~\keV)
`excess' above the X-ray power-law  [\eg Arnaud \etal 1985; Turner \& Pounds
1989; Walter \& Fink 1993 (hereafter WF)]. The BBB generally  dominates the
bolometric luminosity of non-blazar AGN spectra and the soft X-ray excess may
be its high energy tail. The variability of the BBB indicates that it is
emitted from the nuclear regions of AGN (\eg Clavel \etal\ 1991) and a popular
model is that of blackbody-like emission from  an optically-thick accretion
disc (AD; \eg Sun \& Malkan 1989; Czerny \& Elvis 1987; Madau 1988). 

Measurements of the BBB's spectral form, and consequently model development,
are limited however because the peak of the BBB, and much of its frequency
coverage are lost in the EUV where only a handful of AGN may be detected. 
Nonetheless, using UV and soft X-ray data from X-ray bright Seyferts, WF
reported a correlation between the 1375~\AA-to-2~\keV\ flux ratio and the slope
of the soft X-ray (0.2-2.0~\keV) spectrum, in the sense that AGN with 
relatively soft X-ray spectra also have a high UV-to-X-ray ratio. They
suggested that the correlation was due to changes in the flux of a single
optical-to-soft X-ray `big bump' relative to an underlying continuum. Laor
\etal\ (1994; hereafter L94) have reported a similar correlation in their
sample of UV-excess quasars. This has provided the first direct evidence of a
link between the BBB and the soft X-ray excess, although there still remains
doubt as to whether they are the same component, or separate but related (Fiore
\etal\ 1995).

One restriction of the three samples mentioned previously (\ie WF, L94 and
Fiore \etal 1995) is that they are dominated by low-redshift ($z<0.4$) AGN
whose properties may differ from relatively high-luminosity quasars. The L94
sample is made up of nearby ($z\le 0.4$) quasars with strong, blue optical/UV
continua  and the Fiore AGN have a relatively high optical-to-hard-X-ray
(2500~\AA-to-2~\keV) luminosity ratio; these criteria favour AGN with strong
BBBs.  The WF sample contains 58 soft X-ray (0.08-2.4~\keV) bright Seyferts
which may also favour AGN with strong BBBs. A sample of AGN which spans a wider
range in X-ray slope and optical to X-ray ratio  would be better suited to a
more thorough investigation of the soft X-ray excess and the optical/UV BBB.

We have used AGN (\ie the Seyfert 1s and quasars) detected as part of the RIXOS
survey (Mason \etal 1996) to extend this research. The RIXOS survey, which is
described in Section 2, is made up of serendipitous sources detected in
long-exposure pointed observations made with the {\sl ROSAT} Position Sensitive
Proportional Counter (PSPC; Pfeffermann \etal\ 1986). RIXOS sources are
selected according to their flux {\sl above} 0.4~\keV, thereby reducing the
preference for AGN with a strong soft X-ray excess. They cover a wide range in
redshift ($0<z<3.3$), X-ray spectral slope\break (--1.5$<$\ax$<$2.6) and in the
optical-to-X-ray flux ratio\break (0.4$<$\aox$<$1.5), providing an extensive
coverage of parameter space in which to search for trends and correlations. 

In Section 2 we outline the essential details of the survey and the process for
extracting and measuring the X-ray and optical data.  The overall properties of
the sample,  correlations and `mean' optical-to-X-ray spectra are presented in
Section 3 while in Section 4 we present a model for the RIXOS AGN continua. The 
implications for the big bump and for the structure and geometry of AGN are
discussed in Section 5.

\beginfigure{1} 
\psfig{figure=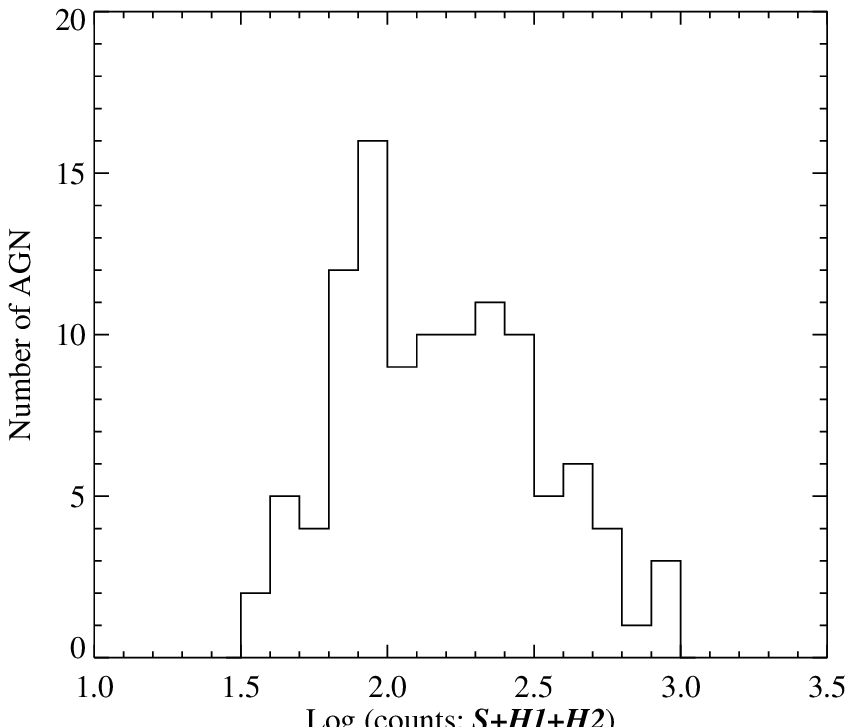,height=3in,width=3.3in,angle=0}
\caption{{\bf Figure 1.} The distribution of total source counts for the
spectra used in this analysis}
\endfigure

In total, 108 AGN are included in this analysis; the full source list giving
X-ray and optical positions, fluxes etc., will be presented in a future paper
(Mason \etal 1996) when the survey is complete. A companion paper to this
(Puchnarewicz \etal\ 1996) will discuss the relationships between the optical
and UV lines and the continuum parameters.

\section{The RIXOS Survey}

The RIXOS survey is compiled from serendipitous sources detected in pointed
PSPC observations. To be included in the survey, an observation must have an
exposure time of at least 8~ksec and it must have been taken at a Galactic
latitude greater than 28$^\circ$ (where the Galactic absorbing column density
is low). Only sources within 17~arcmin of the centre of the field and with a
flux greater than 3$\times10^{-14}$ erg sec$^{-1}$ in the {\sl ROSAT} `hard'
band (0.4-2.0~\keV) are used. These criteria ensure that source selection is
made irrespective of the strength of any soft X-ray excess which dominates at
low energies ($E<0.5~$\keV), however by ignoring any counts below 0.4~\keV,
there will be a bias against  ultrasoft X-ray AGN like E~1346+266 and
RE~J1034+396 (Puchnarewicz, Mason \& C\'ordova 1994; Puchnarewicz \etal\
1995), which have relatively little flux in the `hard' band. Most sources lie
behind relatively low Galactic \nh\ (\nhgal), 60 per cent of the RIXOS AGN
have an \nhgal$\le2\times10^{20}$ cm$^{-2}$  and all have an \nhgal\ below
$5\times10^{20}$ cm$^{-2}$.   

The RIXOS survey is 100\% complete over 20 sq degrees to a flux of
8.4$\times10^{-14}$ erg cm$^{-2}$ s$^{-1}$ and 93\% complete over 15 sq degrees
to a flux of 3$\times10^{-14}$ erg cm$^{-2}$ s$^{-1}$ (Mason \etal 1996).  The
subsample analysed for this paper is drawn from the first objects to be
identified from the whole survey: confused sources (\ie those which were less
than 30 arcsec from other RIXOS sources), sources with poorly constrained fits
to the X-ray data and objects with very weak optical spectra have been
discarded. We find no significant differences between the X-ray parameters
of the AGN presented here and those  from the 100\% complete sample (see 
Mittaz \etal 1996), thus we expect that the properties of this subsample are
representative of the final survey. 

\subsection{X-ray data}

To measure the X-ray spectral distribution of each source, the PSPC counts are
divided into three bands, {\sl S} (0.1 to 0.4~\keV; channels 8 to 41), {\sl H1}
(0.5 to 0.9~\keV; channels 52 to 90) and {\sl H2} (0.9 to 2.0~\keV; channels 91
to 201).   At the first stage in the reduction process, periods of high count
rate in the time series of the background  were removed. The source signal in
each band was measured by summing the counts in a circle centred on the
source's centroid position: in most cases the radius of the extraction circle
was 54~arcsec, however where other sources lay closer than 54~arcsec, a circle
with a radius of one-half of the distance between the source and its nearest
neighbour was used (in these cases, extraction radii are indicated in column 2
of Table 1). The 54~arcsec radius was chosen  by simulation to optimize the
signal-to-background ratio for the generally faint sources in the RIXOS survey.
All source fluxes were corrected  for counts falling outside the extraction
circle using the formulae in Hasinger \etal (1994), out to a radius of 3
arcmin. A large, annular region, centred on the pointing position of the field,
was used to calculate a model for the background, after correcting for
vignetting and having masked out any source emission.  

The broadband measurements were combined to produce `spectra' with three data
points for each source. A histogram illustrating  the distribution of the total
counts in the sources used in the present analysis is plotted in Fig 1 and
shows that one half have  at least 150 counts.  The `3-colour' X-ray spectra
were fitted using single power-law models, fixing the absorption to the
Galactic column density [this was  calculated by interpolating between the
21~cm measurements of Stark \etal (1992)]. With two free parameters (power-law
index and normalization), this left one degree of freedom for the fits. 

Many of the RIXOS sources have less than 15 counts in at least one of the {\sl
S, H1} or {\sl H2} bands; at this level, the assumption of a Gaussian
probability distribution for the data is inappropriate and Poissonian
statistics must be used. Thus instead of finding the best-fitting model for
each source by minimizing $\chi^2$ ($\chi^2$ is normally used when fitting
X-ray data with a higher signal-to-noise ratio and assumes Gaussian
statistics), the best-fit parameters (normalization and index) were calculated
by minimizing the Cash statistic (Cash 1979) which is appropriate when
considering data with relatively few counts. The errors were calculated from
the 90 percent confidence contours on the fit, using a method identical to that
for $\chi^2$ (see Cash 1979). A model for the total number of counts in each
band, which included the model of the background, was constructed for each
source and compared with the actual data (\ie source plus background). The
detector response function and corrections for vignetting effects and particle
contamination were folded into the model. The results of this spectral fitting
procedure have been verified by extensive simulation. A complete discussion of
the X-ray spectral fitting procedure for AGN and other sources in the RIXOS
survey (\ie including AGN, emission line galaxies, clusters and stars) is
presented in Mittaz \etal\ (1996).

\beginfigure{2} 
\psfig{figure=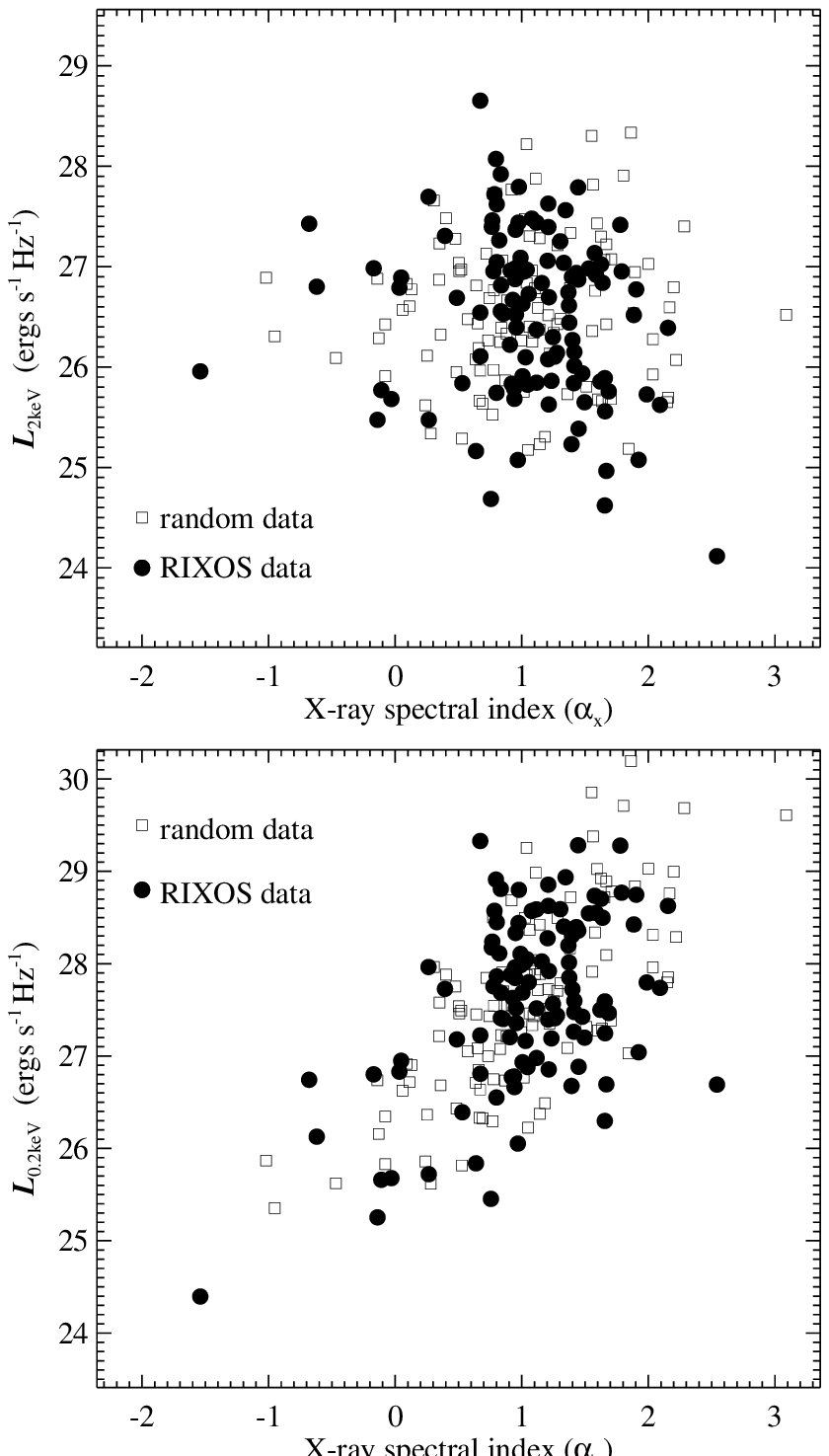,height=6in,width=3.3in,angle=0}
\caption{{\bf Figure 2.} The X-ray luminosities at (a) 2.0~\keV\ (\lhard)  and
(b) 0.2~\keV\ (\lsoft) plotted as a function of X-ray spectral index, \ax\
(filled circles). Also plotted (as open squares) are comparison, random
datasets (see Section 2.1.2) to illustrate the spurious correlation between
\lsoft\ and \ax.}
\endfigure

The best-fitting X-ray spectral indices and their 90\% (Cash) errors are given
in Table~1; \ax, and indeed {\sl all} spectral indices ($\alpha$) are 
defined throughout such that F$_\nu\propto\nu^{-\alpha}$. 

\subsubsection{Luminosities}

The logarithms of the luminosities at 0.2~\keV\ (\lsoft) and 2~\keV\ (\lhard) 
in erg s$^{-1}$ Hz$^{-1}$ in the rest-frame of the quasar were calculated
using the best-fit power-law model for each individual source, assuming a value
of 50 km s$^{-1}$ Mpc$^{-1}$ for the Hubble constant ($H_0$) and 0 for the
deceleration parameter ($q_0$; these values are assumed throughout this paper).
Errors on the X-ray luminosities have been calculated from the Cash 90\% limits
on the best-fitting power-law models (including uncertainties on both the slope
and normalization); \lsoft, \lhard\ and their respective errors are listed in
Table~1.

\subsubsection{Spurious correlations}

Because the RIXOS sample is flux-limited, spurious correlations may be
introduced  between the various continuum parameters (slopes and luminosities)
and these must be borne in mind when considering the viability of trends and
relationships in the data. Sources are selected according to their flux in the
hard (0.4-2.0~\keV) band in the observer frame, thus 2~\keV\ in the rest-frame
of the AGN is contained within the observed range for all of the objects in
this sample. In contrast, the rest-frame 0.2~\keV\ flux is derived from an
extrapolation of the fitted data at all but the lowest redshifts, therefore the
inferred flux at 0.2~\keV\  depends upon the X-ray slope  and \lsoft\ may show
an induced correlation with \ax. 

To illustrate this problem, \lhard\ and \lsoft\ are plotted as a function of
\ax\ in Fig 2.  There is no apparent relationship  between the \lhard\ and the
X-ray slope (\ax), but there is a strong correlation between \ax\ and \lsoft, 
\ie when the X-ray spectrum is very soft,  \lsoft\ is high and when \ax\ is
hard, \lsoft\ is low. This is consistent with the anticipated selection effect.
To verify that the trends in the observed data are similar to those in  a
random distribution, random samples of \ax\ and \lhard\ were generated. The
random data were created by calculating the mean and standard deviations of the
observed \ax\ and \lhard\ and using  these to generate random samples with a
normal (Gaussian) probability distribution. The \lsoft\ of this comparison
sample is the sum of the random \ax\ and \lhard\ (recalling that \lsoft\ and
\lhard\ are defined here as logarithmic quantities).

The random data are compared with the observed quantities in Fig~2; both the
\lhard\ and \lsoft\ distributions with \ax\ are similar in the observed and
simulated datasets.  This shows  that the apparent correlation between \ax\ and
\lsoft\ is consistent with the limitations of the sample's selection criteria
and may not therefore reflect a `true' property of AGN. 

The spurious dependence between \lsoft\ and \ax\ might also affect 
relationships with \lsoft\ and with \aos\ (the 2500~\AA-to-0.2~\keV\ ratio) and
the significance of this effect is discussed where appropriate in Section 3.2.
We do not expect any spurious dependences between the optical parameters (\ie
\aopt\ and \luv) and \lhard\ or \ax.

\subsection{Optical data}

The optical spectra were obtained over several observing runs with the Isaac
Newton (INT) and William Herschel Telescopes (WHT) at the Observatorio del
Roque de los Muchachos, La Palma.  Two different instruments were used, the
Faint Object Spectrograph (FOS) on the INT and the Intermediate-Dispersion
Spectrograph and Imaging System (ISIS) on the WHT (these are indicated in
Table~1). The FOS spectra cover a range of 3500\AA\ to 10000\AA\ with a
resolution of  15-20\AA\ FWHM in the red and 8-10\AA\ FWHM in the blue,  while
ISIS spectra cover 3000\AA\ to 9000\AA\ with a resolution of 3\AA\ FWHM in the
red and 2\AA\ in the blue.  All spectra were taken using a narrow slit and 
with the slit positioned at the parallactic angle except where indicated in
Table~1. 

CCD images of 23 of these AGN were also obtained at the Jacobus Kapteyn (JKT)
and Nordic Optical (NOT) Telescopes and these were used to check for the amount
of light typically lost around the narrow slit. For these AGN,  the average
ratio of fluxes measured from the CCD images to fluxes measured from the
optical spectra was 1.23$\pm0.12$. Objects which had a strong galactic
contribution in the CCD images were not included in this calculation. 

\subsubsection{Optical slope}

Measurements of the optical continuum slope were made by fitting a single
power-law to the spectrum, having first removed all emission and absorption
features. The full wavelength coverage of the spectrum was used, but ignoring
regions of poor data. No correction was made for any host galaxy contamination,
although we expect that this effect is small in AGN with a redshift, $z>0.25$.
It has been demonstrated that the combination of Balmer continuum emission and
blended FeII lines may produce a `quasi'-continuum that may affect the
measurement of the slope of any underlying continuum (\eg Wills, Netzer and
Wills 1985). In this analysis, no correction for the Balmer continuum or FeII
emission has been made unless they have formed a distinct emission feature; in
this case  the features were removed before fitting the overall continuum.

The best-fit observer-frame optical power-law indices (\aopt) are listed in
Table 1. Indices were not measured from spectra not taken at the parallactic
angle. Errors on the optical slopes are dominated by systematics and are
difficult to determine for individual spectra. We have assessed the typical
error expected on the slopes from the dispersion in the measured values for
source observed more than once and estimate that the uncertainty in \aopt\ is
conservatively $\sim\pm$0.5.

\subsubsection{Luminosities}

The logarithm of the optical continuum luminosity in the quasar rest-frame at
2500\AA\ (\luv; also given in Table~1) was calculated from the best-fitting
power-law model fit to the spectra. Luminosities of sources for which CCD data
were also available were explicitly corrected for any light lost around the
slit; these are indicated in Table~1.  Where CCD images are not available, the
luminosities have been increased by the mean factor of 1.23 (see Section 2.2).
For spectra not taken at the parallactic angle, luminosities were calculated
from a power law spectrum which was fitted to the red part of the spectrum only
(these are also indicated in Table 1). Again, errors on \luv\ are dominated by
systematics; taking into account possible variability, light losses around the
slit and errors on the power-law fits, we estimate that uncertainties on
optical luminosities are typically $\sim$50\%.

\begintable*{1}
\caption{{\bf Table 1}: Optical and X-ray continuum parameters.}
\halign{#\hfil
       &\quad#\hfil
       &\hskip0.2truecm\hfil#\hfil\hskip0.2truecm
       &\hskip0.2truecm\hfil#\hfil\hskip0.2truecm
       &\hskip0.2truecm\hfil#\hfil\hskip0.2truecm
       &\hskip0.2truecm#\hfil\hskip0.2truecm
       &\hskip0.2truecm\hfil#\hfil\hskip0.2truecm
       &\hskip0.2truecm\hfil#\hfil\hskip0.2truecm
       &\hskip0.2truecm\hfil#\hskip0.2truecm
       &\hskip0.2truecm\hfil#\hskip0.2truecm
       &\hskip0.2truecm\hfil#\hfil\hskip0.2truecm
       &\hskip0.2truecm\hfil#\hfil
\cr
\noalign{\bigskip}
    FID   &
    S No  &
    Tel   &
     z    &
    \nhgal\ &
    \hfil $L_{\rm 2500}$ &
    $L_{\rm 0.2keV}$ &
    $L_{\rm 2keV}$ &
    $\alpha_{\rm opt}$ &
    $\alpha_{\rm x}$\hfil &
    $\alpha_{\rm os}$ &
    $\alpha_{\rm ox}$ 
 \cr
\noalign{\smallskip}
      (1)\hfil 
    & (2)\hfil 
    & (3)
    & (4)
    & (5)
    & \hfil (6)
    & (7)
    & (8)
    & (9)\hfil
    & (10)\hfil
    & (11)
    & (12) \cr
\noalign{\medskip}
 110 &   1 & WHT & 0.364 & 1.1 & 28.9     & 27.46$^{+0.03}_{-0.03}$ & 25.75$^{+0.05}_{-0.06}$ &  1.3 &  1.69$^{+0.06}_{-0.09}$ & 0.9$^{+0.2}_{-0.2}$ & 1.19$^{+0.09}_{-0.13}$ \cr 
 110 &   8 & WHT & 0.938 & 1.1 & 29.0     & 27.68$^{+0.10}_{-0.14}$ & 26.62$^{+0.06}_{-0.07}$ &  2.0 &  1.01$^{+0.11}_{-0.19}$ & 0.8$^{+0.3}_{-0.3}$ & 0.91$^{+0.08}_{-0.13}$ \cr 
 110 &  34 & WHT & 1.235 & 1.1 & 29.6     & 26.83$^{+0.25}_{-0.70}$ & 26.79$^{+0.08}_{-0.09}$ &  2.9 &  0.03$^{+0.46}_{-0.37}$ & 1.7$^{+0.6}_{-0.3}$ & 1.08$^{+0.07}_{-0.14}$ \cr 
 110 &  35 & WHT & 0.582 & 1.1 & 28.8     & 27.59$^{+0.06}_{-0.07}$ & 25.88$^{+0.08}_{-0.10}$ &  0.5 &  1.66$^{+0.11}_{-0.17}$ & 0.8$^{+0.2}_{-0.2}$ & 1.12$^{+0.07}_{-0.14}$ \cr 
 110 &  50 & WHT & 1.335 & 1.1 & 29.4     & 27.98$^{+0.11}_{-0.15}$ & 26.94$^{+0.06}_{-0.06}$ & --0.3 & 0.99$^{+0.12}_{-0.19}$ & 0.9$^{+0.3}_{-0.3}$ & 0.94$^{+0.14}_{-0.13}$ \cr 
 122 &   1 & WHT & 1.134 & 4.1 & 30.3     & 28.30$^{+0.18}_{-0.33}$ & 26.89$^{+0.05}_{-0.06}$ &  0.4 &  1.40$^{+0.28}_{-0.22}$ & 1.2$^{+0.4}_{-0.3}$ & 1.29$^{+0.09}_{-0.13}$ \cr 
 122 &  14$^{(45)}$ & WHT & 0.379 & 4.1 & 29.3$^i$ & 27.50$^{+0.07}_{-0.09}$ & 25.85$^{+0.05}_{-0.05}$ &  0.4 &  1.62$^{+0.11}_{-0.13}$ & 1.1$^{+0.2}_{-0.2}$ & 1.30$^{+0.09}_{-0.13}$ \cr 
 123 &   1 & WHT & 0.282 & 1.2 & 28.7$^i$ & 27.42$^{+0.02}_{-0.02}$ & 25.93$^{+0.04}_{-0.04}$ &  1.0 &  1.48$^{+0.05}_{-0.06}$ & 0.8$^{+0.2}_{-0.2}$ & 1.07$^{+0.10}_{-0.12}$ \cr 
 123 &  27 & WHT & 0.351 & 1.2 & 28.5$^i$ & 26.66$^{+0.07}_{-0.09}$ & 25.68$^{+0.07}_{-0.08}$ &  1.4 &  0.94$^{+0.10}_{-0.16}$ & 1.2$^{+0.2}_{-0.2}$ & 1.09$^{+0.08}_{-0.14}$ \cr 
 123 &  41 & WHT & 1.818 & 1.2 & 30.5$^i$ & 28.17$^{+0.14}_{-0.21}$ & 27.39$^{+0.05}_{-0.06}$ &  1.8 &  0.76$^{+0.17}_{-0.17}$ & 1.4$^{+0.3}_{-0.3}$ & 1.18$^{+0.09}_{-0.13}$ \cr 
 123 &  42 & WHT & 0.477 & 1.2 & 28.9$^i$ & 26.88$^{+0.09}_{-0.12}$ & 25.82$^{+0.08}_{-0.09}$ &  1.0 &  1.05$^{+0.18}_{-0.14}$ & 1.3$^{+0.3}_{-0.2}$ & 1.17$^{+0.07}_{-0.14}$ \cr 
 123 &  46 & WHT & 1.288 & 1.2 & 29.7$^i$ & 26.94$^{+0.24}_{-0.59}$ & 26.89$^{+0.07}_{-0.08}$ &  1.7 &  0.05$^{+0.36}_{-0.30}$ & 1.7$^{+0.6}_{-0.3}$ & 1.09$^{+0.08}_{-0.14}$ \cr 
 123 &  66 & WHT & 0.494 & 1.2 & 28.9     & 27.16$^{+0.08}_{-0.09}$ & 26.09$^{+0.06}_{-0.08}$ &  1.1 &  1.03$^{+0.10}_{-0.16}$ & 1.1$^{+0.2}_{-0.2}$ & 1.07$^{+0.08}_{-0.13}$ \cr 
 123 &  85 & WHT & 0.652 & 1.2 & 29.8     & 28.42$^{+0.03}_{-0.03}$ & 26.51$^{+0.05}_{-0.06}$ & --0.1 & 1.89$^{+0.06}_{-0.10}$ & 0.9$^{+0.2}_{-0.2}$ & 1.25$^{+0.11}_{-0.11}$ \cr 
 126 &  27$^{(36)}$ & WHT & 3.305 & 2.0 & 30.3$^i$ & 29.32$^{+0.15}_{-0.23}$ & 28.65$^{+0.05}_{-0.05}$ &  0.8 &  0.67$^{+0.16}_{-0.15}$ & 0.7$^{+0.3}_{-0.3}$ & 0.67$^{+0.09}_{-0.13}$ \cr 
 127 &   3 & WHT & 1.038 & 1.7 & 30.2     & 28.76$^{+0.05}_{-0.06}$ & 26.95$^{+0.05}_{-0.06}$ &  0.6 &  1.79$^{+0.08}_{-0.12}$ & 0.9$^{+0.2}_{-0.2}$ & 1.26$^{+0.09}_{-0.13}$ \cr 
 127 &   4 & INT & 0.973 & 1.7 & 30.2     & 28.35$^{+0.07}_{-0.09}$ & 26.87$^{+0.05}_{-0.06}$ &  0.1 &  1.45$^{+0.09}_{-0.15}$ & 1.1$^{+0.2}_{-0.2}$ & 1.26$^{+0.09}_{-0.13}$ \cr 
 127 &  21 & WHT & 0.152 & 1.7 & 27.2   & 26.05$^{+0.06}_{-0.08}$ & 25.07$^{+0.06}_{-0.07}$ & np & 0.97$^{+0.12}_{-0.12}$ & 0.7$^{+0.2}_{-0.2}$ & 0.82$^{+0.09}_{-0.14}$ \cr 
 133 &  22 & WHT & 1.788 & 1.2 & 31.1     & 28.45$^{+0.15}_{-0.23}$ & 27.62$^{+0.06}_{-0.07}$ &  0.5 &  0.80$^{+0.20}_{-0.22}$ & 1.6$^{+0.3}_{-0.3}$ & 1.33$^{+0.08}_{-0.13}$ \cr 
 206 &   6 & WHT & 0.690 & 3.7 & 29.7     & 27.22$^{+0.21}_{-0.44}$ & 26.54$^{+0.06}_{-0.07}$ &  0.6 &  0.67$^{+0.35}_{-0.29}$ & 1.5$^{+0.5}_{-0.3}$ & 1.19$^{+0.08}_{-0.13}$ \cr 
 206 &   9 & WHT & 0.801 & 3.7 & 28.9     & 27.18$^{+0.22}_{-0.51}$ & 26.68$^{+0.06}_{-0.06}$ &  0.7 &  0.49$^{+0.38}_{-0.31}$ & 1.1$^{+0.5}_{-0.3}$ & 0.88$^{+0.09}_{-0.13}$ \cr 
 208 &   2 & WHT & 0.387 & 0.7 & 28.8     & 27.44$^{+0.03}_{-0.03}$ & 26.14$^{+0.04}_{-0.05}$ &  2.0 &  1.28$^{+0.05}_{-0.08}$ & 0.8$^{+0.2}_{-0.2}$ & 1.01$^{+0.09}_{-0.13}$ \cr 
 208 &  55 & WHT & 1.718 & 0.7 & 30.1     & 28.23$^{+0.10}_{-0.14}$ & 27.46$^{+0.05}_{-0.05}$ &  1.6 &  0.77$^{+0.14}_{-0.14}$ & 1.2$^{+0.3}_{-0.3}$ & 1.03$^{+0.09}_{-0.13}$ \cr 
 211 &  30 & INT & 1.420 & 4.0 & 30.7     & 28.33$^{+0.21}_{-0.46}$ & 27.36$^{+0.05}_{-0.06}$ &  0.8 &  0.95$^{+0.34}_{-0.26}$ & 1.4$^{+0.5}_{-0.3}$ & 1.26$^{+0.09}_{-0.13}$ \cr 
 211 &  42 & INT & 0.232 & 4.0 & 28.9     & 26.88$^{+0.10}_{-0.13}$ & 25.38$^{+0.07}_{-0.09}$ &  1.9 &  1.45$^{+0.13}_{-0.21}$ & 1.2$^{+0.3}_{-0.3}$ & 1.33$^{+0.08}_{-0.14}$ \cr 
 212 &   6 & WHT & 1.004 & 1.2 & 29.8     & 27.68$^{+0.17}_{-0.29}$ & 26.81$^{+0.09}_{-0.11}$ &  0.2 &  0.84$^{+0.27}_{-0.25}$ & 1.3$^{+0.4}_{-0.3}$ & 1.14$^{+0.07}_{-0.14}$ \cr 
 212 &  16 & WHT & 0.843 & 1.2 & 30.1     & 27.51$^{+0.16}_{-0.27}$ & 26.52$^{+0.10}_{-0.13}$ &  0.3 &  0.95$^{+0.28}_{-0.26}$ & 1.6$^{+0.4}_{-0.3}$ & 1.37$^{+0.06}_{-0.15}$ \cr 
 212 &  25 & WHT & 0.801 & 1.2 & 29.7$^i$ & 28.01$^{+0.09}_{-0.12}$ & 26.61$^{+0.09}_{-0.10}$ &  0.2 &  1.37$^{+0.18}_{-0.17}$ & 1.1$^{+0.3}_{-0.2}$ & 1.20$^{+0.07}_{-0.14}$ \cr 
 213 &  17 & WHT & 0.438 & 4.4 & 29.0     & 27.47$^{+0.21}_{-0.44}$ & 26.01$^{+0.12}_{-0.16}$ &  0.3 &  1.42$^{+0.42}_{-0.34}$ & 1.0$^{+0.5}_{-0.3}$ & 1.15$^{+0.05}_{-0.16}$ \cr 
 213 &  19 & WHT & 0.467 & 4.4 & 29.0     & 27.59$^{+0.18}_{-0.31}$ & 26.15$^{+0.09}_{-0.12}$ &  1.1 &  1.42$^{+0.31}_{-0.29}$ & 0.9$^{+0.4}_{-0.3}$ & 1.09$^{+0.06}_{-0.15}$ \cr 
 215 &   1 & INT & 2.248 & 1.2 & 31.1     & 28.80$^{+0.15}_{-0.24}$ & 27.79$^{+0.06}_{-0.07}$ &  0.6 &  0.98$^{+0.22}_{-0.19}$ & 1.4$^{+0.3}_{-0.3}$ & 1.27$^{+0.08}_{-0.13}$ \cr 
 215 &  19 & WHT & 0.584 & 1.2 & 29.5     & 27.85$^{+0.05}_{-0.06}$ & 26.44$^{+0.06}_{-0.08}$ &  1.1 &  1.38$^{+0.09}_{-0.14}$ & 1.0$^{+0.2}_{-0.2}$ & 1.18$^{+0.08}_{-0.14}$ \cr 
 215 &  32 & WHT & 0.613 & 1.2 & 29.4     & 27.39$^{+0.12}_{-0.18}$ & 26.07$^{+0.12}_{-0.17}$ & --0.1 & 1.21$^{+0.19}_{-0.30}$ & 1.2$^{+0.3}_{-0.3}$ & 1.28$^{+0.18}_{-0.14}$ \cr 
 216 &   7 & WHT & 0.804 & 3.5 & 29.5$^i$ & 27.41$^{+0.22}_{-0.49}$ & 26.55$^{+0.06}_{-0.08}$ & --1.1 & 0.84$^{+0.37}_{-0.30}$ & 1.3$^{+0.5}_{-0.3}$ & 1.12$^{+0.05}_{-0.12}$ \cr 
 217 &   3 & WHT & 0.989 & 1.1 & 29.7$^i$ & 27.92$^{+0.08}_{-0.10}$ & 26.69$^{+0.06}_{-0.07}$ &  0.3 &  1.22$^{+0.13}_{-0.12}$ & 1.1$^{+0.2}_{-0.2}$ & 1.15$^{+0.08}_{-0.13}$ \cr 
 217 &  21$^{(52)}$ & WHT & 0.562 & 1.1 & 28.6     & 27.40$^{+0.07}_{-0.08}$ & 26.10$^{+0.09}_{-0.08}$ &  1.5 &  1.26$^{+0.10}_{-0.16}$ & 0.7$^{+0.2}_{-0.2}$ & 0.95$^{+0.08}_{-0.14}$ \cr 
 217 &  34 & WHT & 1.200 & 1.1 & 30.3  & 28.40$^{+0.06}_{-0.07}$ & 27.03$^{+0.05}_{-0.14}$ & np & 1.33$^{+0.08}_{-0.12}$ & 1.2$^{+0.2}_{-0.2}$ & 1.26$^{+0.16}_{-0.17}$ \cr 
 217 &  35 & WHT & 0.435 & 1.1 & 28.6$^i$ & 26.77$^{+0.08}_{-0.11}$ & 25.83$^{+0.07}_{-0.08}$ &  1.8 &  0.92$^{+0.15}_{-0.14}$ & 1.2$^{+0.3}_{-0.2}$ & 1.07$^{+0.08}_{-0.14}$ \cr 
 217 &  59 & WHT & 0.587 & 1.1 & 28.6     & 27.51$^{+0.06}_{-0.07}$ & 26.37$^{+0.05}_{-0.06}$ &  1.6 &  1.12$^{+0.09}_{-0.12}$ & 0.7$^{+0.2}_{-0.2}$ & 0.84$^{+0.09}_{-0.13}$ \cr 
 218 &   1 & WHT & 0.545 & 3.0 & 29.5$^i$ & 27.40$^{+0.17}_{-0.28}$ & 26.53$^{+0.08}_{-0.09}$ & --0.7 & 0.85$^{+0.28}_{-0.25}$ & 1.3$^{+0.4}_{-0.3}$ & 1.15$^{+0.07}_{-0.14}$ \cr 
 218 &  27 & WHT & 0.631 & 3.0 & 29.8   & 27.80$^{+0.14}_{-0.21}$ & 26.72$^{+0.07}_{-0.08}$ & np & 1.06$^{+0.22}_{-0.21}$ & 1.3$^{+0.3}_{-0.3}$ & 1.18$^{+0.08}_{-0.14}$ \cr 
 219 &  15 & WHT & 1.190 & 1.3 & 29.6$^i$ & 28.55$^{+0.07}_{-0.09}$ & 26.91$^{+0.06}_{-0.07}$ &  1.9 &  1.59$^{+0.10}_{-0.16}$ & 0.6$^{+0.2}_{-0.2}$ & 1.01$^{+0.08}_{-0.13}$ \cr 
 219 &  48 & WHT & 1.367 & 1.3 & 29.5     & 27.96$^{+0.17}_{-0.30}$ & 26.98$^{+0.07}_{-0.09}$ &  0.5 &  0.95$^{+0.26}_{-0.23}$ & 1.0$^{+0.4}_{-0.3}$ & 0.98$^{+0.08}_{-0.15}$ \cr 
 220 &  23 & WHT & 0.193 & 3.9 & 27.6$^i$ & 25.25$^{+0.14}_{-0.21}$ & 25.47$^{+0.14}_{-0.21}$ & 3.8 & --0.14$^{+1.04}_{-0.75}$ & 1.4$^{+0.3}_{-0.3}$ & 0.80$^{+0.03}_{-0.16}$ \cr 
 220 &  25 & WHT & 0.210 & 3.9 & 28.0$^i$ & 27.04$^{+0.13}_{-0.18}$ & 25.07$^{+0.12}_{-0.18}$ &  1.3 &  1.92$^{+0.27}_{-0.26}$ & 0.6$^{+0.3}_{-0.3}$ & 1.13$^{+0.04}_{-0.16}$ \cr 
 221 &  16 & WHT & 0.184 & 2.9 & 28.4     & 26.69$^{+0.08}_{-0.10}$ & 24.96$^{+0.10}_{-0.12}$ &  1.4 &  1.67$^{+0.13}_{-0.21}$ & 1.0$^{+0.3}_{-0.2}$ & 1.31$^{+0.06}_{-0.15}$ \cr 
 222 &  20 & WHT & 1.068 & 2.4 & 29.7     & 28.19$^{+0.15}_{-0.23}$ & 26.74$^{+0.08}_{-0.09}$ &  0.6 &  1.37$^{+0.17}_{-0.28}$ & 1.0$^{+0.3}_{-0.3}$ & 1.14$^{+0.07}_{-0.14}$ \cr 
 223 &  17 & WHT & 0.288 & 1.8 & 28.7$^i$ & 27.26$^{+0.02}_{-0.03}$ & 25.83$^{+0.03}_{-0.03}$ &  2.1 &  1.41$^{+0.04}_{-0.07}$ & 0.9$^{+0.2}_{-0.2}$ & 1.10$^{+0.10}_{-0.12}$ \cr 
 225 &   1$^{(52)}$ & INT & 0.488 & 2.2 & 28.9$^i$ & 26.97$^{+0.11}_{-0.15}$ & 25.84$^{+0.08}_{-0.10}$ &  1.1 &  1.12$^{+0.19}_{-0.18}$ & 1.2$^{+0.3}_{-0.3}$ & 1.17$^{+0.07}_{-0.14}$ \cr 
}          
\endtable           

\begintable*{2}
\caption{Table 1 (continued): Optical and X-ray continuum parameters.}
\halign{#\hfil
       &\quad#\hfil
       &\hskip0.2truecm\hfil#\hfil\hskip0.2truecm
       &\hskip0.2truecm\hfil#\hfil\hskip0.2truecm
       &\hskip0.2truecm\hfil#\hfil\hskip0.2truecm
       &\hskip0.2truecm#\hfil\hskip0.2truecm
       &\hskip0.2truecm\hfil#\hfil\hskip0.2truecm
       &\hskip0.2truecm\hfil#\hfil\hskip0.2truecm
       &\hskip0.2truecm\hfil#\hskip0.2truecm
       &\hskip0.2truecm\hfil#\hskip0.2truecm
       &\hskip0.2truecm\hfil#\hfil\hskip0.2truecm
       &\hskip0.2truecm\hfil#\hfil
\cr
\noalign{\bigskip}
    FID   &
    S No  &
    Tel   &
     z    &
    \nhgal\ &
    \hfil $L_{\rm 2500}$ &
    $L_{\rm 0.2keV}$ &
    $L_{\rm 2keV}$ &
    $\alpha_{\rm opt}$ &
    $\alpha_{\rm x}$\hfil &
    $\alpha_{\rm os}$ &
    $\alpha_{\rm ox}$ 
 \cr
\noalign{\smallskip}
      (1)\hfil 
    & (2)\hfil 
    & (3)
    & (4)
    & (5)
    & \hfil (6)
    & (7)
    & (8)
    & (9)\hfil
    & (10)\hfil
    & (11)
    & (12) \cr
\noalign{\medskip}
 226 &  41 & WHT & 1.315 & 1.2 & 30.1$^i$ & 28.57$^{+0.04}_{-0.04}$ & 27.47$^{+0.02}_{-0.02}$ & np & 1.08$^{+0.04}_{-0.07}$ & 0.9$^{+0.2}_{-0.2}$ & 0.98$^{+0.10}_{-0.12}$ \cr 
 226 & 114$^{(49)}$ & WHT & 0.113 & 1.2 & 26.8     & 25.45$^{+0.10}_{-0.14}$ & 24.68$^{+0.04}_{-0.05}$ &  0.6 &  0.76$^{+0.14}_{-0.13}$ & 0.8$^{+0.3}_{-0.3}$ & 0.81$^{+0.09}_{-0.13}$ \cr 
 227 &  19 & WHT & 1.861 & 1.8 & 30.5$^i$ & 28.93$^{+0.09}_{-0.11}$ & 27.56$^{+0.04}_{-0.04}$ &  1.0 &  1.35$^{+0.10}_{-0.13}$ & 1.0$^{+0.3}_{-0.2}$ & 1.11$^{+0.09}_{-0.13}$ \cr 
 227 &  37 & WHT & 1.413 & 1.8 & 30.6$^i$ & 28.85$^{+0.06}_{-0.07}$ & 27.62$^{+0.03}_{-0.03}$ &  1.4 &  1.21$^{+0.06}_{-0.10}$ & 1.1$^{+0.2}_{-0.2}$ & 1.14$^{+0.10}_{-0.12}$ \cr 
 227 & 301 & WHT & 0.114 & 1.8 & 27.6$^i$ & 26.69$^{+0.02}_{-0.02}$ & 24.11$^{+0.08}_{-0.09}$ &  1.7 &  2.54$^{+0.07}_{-0.12}$ & 0.6$^{+0.2}_{-0.2}$ & 1.33$^{+0.08}_{-0.14}$ \cr 
 227 & 513 & WHT & 0.959 & 1.8 & 30.0$^i$ & 27.87$^{+0.07}_{-0.09}$ & 26.96$^{+0.04}_{-0.04}$ & --0.1 &  0.91$^{+0.10}_{-0.10}$ & 1.3$^{+0.2}_{-0.2}$ & 1.17$^{+0.10}_{-0.12}$ \cr 
 228 &   1 & INT & 1.726 & 3.7 & 30.8     & 27.96$^{+0.26}_{-0.77}$ & 27.69$^{+0.07}_{-0.09}$ &  0.4 &  0.26$^{+0.40}_{-0.35}$ & 1.8$^{+0.7}_{-0.4}$ & 1.18$^{+0.08}_{-0.14}$ \cr 
 231 & 301 & WHT & 0.783 & 0.7 & 30.2     & 28.49$^{+0.04}_{-0.04}$ & 26.83$^{+0.10}_{-0.12}$ &  0.5 &  1.64$^{+0.08}_{-0.10}$ & 1.0$^{+0.2}_{-0.2}$ & 1.27$^{+0.11}_{-0.11}$ \cr 
 231 & 302 & INT & 1.572 & 0.7 & 30.5     & 28.58$^{+0.08}_{-0.10}$ & 27.43$^{+0.05}_{-0.06}$ &  0.6 &  1.12$^{+0.12}_{-0.14}$ & 1.2$^{+0.3}_{-0.2}$ & 1.19$^{+0.09}_{-0.13}$ \cr 
 231 & 503 & INT & 0.234 & 0.7 & 28.3     & 25.66$^{+0.09}_{-0.12}$ & 25.77$^{+0.06}_{-0.08}$ &  2.1 &  -0.11$^{+0.15}_{-0.13}$ & 1.7$^{+0.3}_{-0.2}$ & 0.97$^{+0.08}_{-0.13}$ \cr 
 232 &  16 & WHT & 0.227 & 0.8 & 28.3     & 25.83$^{+0.14}_{-0.22}$ & 25.16$^{+0.11}_{-0.15}$ &  2.2 &  0.64$^{+0.28}_{-0.24}$ & 1.5$^{+0.3}_{-0.3}$ & 1.20$^{+0.05}_{-0.15}$ \cr 
 232 & 301 & INT & 0.385 & 0.8 & 29.3     & 26.77$^{+0.07}_{-0.09}$ & 25.78$^{+0.08}_{-0.10}$ &  0.6 &  0.94$^{+0.11}_{-0.18}$ & 1.6$^{+0.2}_{-0.2}$ & 1.35$^{+0.07}_{-0.14}$ \cr 
 234 &   1 & INT & 1.666 & 4.1 & 30.7     & 29.28$^{+0.13}_{-0.20}$ & 27.78$^{+0.04}_{-0.05}$ &  1.0 &  1.45$^{+0.13}_{-0.21}$ & 0.9$^{+0.3}_{-0.3}$ & 1.12$^{+0.09}_{-0.13}$ \cr 
 234 &  33 & WHT & 1.019 & 4.1 & 29.8     & 28.74$^{+0.13}_{-0.19}$ & 26.77$^{+0.07}_{-0.08}$ &  0.1 &  1.90$^{+0.16}_{-0.26}$ & 0.6$^{+0.3}_{-0.3}$ & 1.15$^{+0.08}_{-0.14}$ \cr 
 240 &  15 & WHT & 1.263 & 1.2 & 29.7     & 28.39$^{+0.05}_{-0.06}$ & 26.93$^{+0.04}_{-0.04}$ &  0.5 &  1.43$^{+0.067}_{-0.11}$ & 0.8$^{+0.2}_{-0.2}$ & 1.07$^{+0.09}_{-0.13}$ \cr 
 240 &  82 & WHT & 0.518 & 1.2 & 29.4     & 26.80$^{+0.07}_{-0.09}$ & 26.10$^{+0.04}_{-0.05}$ & --0.1 &  0.67$^{+0.08}_{-0.13}$ & 1.6$^{+0.2}_{-0.2}$ & 1.25$^{+0.09}_{-0.13}$ \cr 
 248 &   2 & INT & 0.274 & 1.5 & 27.7     & 26.55$^{+0.06}_{-0.07}$ & 25.74$^{+0.05}_{-0.06}$ &  2.4 &  0.80$^{+0.11}_{-0.11}$ & 0.7$^{+0.2}_{-0.2}$ & 0.75$^{+0.09}_{-0.13}$ \cr 
 252 &   9 & WHT & 0.673 & 0.8 & 29.3$^i$ & 27.56$^{+0.07}_{-0.09}$ & 26.29$^{+0.08}_{-0.10}$ & --0.3 &  1.25$^{+0.15}_{-0.14}$ & 1.1$^{+0.2}_{-0.2}$ & 1.15$^{+0.07}_{-0.14}$ \cr 
 252 &  34 & WHT & 0.680 & 0.8 & 28.9     & 27.20$^{+0.11}_{-0.15}$ & 26.22$^{+0.09}_{-0.11}$ &  0.5 &  0.91$^{+0.14}_{-0.24}$ & 1.0$^{+0.3}_{-0.3}$ & 1.00$^{+0.07}_{-0.14}$ \cr 
 252 &  36 & WHT & 1.037 & 0.8 & 29.3     & 28.04$^{+0.07}_{-0.09}$ & 26.96$^{+0.06}_{-0.07}$ &  0.7 &  1.04$^{+0.09}_{-0.15}$ & 0.8$^{+0.2}_{-0.2}$ & 0.90$^{+0.08}_{-0.13}$ \cr 
 252 &  38 & WHT & 0.216 & 0.8 & 28.7     & 26.85$^{+0.03}_{-0.03}$ & 25.62$^{+0.05}_{-0.06}$ &  0.6 &  1.21$^{+0.07}_{-0.09}$ & 1.1$^{+0.2}_{-0.2}$ & 1.17$^{+0.09}_{-0.13}$ \cr 
 253 &   5 & WHT & 1.211 & 1.6 & 29.6     & 28.02$^{+0.13}_{-0.20}$ & 26.83$^{+0.07}_{-0.09}$ &  0.5 &  1.16$^{+0.21}_{-0.19}$ & 1.0$^{+0.3}_{-0.3}$ & 1.04$^{+0.08}_{-0.14}$ \cr 
 253 &  32 & WHT & 0.237 & 1.6 & 28.5     & 26.67$^{+0.07}_{-0.09}$ & 25.23$^{+0.09}_{-0.12}$ &  0.2 &  1.39$^{+0.12}_{-0.19}$ & 1.1$^{+0.2}_{-0.2}$ & 1.24$^{+0.07}_{-0.15}$ \cr 
 254 &  10 & INT & 0.936 & 1.1 & 31.1     & 28.73$^{+0.04}_{-0.05}$ & 27.13$^{+0.05}_{-0.05}$ &  0.6 &  1.58$^{+0.07}_{-0.11}$ & 1.5$^{+0.2}_{-0.2}$ & 1.53$^{+0.09}_{-0.13}$ \cr 
 254 &  11 & INT & 1.166 & 1.1 & 30.9     & 28.59$^{+0.07}_{-0.08}$ & 27.25$^{+0.05}_{-0.06}$ &  0.7 &  1.30$^{+0.09}_{-0.14}$ & 1.4$^{+0.2}_{-0.2}$ & 1.39$^{+0.09}_{-0.13}$ \cr 
 254 &  41 & WHT & 0.486 & 1.1 & 29.2     & 27.19$^{+0.11}_{-0.14}$ & 25.86$^{+0.11}_{-0.15}$ &  0.3 &  1.23$^{+0.17}_{-0.27}$ & 1.3$^{+0.3}_{-0.3}$ & 1.28$^{+0.05}_{-0.15}$ \cr 
 255 &  19 & WHT & 0.862 & 5.1 & 29.6     & 28.62$^{+0.21}_{-0.46}$ & 26.39$^{+0.12}_{-0.17}$ &  0.6 &  2.16$^{+0.41}_{-0.38}$ & 0.6$^{+0.5}_{-0.3}$ & 1.23$^{+0.04}_{-0.16}$ \cr 
 257 &   1$^{(53)}$ & INT & 1.021 & 2.2 & 30.8     & 28.62$^{+0.06}_{-0.07}$ & 27.39$^{+0.04}_{-0.04}$ &  0.2 &  1.21$^{+0.07}_{-0.11}$ & 1.3$^{+0.2}_{-0.2}$ & 1.30$^{+0.10}_{-0.12}$ \cr 
 257 &  14 & WHT & 1.096 & 2.2 & 28.6     & 27.75$^{+0.17}_{-0.31}$ & 26.95$^{+0.06}_{-0.08}$ &  1.0 &  0.78$^{+0.25}_{-0.24}$ & 0.5$^{+0.4}_{-0.3}$ & 0.62$^{+0.08}_{-0.13}$ \cr 
 257 &  20 & INT & 1.304 & 2.2 & 30.2     & 28.01$^{+0.20}_{-0.39}$ & 26.95$^{+0.08}_{-0.10}$ &  0.3 &  1.02$^{+0.29}_{-0.30}$ & 1.4$^{+0.4}_{-0.3}$ & 1.25$^{+0.07}_{-0.14}$ \cr 
 257 &  38 & INT & 1.260 & 2.2 & 30.6     & 28.10$^{+0.18}_{-0.34}$ & 27.08$^{+0.07}_{-0.09}$ &  0.0 &  0.99$^{+0.27}_{-0.26}$ & 1.6$^{+0.4}_{-0.3}$ & 1.34$^{+0.08}_{-0.14}$ \cr 
 258 &   1 & WHT & 0.698 & 3.4 & 28.1     & 26.80$^{+0.18}_{-0.34}$ & 26.98$^{+0.04}_{-0.04}$ &  2.4 & --0.17$^{+0.27}_{-0.26}$ & 0.8$^{+0.4}_{-0.3}$ & 0.41$^{+0.09}_{-0.12}$ \cr 
 258 &  32 & WHT & 1.618 & 3.4 & 30.0     & 28.11$^{+0.27}_{-0.93}$ & 27.26$^{+0.08}_{-0.10}$ &  0.7 &  0.82$^{+0.49}_{-0.36}$ & 1.2$^{+0.8}_{-0.4}$ & 1.03$^{+0.07}_{-0.14}$ \cr 
 259 &   5 & WHT & 0.977 & 2.0 & 29.6     & 27.84$^{+0.14}_{-0.20}$ & 26.87$^{+0.06}_{-0.07}$ &  0.4 &  0.95$^{+0.19}_{-0.20}$ & 1.1$^{+0.3}_{-0.3}$ & 1.05$^{+0.08}_{-0.13}$ \cr 
 259 &   7$^{(52)}$ & WHT & 0.408 & 2.0 & 28.6     & 26.39$^{+0.21}_{-0.44}$ & 25.83$^{+0.09}_{-0.12}$ &  1.5 &  0.53$^{+0.40}_{-0.30}$ & 1.4$^{+0.5}_{-0.3}$ & 1.07$^{+0.06}_{-0.15}$ \cr 
 259 &  11$^{(37)}$ & WHT & 0.995 & 2.0 & 29.3     & 27.63$^{+0.18}_{-0.33}$ & 26.66$^{+0.08}_{-0.11}$ &  0.2 &  0.93$^{+0.27}_{-0.28}$ & 1.0$^{+0.4}_{-0.3}$ & 1.00$^{+0.07}_{-0.14}$ \cr 
 260 &   8 & INT & 1.823 & 0.9 & 31.1     & 28.44$^{+0.13}_{-0.19}$ & 27.43$^{+0.06}_{-0.08}$ &  1.3 &  0.97$^{+0.18}_{-0.19}$ & 1.7$^{+0.3}_{-0.3}$ & 1.41$^{+0.08}_{-0.13}$ \cr 
 260 &  44 & WHT & 1.504 & 0.9 & 30.2     & 27.72$^{+0.21}_{-0.42}$ & 27.30$^{+0.07}_{-0.08}$ &  1.4 &  0.39$^{+0.30}_{-0.25}$ & 1.5$^{+0.5}_{-0.3}$ & 1.09$^{+0.08}_{-0.14}$ \cr 
 265 &   1 & WHT & 2.336 & 1.1 & 31.7     & 28.91$^{+0.11}_{-0.15}$ & 28.07$^{+0.04}_{-0.04}$ &  1.5 &  0.80$^{+0.10}_{-0.16}$ & 1.8$^{+0.3}_{-0.3}$ & 1.40$^{+0.09}_{-0.13}$ \cr 
 265 &  17 & WHT & 0.448 & 1.1 & 29.0     & 26.93$^{+0.09}_{-0.11}$ & 25.90$^{+0.08}_{-0.10}$ &  0.4 &  1.01$^{+0.16}_{-0.16}$ & 1.3$^{+0.3}_{-0.2}$ & 1.16$^{+0.07}_{-0.14}$ \cr 
 266 &  32 & INT & 2.460 & 2.1 & 31.0     & 28.57$^{+0.16}_{-0.27}$ & 27.72$^{+0.20}_{-0.39}$ &  0.8 &  0.78$^{+0.98}_{-0.57}$ & 1.5$^{+0.4}_{-0.3}$ & 1.25$^{+-0.0}_{-0.19}$ \cr 
 268 &  11 & WHT & 1.196 & 2.1 & 28.9     & 26.12$^{+0.15}_{-0.23}$ & 26.80$^{+0.24}_{-0.10}$ &  1.2 & --0.62$^{+0.85}_{-0.68}$ & 1.7$^{+0.3}_{-0.3}$ & 0.78$^{+0.11}_{-0.11}$ \cr 
 268 &  24 & INT & 0.251 & 2.1 & 29.2     & 27.73$^{+0.03}_{-0.03}$ & 25.62$^{+0.06}_{-0.07}$ &  1.4 &  2.09$^{+0.06}_{-0.10}$ & 0.9$^{+0.2}_{-0.2}$ & 1.35$^{+0.08}_{-0.13}$ \cr 
 271 &   2 & INT & 0.446 & 2.1 & 29.0     & 27.79$^{+0.08}_{-0.10}$ & 25.72$^{+0.12}_{-0.17}$ &  1.6 &  1.99$^{+0.16}_{-0.25}$ & 0.8$^{+0.3}_{-0.2}$ & 1.26$^{+0.04}_{-0.16}$ \cr 
 271 &   7 & WHT & 1.039 & 2.1 & 30.0     & 28.70$^{+0.10}_{-0.14}$ & 27.02$^{+0.08}_{-0.10}$ &  0.4 &  1.63$^{+0.16}_{-0.20}$ & 0.8$^{+0.3}_{-0.3}$ & 1.12$^{+0.07}_{-0.14}$ \cr 
 272 &  23 & INT & 0.095 & 4.7 & 27.6     & 26.29$^{+0.11}_{-0.15}$ & 24.62$^{+0.09}_{-0.12}$ &  2.8 &  1.65$^{+0.20}_{-0.21}$ & 0.8$^{+0.3}_{-0.3}$ & 1.15$^{+0.06}_{-0.15}$ \cr 
 273 &   4 & INT & 1.046 & 2.8 & 30.5     & 28.54$^{+0.15}_{-0.23}$ & 26.98$^{+0.09}_{-0.11}$ &  0.7 &  1.53$^{+0.25}_{-0.23}$ & 1.2$^{+0.3}_{-0.3}$ & 1.33$^{+0.07}_{-0.14}$ \cr 
 273 &   6 & INT & 0.270 & 2.8 & 29.4     & 27.35$^{+0.06}_{-0.08}$ & 26.39$^{+0.05}_{-0.05}$ &  1.0 &  0.96$^{+0.11}_{-0.11}$ & 1.3$^{+0.2}_{-0.2}$ & 1.16$^{+0.09}_{-0.13}$ \cr 
 273 &  18 & WHT & 0.361 & 2.8 & 28.7     & 27.19$^{+0.15}_{-0.24}$ & 25.64$^{+0.14}_{-0.20}$ &  0.8 &  1.50$^{+0.31}_{-0.28}$ & 0.9$^{+0.3}_{-0.3}$ & 1.15$^{+0.03}_{-0.16}$ \cr 
}
\endtable

\begintable*{3}
\caption{Table 1 (continued): Optical and X-ray continuum parameters.}
\halign{#\hfil
       &\quad#\hfil
       &\hskip0.2truecm\hfil#\hfil\hskip0.2truecm
       &\hskip0.2truecm\hfil#\hfil\hskip0.2truecm
       &\hskip0.2truecm\hfil#\hfil\hskip0.2truecm
       &\hskip0.2truecm#\hfil\hskip0.2truecm
       &\hskip0.2truecm\hfil#\hfil\hskip0.2truecm
       &\hskip0.2truecm\hfil#\hfil\hskip0.2truecm
       &\hskip0.2truecm\hfil#\hskip0.2truecm
       &\hskip0.2truecm\hfil#\hskip0.2truecm
       &\hskip0.2truecm\hfil#\hfil\hskip0.2truecm
       &\hskip0.2truecm\hfil#\hfil
\cr
\noalign{\bigskip}
    FID   &
    S No  &
    Tel   &
     z    &
    \nhgal\ &
    \hfil $L_{\rm 2500}$ &
    $L_{\rm 0.2keV}$ &
    $L_{\rm 2keV}$ &
    $\alpha_{\rm opt}$ &
    $\alpha_{\rm x}$\hfil &
    $\alpha_{\rm os}$ &
    $\alpha_{\rm ox}$ 
 \cr
\noalign{\smallskip}
      (1)\hfil 
    & (2)\hfil 
    & (3)
    & (4)
    & (5)
    & \hfil (6)
    & (7)
    & (8)
    & (9)\hfil
    & (10)\hfil
    & (11)
    & (12) \cr
\noalign{\medskip}
 274 &   8 & INT & 0.156 & 2.1 & 29.2 & 27.24$^{+0.04}_{-0.04}$ & 25.55$^{+0.06}_{-0.07}$ &  1.1 &  1.66$^{+0.07}_{-0.12}$ & 1.2$^{+0.2}_{-0.2}$ & 1.40$^{+0.08}_{-0.13}$ \cr 
 277 &   1 & WHT & 0.595 & 1.7 & 29.8 & 27.72$^{+0.09}_{-0.11}$ & 26.26$^{+0.08}_{-0.10}$ &  0.2 &  1.40$^{+0.12}_{-0.20}$ & 1.3$^{+0.3}_{-0.2}$ & 1.33$^{+0.07}_{-0.14}$ \cr 
 278 &   9 & INT & 0.949 & 1.9 & 30.5 & 27.86$^{+0.13}_{-0.19}$ & 27.04$^{+0.06}_{-0.06}$ & --0.1 &  0.80$^{+0.19}_{-0.18}$ & 1.6$^{+0.3}_{-0.3}$ & 1.32$^{+0.09}_{-0.13}$ \cr 
 278 &  10 & INT & 0.090 & 1.9 & 28.5 & 24.39$^{+0.18}_{-0.32}$ & 25.95$^{+0.05}_{-0.06}$ &  3.0 & --1.54$^{+0.29}_{-0.29}$ & 2.5$^{+0.4}_{-0.3}$ & 0.96$^{+0.09}_{-0.13}$ \cr 
 283 &   6 & INT & 1.219 & 0.5 & 30.5 & 26.74$^{+0.20}_{-0.39}$ & 27.42$^{+0.05}_{-0.06}$ &  1.0 & --0.68$^{+0.27}_{-0.24}$ & 2.4$^{+0.4}_{-0.3}$ & 1.18$^{+0.09}_{-0.13}$ \cr 
 283 &  14 & INT & 0.284 & 0.5 & 29.0 & 25.67$^{+0.20}_{-0.39}$ & 25.68$^{+0.11}_{-0.15}$ &  2.0 & --0.03$^{+0.36}_{-0.31}$ & 2.0$^{+0.4}_{-0.3}$ & 1.25$^{+0.05}_{-0.15}$ \cr 
 286 &   2 & INT & 1.498 & 2.3 & 31.0 & 29.27$^{+0.12}_{-0.18}$ & 27.41$^{+0.08}_{-0.09}$ &  0.5 &  1.78$^{+0.16}_{-0.25}$ & 1.1$^{+0.3}_{-0.3}$ & 1.38$^{+0.07}_{-0.14}$ \cr 
 290 &  21$^{(43)}$ & INT & 2.575 & 1.5 & 31.1 & 28.81$^{+0.25}_{-0.65}$ & 27.92$^{+0.10}_{-0.12}$ &  0.6 &  0.83$^{+0.38}_{-0.31}$ & 1.4$^{+0.6}_{-0.3}$ & 1.20$^{+0.06}_{-0.15}$ \cr 
 293 &  10 & WHT & 0.760 & 4.6 & 29.7 & 28.27$^{+0.16}_{-0.26}$ & 27.05$^{+0.05}_{-0.06}$ &  0.4 &  1.20$^{+0.24}_{-0.22}$ & 0.9$^{+0.4}_{-0.3}$ & 1.03$^{+0.09}_{-0.13}$ \cr 
 293 &  13 & WHT & 0.189 & 4.6 & 27.4 & 25.72$^{+0.27}_{-0.95}$ & 25.47$^{+0.12}_{-0.16}$ &  2.6 &  0.27$^{+0.69}_{-0.60}$ & 1.0$^{+0.8}_{-0.4}$ & 0.73$^{+0.05}_{-0.16}$ \cr 
}

\tabletext{{\sl (1)} RIXOS field number (see Mason et al 1996); {\sl (2)} RIXOS
source number (Mason et al 1996) - the number in brackets is the  radius of the
extraction circle used for the X-ray data (in arcseconds) where it is less than
54 arcsec (see Section 2.1); {\sl (3)} the telescope at which the spectrum was
taken - see Section 2.2; {\sl (4)} redshift;  {\sl (5)} Galactic column density
(10$^{20}$ cm$^{-2}$) - errors are $\sim$10 per cent (see also Section 4.2.1);
{\sl (6)} log of the monochromatic optical luminosity at 2500~\AA\ (erg
s$^{-1}$ Hz$^{-1}$) - error is estimated to be $\sim$50 percent (Section
2.2.2);   {\sl (7)} log of the monochromatic X-ray luminosity at 0.2~\keV\
(erg s$^{-1}$ Hz$^{-1}$) - errors are calculated from the 90\% errors on the
fits (Section 2.1.1);   {\sl (8)} log of the monochromatic X-ray luminosity at
2.0~\keV\ (erg s$^{-1}$ Hz$^{-1}$)  - errors are calculated from the 90\%
errors on  the fits (Section 2.1.1);   {\sl (9)} energy index of the
best-fitting power-law to the optical continuum - error is estimated to be
$\pm$0.5 (Section 2.2.1);  {\sl (10)} energy index of the best-fitting
power-law to the X-ray data - errors are 90\% (Section 2.1); {\sl (11)} and
{\sl (12)}  for definitions see Section 3.2.3 - errors are calculated from the
quoted errors on \luv, \lsoft\ and \lhard; $^i$ - CCD image taken; np -
spectrum not taken at the parallactic angle.} 

\endtable

\section{Results}

In the remainder of this paper, we shall include a discussion of the optical/UV
BBB and the soft X-ray excess in AGN continua, thus it is important to
differentiate between these. Therefore we use the term `BBB' to refer to the
rise in the optical/UV continuum towards the blue, the `soft X-ray excess' to
the excess above an extrapolation of the hard X-ray power-law at energies below 
$\sim$1~\keV, and the `big bump' refers to an optical-to-soft X-ray component
which incorporates the two former features.

We shall be describing much of the changes observed in the RIXOS optical and
X-ray spectra in terms of the gradients of the continuum slopes. Different
terminologies are commonly used in the literature for the slopes in the optical
and the X-ray  ranges, thus to avoid any confusion, we have adopted a single
convention for both, \ie using the terms ``soft'' and ``hard''. A ``soft''
slope falls towards high energies and has a relatively high energy index
$\alpha$ (recalling here that the negative sign is implicit in our definition
of $\alpha$); in the optical region, a soft slope corresponds to a ``red''
continuum. A ``hard'' slope {\sl rises} towards high energy and has a low or
negative $\alpha$; this corresponds to a ``blue'' slope in the optical. To
describe changes in slope, we will use the terms ``soften'' (\ie $\alpha$
increasing) and ``harden'' (where $\alpha$ is decreasing).

\subsection{Sample properties}

\subsubsection{X-ray power-law slope}

The average \ax\ for the RIXOS AGN is 1.07$\pm$0.63 (1$\sigma$ standard
deviation); this is harder than that of WF (1.50$\pm$0.48)  and suggests that
the WF sample has a bias towards AGN with strong soft X-ray excesses and  thus
perhaps also strong BBBs. The RIXOS mean \ax\ is also harder than for the L94
sample of optically-selected quasars (1.50$\pm$0.40).

\beginfigure{3} 
\psfig{figure=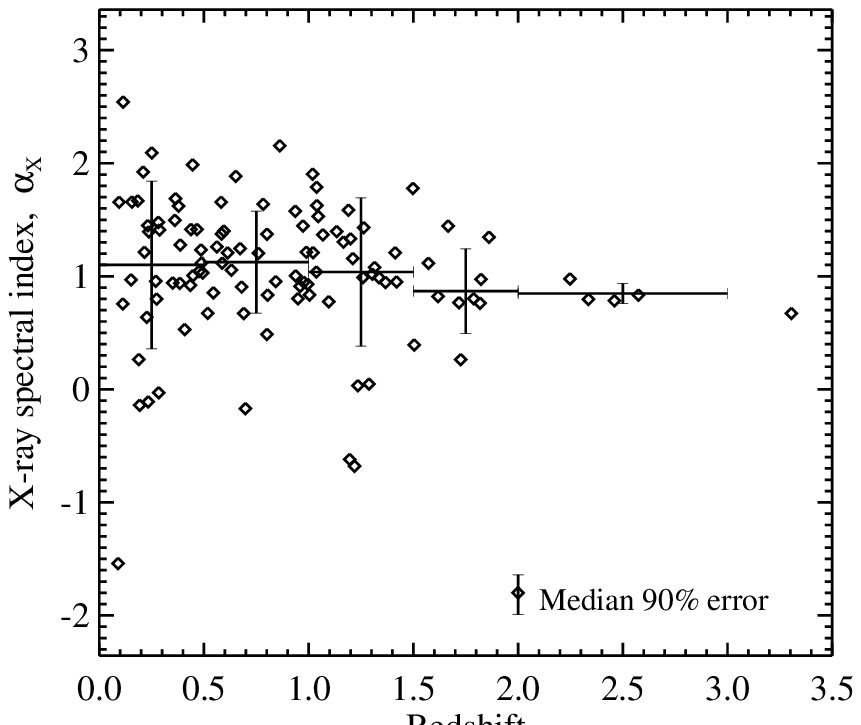,height=3in,width=3.3in,angle=0}
\caption{{\bf Figure 3.} The X-ray spectral index, \ax, 
plotted as a function of redshift.}
\endfigure

We find no significant dependence of \ax\ on redshift for this sample of the
RIXOS AGN (see Fig 3), neither is there any such dependence in the complete
sample  (Mittaz \etal 1996). However the scatter in \ax\ may lessen at high
redshifts, \eg for $2<z<3$, the mean \ax=0.85$\pm$0.09 (standard deviation),
which is similar to the mean slope of \ax=0.89$\pm$0.25 measured at higher
energies in low-$z$ AGN by Comastri \etal (1992). This would be consistent with
two competing trends, {\sl 1.}  a tendency for the soft X-ray excess to be
shifted out of the observed range as the redshift increases, and {\sl 2.} a
reduced sensitivity to absorption in the rest-frame at higher redshifts, as the
most strongly affected regions (at energies below $\sim0.5$\keV) also move out
of range.  

\subsubsection{Optical continuum slope}

The best-fitting optical power-law slopes have been calculated over the full
range of the {\sl observed} spectra and therefore correspond to different
rest-frame wavelength ranges at different redshifts. This must be taken into
account when considering  \aopt, as previous studies have shown that the
optical/UV continua of individual AGN soften towards the UV (\eg Kinney \etal
1987;  O'Brien, Gondhalekar \& Wilson 1988; Zheng \& Malkan 1993), thus we
might expect the RIXOS \aopt\ distribution to show a general softening with
redshift.  The optical indices are plotted in Fig~4 as a function of $z$, but
at $z>0.25$ (where galactic contamination is small), there are no changes with
redshift within 1$\sigma$. This appears to be contrary  to the previous studies
mentioned, but requires further investigation with spectrophotometric
optical-to-UV coverage in the rest-frame to assess its significance. 

\beginfigure{4} 
\psfig{figure=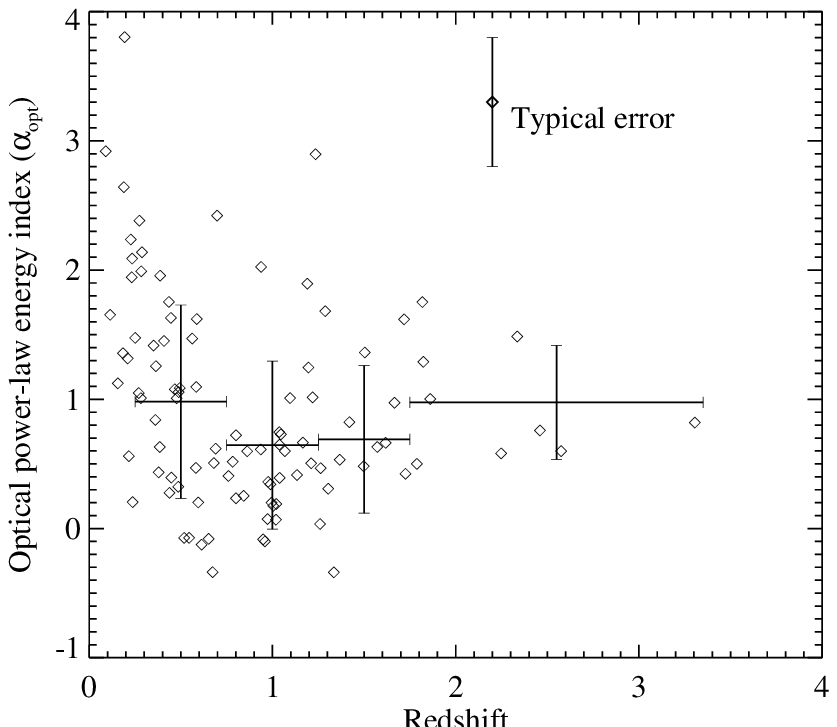,height=3in,width=3.3in,angle=0}
\caption{{\bf Figure 4.} The slope of the best-fitting power law to the optical
continuum, \aopt, plotted as a function of redshift.}
\endfigure

The  mean \aopt\ for the RIXOS AGN is 0.92$\pm$0.76 (1$\sigma$ standard
deviation). This is softer than the mean slope measured for samples of
optically-selected quasars, \eg Neugebauer \etal (1987) found a median \aopt\
of 0.2 while Francis \etal (1991) measured a median index of 0.3. This suggests
that the optically-selected samples may be biased towards objects with
relatively strong optical/UV BBBs.
                                                                           
\beginfigure*{5} 
\psfig{figure=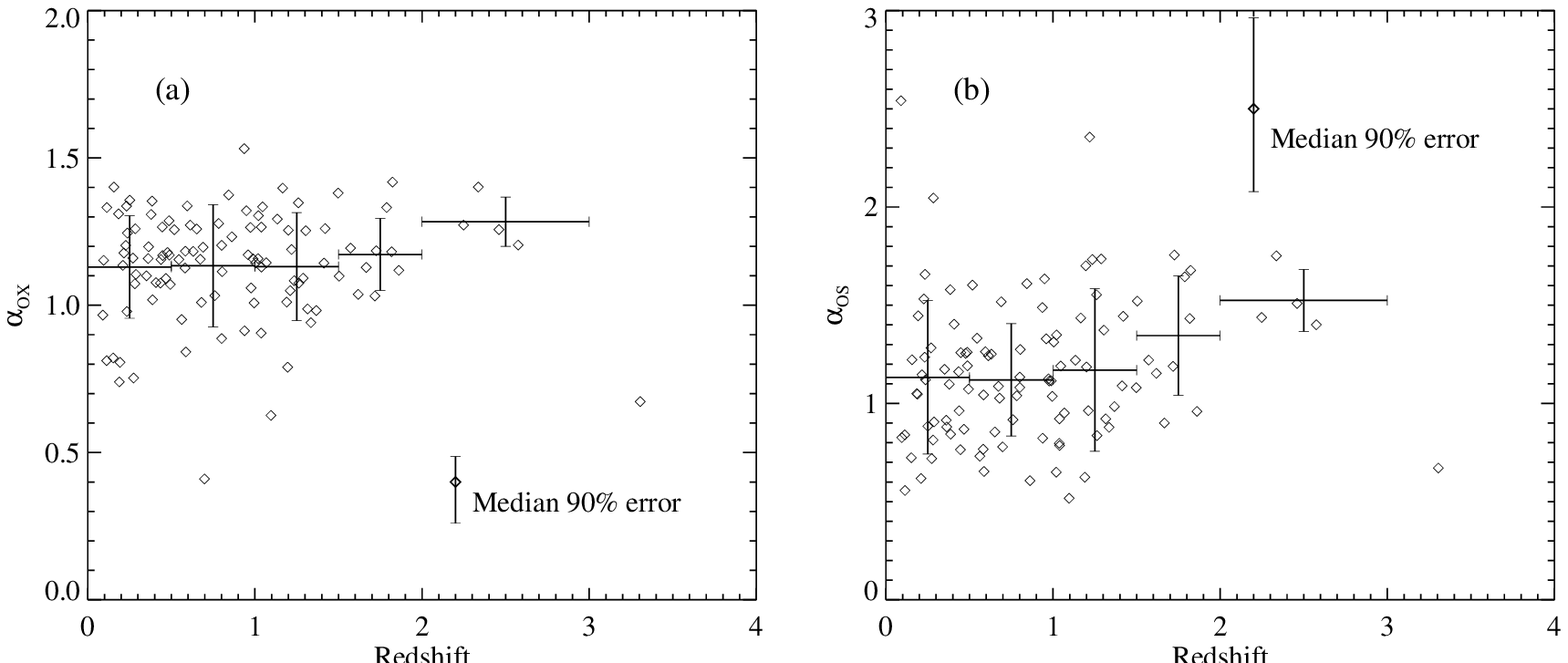,height=3in,width=7.0in,angle=0}
\caption{{\bf Figure 5.} The ratios of optical-to-X-ray luminosity at 2.0~\keV\
and 0.2~\keV, (a) \aox\ and (b) \aos\ respectively, plotted as a function of
redshift. Also shown are the mean indices in intervals of $\Delta z=0.5$ for
$0<z<2$ and the mean for $2<z<3$; vertical error bars represent the
standard deviation ($\sigma$) on the means.} 
\endfigure

\subsubsection{Ratio of optical to X-ray fluxes}

The ratios of optical (\ie 2500~\AA) to X-ray flux at 2.0~\keV\ and  0.2~\keV\
are parameterized using \aox\ (Tananbaum \etal 1979)  and \aos\ (Puchnarewicz
\etal 1992) respectively. The \aox\ and \aos\ are plotted as a function of
redshift in Fig 5 (and listed in Table 1) and show no significant changes
with $z$ (although there is the suggestion of a possible increase in \aos\ at
$z>1.5$).

The mean \aox\ for the RIXOS AGN is 1.14$\pm$0.18 (1$\sigma$ standard
deviation). This is lower than for optically-selected objects, \eg for the
complete sample  of PG quasars, the mean \aox\ is 1.50 with a range of 1.0 to
2.0 (Wilkes \etal 1994). It is also lower than the mean which we calculate for
the WF sample of X-ray-bright AGN, \ie 1.37$\pm$0.23 (1$\sigma$ standard
deviation). Note that the values of \aox\ for WF have slightly different
definitions to those used for the RIXOS objects because WF measure the optical
continuum flux at 2675~\AA. For WF, we have calculated the optical-to-X-ray
ratio using  \aox(2675~\AA)=($L_{\rm 2675A}$-\lhard)/2.635 (where $L_{\rm
2675A}$ is the logarithm of the rest-frame luminosity at 2675~\AA). By
calculating the \aox(2675~\AA)  for the RIXOS AGN, we estimate that the mean
difference between the two definitions of \aox\ [\ie \aox-\aox(2675~\AA)] is
+0.02.   When taking into account the difference in the definitions, we find
that the `corrected' WF mean would be 1.39$\pm$0.23, increasing the discrepancy
between the samples.

The mean 2500~\AA-to-{\sl 0.2}~\keV\ ratio (\aos) is 1.17$\pm$0.37 (1$\sigma$);
however, note that this value may be biased by the selection effect inherent in
\lsoft\ (see Section 2.1.2).

\subsection{Correlations}

All correlations in this section have been evaluated using the Spearman
rank-order correlation coefficient which is a non-parametric method insensitive
to outlying points. The `correlation probabilities' given in the following
sections, \pcorr, are expressed as percentages and are equal to 100\% minus the
percentage probability of being drawn from a random sample. The correlation
probabilities are given for all continuum parameters in Table~2.

\begintable*{4} 
\caption{{\bf Table 2}: Continuum correlation probabilities (\pcorr; percent)}
\halign{        #\hfil 
     &\quad\hfil(#) &\hfil#\quad
     &\quad\hfil(#) &\hfil#\quad
     &\quad\hfil(#) &\hfil#\quad
     &\quad\hfil(#) &\hfil#\quad
     &\quad\hfil(#) &\hfil#\quad
     &\quad\hfil(#) &\hfil#\quad
     &\quad\hfil(#) &\hfil#\quad \cr
          & \multispan2{\hfil z \hfil} & \multispan2{\hfil\luv\hfil}  
          & \multispan2{\hfil\lsoft\hfil} & \multispan2{\hfil\lhard\hfil} 
          & \multispan2{\hfil\aopt\hfil} & \multispan2{\hfil\ax\hfil}   
          & \multispan2{\hfil\aos\hfil}   \cr
          & $r_S$& \% & $r_S$& \% & $r_S$& \% & $r_S$& \% 
          & $r_S$& \% & $r_S$& \% & $r_S$& \% \cr
\noalign{\medskip} 
\luv\   &   0.8 & $>$99 \cr
\lsoft\ &   0.8 & $>$99 &   0.4 & $>$99 \cr
\lhard\ &   0.9 & $>$99 &   0.5 & $>$99 &   0.4 & $>$99 \cr
\aopt\  & --0.3 &    93 & --0.4 & $>$99 & --0.4 & $>$99 & --0.3 & 92 \cr
\ax\    & --0.2 &    37 &   0.0 &     4 &   0.3 &    99 & --0.2 & 52 & --0.2 & 68 \cr
\aos\   &   0.2 &    55 &   0.4 & $>$99 & --0.1 &     2 &   0.2 & 75 &   0.0 &  4 & --0.6 & $>$99 \cr
\aox\   &   0.1 &     3 &   0.5 & $>$99 &   0.3 &    95 &   0.1 &  3 & --0.3 & 97 &   0.4 & $>$99 & 0.4 & $>$99 \cr
} 
\tabletext{The correlation coefficients between two different  luminosities
were calculated after having adjusted for redshift (see Section 4).}
\endtable 

One problem in searching for correlations in flux-limited samples of this kind,
is that luminosities are dominated by the correction for the distance to the
quasar, so that the continua of low-redshift objects have low luminosities
while high-$z$ objects tend to have high luminosities. This usually results in
very strong correlations between luminosities in different bands and may not
reflect a real dependence.  In order to allow for this, when comparing
parameters which are both luminosities, each pair of parameters is divided by a
normalizing factor which  has the redshift dependence built into it. The factor
used is the rest-frame luminosity equivalent to an observed flux of
1$\times10^{-15}$ erg cm$^{-2}$ s$^{-1}$ (arbitrarily chosen to match the
observed-flux typically measured in the optical emission lines). In Table 2,
correlation probabilities have been calculated after dividing the luminosities
by this factor.

\subsubsection{Luminosity and X-ray slope}

We find no relationship between the X-ray spectral slope and the luminosity at
2~\keV; the apparent correlation between \ax\ and \lsoft\ is due to a selection
effect (see Section 2.1.2). There is no correlation between \ax\ and the
optical luminosity at 2500~\AA.

\subsubsection{Luminosity and optical slope}

Table~2 shows that the slope of the optical continuum is strongly 
anti-correlated with \luv, \ie as the 2500~\AA\ luminosity increases, \aopt\
becomes harder. This could be a consequence of measuring \luv\ from optical
power-law slopes which are measured in the observer's frame; the rest-frame
flux at 2500~\AA\ does not fall within the observed wavelength range below
$z$=0.4 and must be extrapolated from the power-law fit at longer wavelengths.
\luv\ would then be high when \aopt\ was hard which might influence the
correlation. It may also be due to contamination by the host galaxy which would
have the effect of softening \aopt\ when the nuclear emission (and therefore
\luv) is low, indeed, Wilkes \etal (1994) found that starlight contamination
was an important factor for AGN with an \luv\ lower than $\sim$29. However at
$z$=0.4, the rest-frame wavelength range over which \aopt\ is calculated lies
at $\sim$2500-6400~\AA\ where galactic contamination is already relatively low
and should have only a small effect (if any) on the measurement of the optical
slope.

Thus to eliminate these effects, we have excluded the $z\le 0.4$ objects  and
find that while the anti-correlation does disappear for $z>1.0$ ($r_{\rm
S}$=--0.1; \pcorr=2\%), it remains strong at $0.4<z<1.0$ ($r_{\rm S}$=--0.6 and
\pcorr=99.98\%; see also Fig 6). A similar effect is seen in the correlations
between \aopt\ and \lsoft\ and between \aopt\ and \lhard\ (\ie the data are
correlated for $0.4<z<1.0$, but not at $z>1.0$), although the latter
dependences may be induced by strong correlations between \luv, \lsoft\ and
\lhard\ (see Table 2). 

\beginfigure{6} 
\psfig{figure=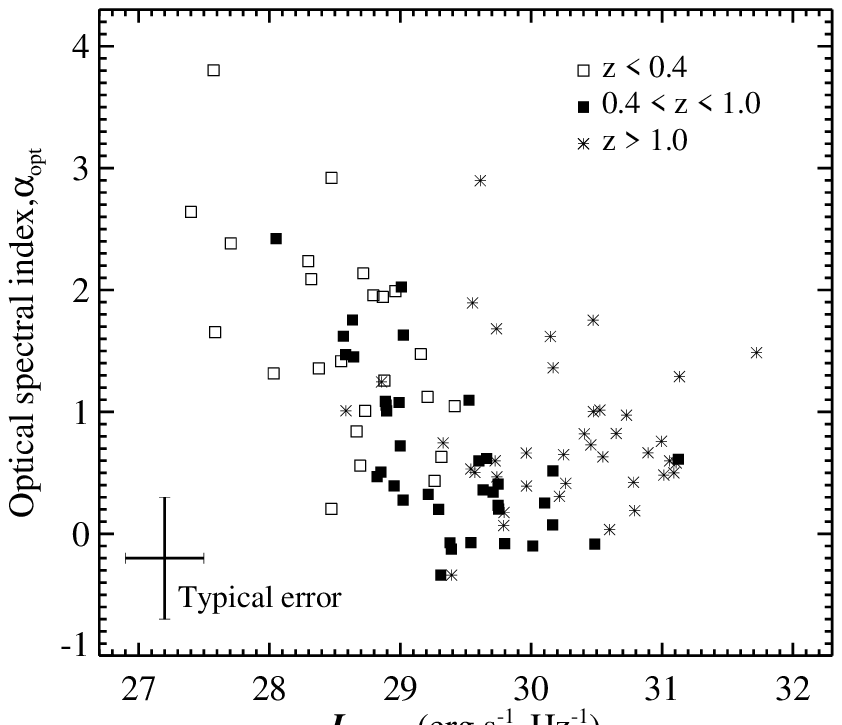,height=3in,width=3.3in,angle=0}
\caption{{\bf Figure 6.} The optical spectral index, \aopt, plotted as a
function of optical luminosity at 2500~\AA, \luv. The AGN are plotted
according to their redshift, \ie AGN with $z<0.4$ are plotted as open squares,
$0.4<z<1.0$ are plotted as solid squares and $z>$1.0 as asterisks.}
\endfigure

This might be a luminosity, rather than a wavelength-dependent effect, but it
is difficult to distinguish between these due to the strong redshift dependence
in the luminosities. Nonetheless, the correlations imply that at relatively low
bolometric luminosity, the $\lambda > 5000$~\AA\ region hardens  as the
luminosity increases, while  at higher bolometric luminosities, the continuum
slope at shorter wavelengths ($\lambda < 5000$~\AA) is independent of
luminosity. A similar effect was reported by Zheng \& Malkan (1993), who found
that the rest-frame  6500-3100~\AA\ continua of their quasars and Seyfert 1s
also harden significantly with luminosity ($\Delta\alpha$=1.1) while at shorter
wavelengths (3100-1250~\AA\ in the rest-frame), the change in slope is smaller
over the same luminosity range ($\Delta\alpha$=0.5). To explain this, they
suggest that  either (1) the BBB shifts from the UV into the optical as the
luminosity increases, or (2)  that the BBBs are stronger in high-luminosity
AGN.

\subsubsection{Optical-to-X-ray luminosity ratios}

We find evidence for  a strong correlation between \aox\ and \ax\
(\pcorr=99.8\%; see Fig 7). The 2500~\AA\ luminosity is measured independently
of the X-ray parameters and \lhard\ is not correlated with \ax, thus we
conclude that this is {\sl not} due to any selection effects. Similar
correlations have also been reported by WF and L94. This relationship suggests
that when the 2500~\AA-to-2~\keV\ ratio is high, the 0.2-2~\keV\ slope is
relatively soft so that it rises steeply towards the EUV, and the overall
2500~\AA-to-2~\keV\ continuum is convex. Conversely, when \aox\ is low, \ax\ is
hard (\ie it  falls towards the EUV) and the 2500~\AA-to-2~\keV\ continuum is
concave. Assuming that \aox\ is an indicator of the strength of the BBB
(because the BBB is strong at 2500~\AA\ but negligible at 2~\keV) and that \ax\
measures the strength of the soft X-ray excess, then this implies that the
optical/UV BBB and the soft excess may be part of the same component. 

\beginfigure{7} 
\psfig{figure=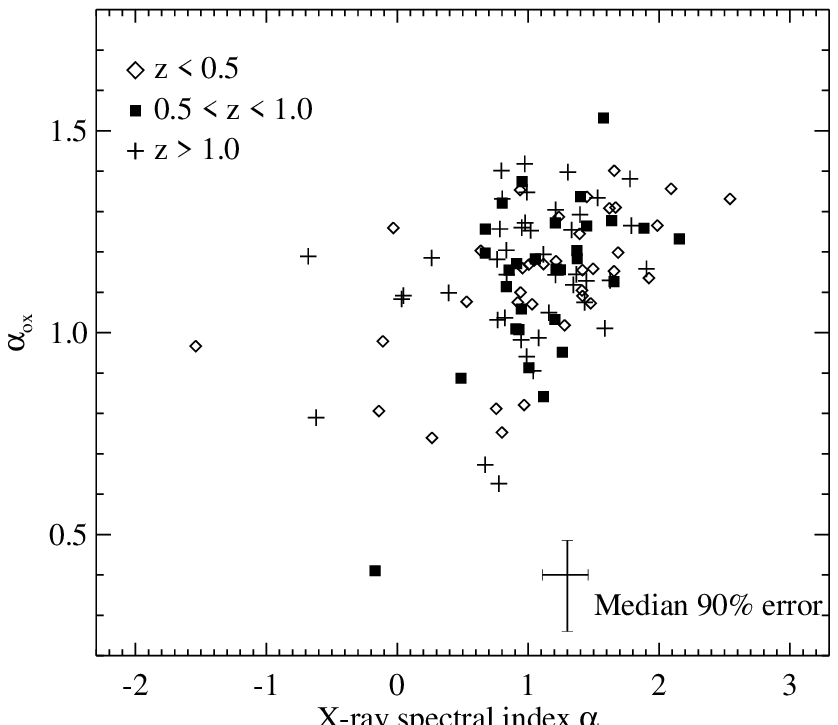,height=3in,width=3.3in,angle=0}
\caption{{\bf Figure 7.} The optical-to-X-ray ratio, \aox,
plotted as a function of  X-ray spectral index, \ax. The data are plotted
according to the redshift of the AGN, \ie $z<0.5$ are plotted as diamonds,
$0.5<z<1.0$ are plotted as filled boxes and $z>1.0$ as crosses.}
\endfigure

It is important to be mindful of possible selection effects when considering
such correlations, particularly in the X-ray data where the spectral resolution
is low, and the response of the detector favours particular energies within the
nominal bandpass. For example, Ulrich \& Molendi (1996) have cautioned that the
2~\keV\ flux inferred from ROSAT spectra is coupled to the spectral slope. They
suggest that this may induce the correlations seen by WF and L94 because
2~\keV, the reference energy for \aox,  is far removed from the centroid energy
of the counts detected by the ROSAT PSPC for a typical low-redshift Seyfert,
which is $\sim$0.4~\keV\ for a slope, \ax$\sim$1.6.  However, there are a
number of reasons why this effect is unlikely to induce the correlations seen
in the RIXOS AGN. Firstly, the  typical error bars, which reflect the coupling
between spectral normalization and slope in the size of the uncertainty in
\aox, are significantly smaller than the observed range of the correlation. 
Secondly,  the X-ray spectra of  objects in our sample are typically harder
than those in  WF and L94 and cover a wider range in \ax; both of these factors
will reduce the  Ulrich \& Molendi effect. Thirdly, the RIXOS AGN extend to
much higher redshifts than WF and L94 moving the equivalent of the 2~\keV\ flux
in the quasar's rest-frame closer to and below the centroid energy of the
photons registered in the detector frame, yet  the correlation is still seen in
the sample of RIXOS objects above $z$=1  (see Fig~7). 

\beginfigure{8} 
\psfig{figure=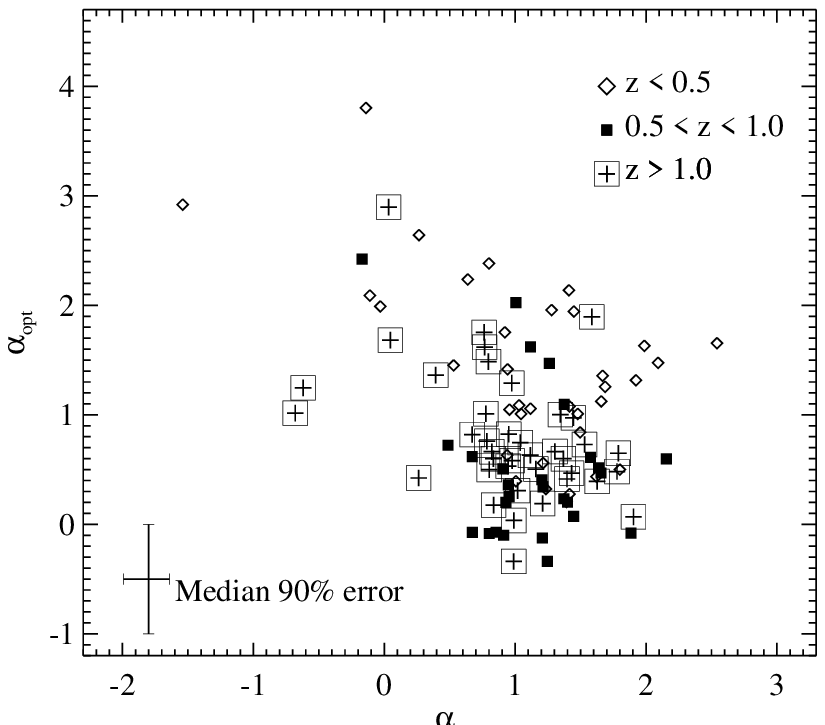,height=3in,width=3.3in,angle=0}
\caption{{\bf Figure 8.} Optical power-law slope, \aopt, plotted as a function
of the X-ray power-law slope, \ax. The data are plotted
according to the redshift of the AGN, \ie $z<0.5$ are plotted as diamonds,
$0.5<z<1.0$ are plotted as filled boxes and $z>1.0$ as crosses within boxes.}
\endfigure

We also find evidence of a correlation between \lsoft\ and \aox\ (\pcorr=95\%),
\ie when the optical-to-X-ray ratio is high, the luminosity at 0.2~\keV\ is
also high, although this may be an induced correlation as both \lsoft\ and
\aox\ are also correlated with \ax\ (and the \lsoft-\ax\ correlation is due to
a selection effect; see Section 2.1.2). 

Table 2 also indicates a correlation between \luv\ and \aox, but this is
expected since  the \lhard\ distribution is random with \ax\ (Section 4.2.1)
and \luv\ is closely  related to \aox\ (\ie \luv=2.605\aox +\lhard). However,
significant dependences have been reported elsewhere between luminosity and
\aox\ (\eg Reichert \etal 1982; Avni \& Tananbaum 1986; Wilkes \etal 1994). The
correlation between \luv\ and \aos\ is consistent with the selection effect
discussed in Section 2.1.2.

\subsubsection{Optical and X-ray slopes}

There is an  anti-correlation between \aopt\ and \aox, implying that when the
2500~\AA-to-2~\keV\ ratio is high, the optical slope is relatively hard. There
is also a weak (\pcorr=68 per cent) anti-correlation between the optical and
X-ray slopes when the sample is considered as a whole, however there appears to
be a redshift dependence in the relationship (see Fig 8).  At low redshifts,
\aopt\ is likely to be affected by the host galaxy and thus may be soft
relative to higher-$z$ AGN (cf.~Fig 4); indeed Fig 8 demonstrates that the
$z<0.5$ AGN are systematically  displaced to softer values of \aopt\ relative
to the higher redshift objects. When the data are divided into redshift bands, 
the correlation strengthens, \eg at $z<0.5$,  \pcorr=80 per cent (r$_S$=--0.4)
while for $z>0.5$, \pcorr\ also rises to 80 per cent (r$_S$=--0.3).  However,
the strongest correlation is found in the high-$z$ AGN, \ie for $z>1.0$,
\pcorr=96 per cent (r$_S$=--0.5). The degree of correlation (as measured using
the Spearman rank-order coefficient) is affected by the tendency of the data to
cluster around \ax=1 (see Fig 3) and the relative paucity of very hard and soft
\ax\ objects which is an expected consequence of the RIXOS selection criteria.
Samples with a more homogeneous dispersion in \ax\ and improved measurements of
the intrinsic slopes will provide a more definitive answer.

Nonetheless, even given the relatively basic methods for measuring \ax\ and
\aopt, the data for these AGN show compelling evidence for a correlation
between the two slopes. The correlation implies that when the optical continuum
is hard, the soft X-ray slope is soft (and vice-versa) so that  both the
optical and X-ray  slopes fall and rise towards the EUV together.  

This is the first time that a relationship between the optical and X-ray slopes
has been found. L94 found no correlation in their PG quasar sample,  but since
the RIXOS data suggest that  \aopt\ changes only very slowly with \ax\ (see Fig
8), a sample covering a wide range in \ax\ and/or \aopt\ is essential to
identify any trend; the L94 AGN were UV-excess quasars whose range in \aopt\
(--0.2 to 0.9) may have been too small for this correlation to be detected.

Together with the dependence of \aox\ on \ax, these correlations are additional
evidence linking the optical/UV BBB to the soft X-ray excess. In short, when
\ax\ is soft, the optical continuum (\aopt) is hard and the  optical-to-X-ray
ratio (\ie \aox) is high.  Conversely, when \ax\ is hard, \aopt\ is soft and
\aox\ is low. 

\begintable{5} 
\caption{{\bf Table 3.} Mean optical and X-ray spectral indices ($0.25\le
z\le3.25$)}
\halign{        #\hfil 
     &\quad\hfil#\hfil\quad    
     &\quad\hfil#\hfil\quad    
     &\quad\hfil#\hfil\quad \cr
\noalign{\medskip} 
       &  --0.1$<$\ax$\le$0.6 & 0.6$<$\ax$\le$1.3 & 1.3$<$\ax$\le$2.0 \cr
\noalign{\smallskip} 
\aox\  &     1.10$\pm$0.05  &  1.13$\pm$0.02  &  1.21$\pm$0.02  \cr 
\aos\  &     1.62$\pm$0.12  &  1.21$\pm$0.04  &  0.99$\pm$0.04  \cr
\aopt\ &     1.50$\pm$0.33  &  0.72$\pm$0.10  &  0.72$\pm$0.10  \cr
} 
\tabletext{Errors quoted are the errors on the mean (\ie $\sigma$/(N-1), where
N is the number of data points).}
\endtable 

\subsection{Mean optical-to-X-ray continua}

To illustrate the changes in the shape of the  optical-to-X-ray spectrum, we
have calculated three `mean' spectra at different values of \ax, covering
6000~\AA\ to 2~\keV. The mean \aox, \aos\ and \aopt\ were calculated in three
bins centred on \ax=0.25, 0.95 and 1.65 and each covers an interval 
$\Delta$\ax=0.7. Only AGN with redshifts in the range $0.25<z<3.25$ were used
to avoid possible host galaxy contamination in  the optical at low redshifts
and the highest-$z$ object whose X-ray and optical continuum slopes have
relatively high errors.

The mean indices and the errors on the means  are shown in Fig~9a-c and
listed in Table 3. These were then used to construct the mean `spectra' at
\ax=0.25, 0.95 and 1.65 (Fig~9d; they have been normalized at 2.0~\keV) and
illustrate more clearly the changes in the spectra suggested by the data. When
\ax\ is very soft, \aopt\ is hard and the spectrum is ``convex'' (indicated by
the dashed line in Fig 9d), while when \ax\ is hard, \aopt\ is soft and the
overall spectrum is ``concave'' (the dot-dashed line in Fig 9d). 


\beginfigure*{9} 
\psfig{figure=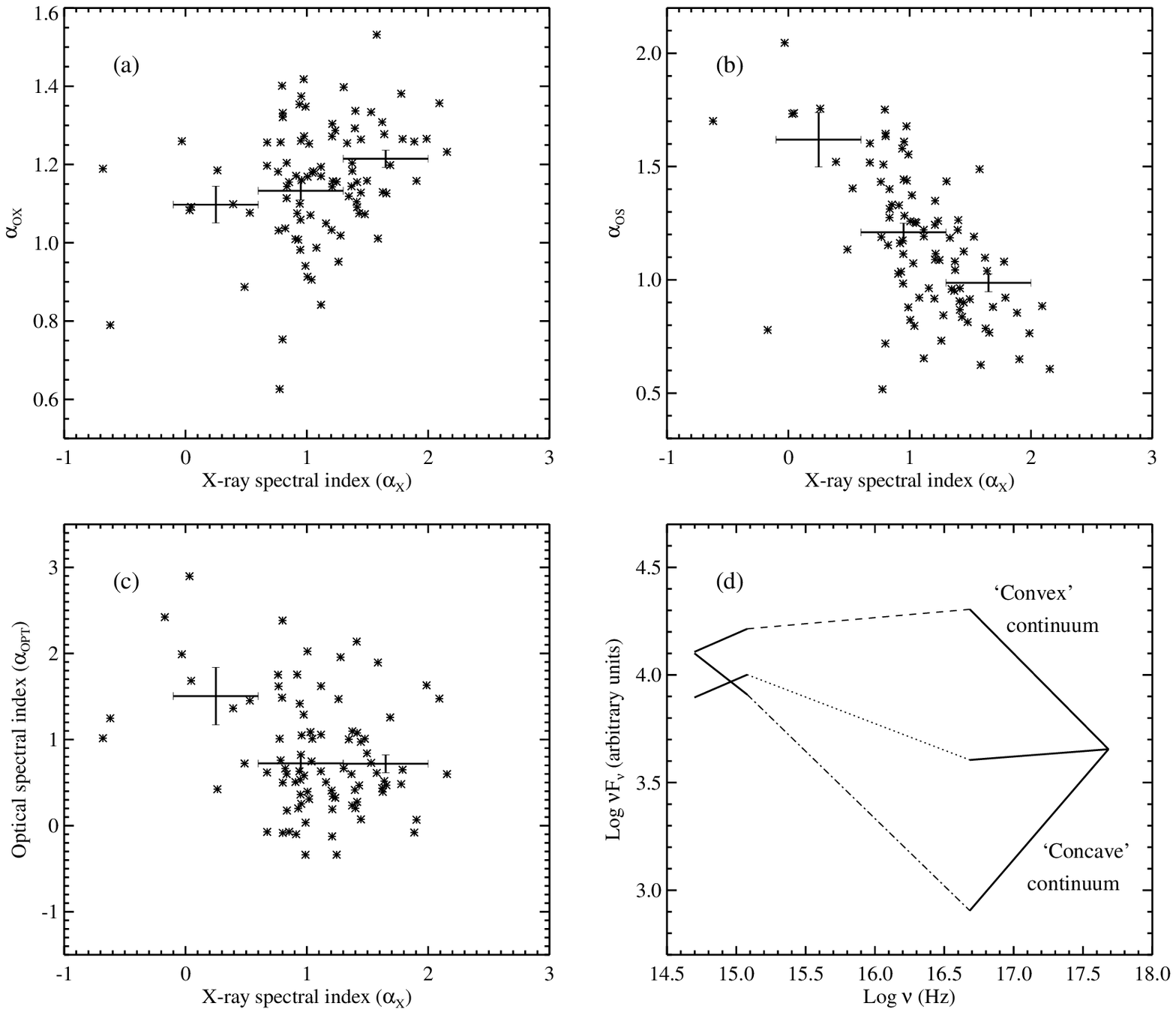,height=6in,width=7.0in,angle=0}
\caption{{\bf Figure 9} The calculation of three `mean' RIXOS AGN spectra from
6000~\AA-to-2~\keV. (a)-(c) Data in the range $z$=0.25 to $z$=3.25.
Also plotted as
horizontal bars are the means  in each of three equally spaced \ax\ bins for
these data (vertical bars indicate the errors on the
means; see Table 3). {\sl (a)} \aox\ plotted against \ax; {\sl (b)} \aos\
plotted against \ax; {\sl (c)} \aopt\ plotted against \ax. {\sl (d)} Three mean
continuum spectra calculated from the means shown in (a)-(c). Only the separate
optical and X-ray slopes and their relative normalization are known; we have no
information in the 2500~\AA-to-0.2~\keV\ region. The spectra may be identified
by the broken lines joining the optical and X-ray regions; \ie \ax=0.25 is
indicated by the dot-dash line (also referred to as the ``concave'' spectrum), 
\ax=0.95 is indicated by the dotted line and \ax=1.65 (referred to as the
``convex'' spectrum) is shown by the dashed line.}  
\endfigure

\section{Modelling the continuum}

Recent work on the continua of Seyfert 1s and quasars has contributed much to
our understanding of AGN spectra, especially with regard to the possibility of
a single `big bump'. Both WF and L94  identified a strong \aox\ vs \ax\
correlation; WF suggested that changes in the flux of a big bump  relative to
an invariant underlying component could explain the spectral differences from
source to source and that the shape of the big bump from the UV to soft X-rays
showed little change. 


The overall changes in  optical-to-X-ray continua of the RIXOS AGN 
(illustrated in Fig 9d) could also be interpreted as being due to the sum of
two components; a big bump spectrum, whose flux varies significantly from
source to source, and a relatively invariant underlying continuum (\ie a 
similar model to that proposed by WF). If this model is to satisfy the full
range of the observed RIXOS spectra, the underlying continuum must be concave
with a break between 2500~\AA\ and 0.2~\keV\  (see the mean spectrum indicated
by the dot-dashed line in Fig 9d; we stress that we have {\sl no} information
about the shape of spectrum between 2500~\AA\ and 0.2~\keV).

However, an underlying continuum of this form is difficult to reconcile with
observations at longer and shorter wavelengths.  The slope of the underlying 
continuum in soft X-rays (\ie based on the RIXOS AGN with the hardest \ax) 
would be equivalent to  \ax$\sim$0, which is significantly harder than the
indices of Seyferts and quasars at higher energies [$\sim$2-50~\keV; for
Seyferts \ax=0.7 while for quasars, \ax\ is closer to 1.0 with small
dispersions about the mean (Mushotzky, Done and Pounds 1993 and references
therein)]. Also, an extrapolation of the rather soft optical continuum  to
longer wavelengths (equivalently \aopt$\sim$2 when \ax$\sim$0) predicts a very
strong flux in the IR, whereas typically the continua of AGN spectra are
generally harder with an $\alpha\sim$1 extending down to $\sim$100$\mu$m (for
examples see Elvis \etal 1994). 

It is possible to have an underlying continuum of this shape {\sl and} remain
consistent with the IR and hard X-ray continua by assuming that the underlying
continuum is a simple power-law which is absorbed predominantly in the
optical/UV and soft  X-ray regions, \eg by cold gas and dust. However, it is in
practice difficult to envisage a physical model where the nuclear power-law
source lies behind significant amounts of cold, dusty gas, leaving the
big bump-emitting region bare. 

Thus we conclude that the changes in continuum shape for the RIXOS AGN cannot
simply be due to either {\sl 1.}  a continuum which is the sum of a power law
and a big bump with different relative strengths, and which is unabsorbed; or
{\sl 2.} a continuum which is the sum of an absorbed power law and an
unabsorbed big bump.

Neither are the observed correlations  consistent with a big bump component
which `shifts'  between the optical and soft X-ray regions. In this case, when
the big bump spectrum lies at relatively low frequencies (\ie it is strong in
the optical but weak in soft X-rays), we would expect a hard \aopt\ and a hard
\ax; conversely when the spectrum lies at relatively high frequencies (\ie weak
in the optical but strong in soft X-rays), both \aopt\ and \ax\ would be soft.
Thus we would expect to see a {\sl correlation} of \aopt\ with \ax\ rather than
the {\sl anti-}correlation observed.  Examples of shifting spectra are AD
models with different black hole masses and bremsstrahlung components with a
range of temperatures. 

\subsection{Cold absorption}

Instead we propose that the changes in continuum shape which are responsible
for producing the observed correlations, may be primarily due to absorption of
the entire intrinsic spectrum by cold gas and dust. The cold gas would be 
responsible for absorption in soft X-rays while the dust would be the source of
absorption in the optical and UV. The intrinsic spectrum may be the sum of a
power-law and a big bump. The big bump temperature, power-law slope and big
bump-to-power-law continuum flux ratio may also vary from source to source,
introducing scatter into the data, however it is the  absorption which results
in the observed correlations.  

To test whether this crude model is consistent with the observed spectra, the
optical and X-ray spectral slopes and ratios were simulated  by constructing a
model for the intrinsic spectrum and modifying this for the effects of
absorption by cold gas and dust.  Throughout the remainder of this section,  we
investigate whether this model can predict the data which is actually observed,
while the physical consequences of such a model for AGN are discussed in
Section 5. This is intended to be a preliminary assessment and consistency
check of the model; a more rigorous treatment awaits a future paper.

The intrinsic spectrum in the model was assumed to be the sum of an underlying
continuum and a big bump and its flux was fixed while the amount of absorption
was varied.  A simple power law with a slope of $\alpha$=1.0 was adopted for
the underlying optical-to-X-ray continuum; the chosen model for the big bump is
discussed in Section 4.1.1. In soft X-rays, the Morrison and M$^c$Cammon (1983)
absorption cross-sections were assumed while in the optical and UV, the
reddening curve of Cardelli, Clayton \& Mathis (1989) was used.  A Galactic
dust-to-gas ratio [where E(B-V)=1.0 corresponds to an \nh\ of 6$\times10^{21}$
cm$^{-2}$; Ryter, Cesarsky \& Audouze 1975; Gorenstein 1975] was assumed for
the dust abundance. The model \aopt\ was calculated by fitting a power-law to
the model spectrum; the 2200~\AA\ dust absorption feature was ignored in the
fitting as this is rarely observed in the UV spectra of AGN (Cardelli \&
Clayton 1991; Mathis \& Cardelli 1992). The X-ray spectral index was derived by
first binning the model spectrum into three `colours' which reflected the
binning used in the analysis of the data (see Section 2.2), then fitting a
power-law to these colours. 

Because the AGN spectra sample such a wide range in $z$, it is necessary to
include the effects of redshift on the model.  The optical and X-ray slopes 
have been measured in fixed ranges in the frame of the {\sl observer},
therefore they  correspond to different parts of the {\sl intrinsic}  quasar
spectrum, depending on $z$, and this must be taken into account when trying to
reproduce the data. Therefore the  models have been calculated at different
redshifts ($z$=0.2, 0.5, 1.0, 1.5 and 2.5) to check that the predictions are
consistent with the data. Both the optical and X-ray ranges were corrected by a
factor of $1+z$ before the model indices were calculated. 

\subsubsection{Choosing a model for the big bump}

To find an appropriate model for the big bump, we have compared the ``convex''
spectrum in Fig 9d (\ie the mean \ax=1.65 RIXOS spectrum, which we assume to be
dominated by the big bump) with several different models including ADs and
bremsstrahlung spectra.  The convex spectrum  is much broader in wavelength
than for many types of AD model, whether geometrically-thick (\eg Madau 1988)
or thin (\eg Sun and Malkan 1989), although it does compare well with the
modified Czerny \& Elvis (1987) model of an AD which is surrounded by a hot
($T=10^8$~K) corona.  

\beginfigure{10} 
\psfig{figure=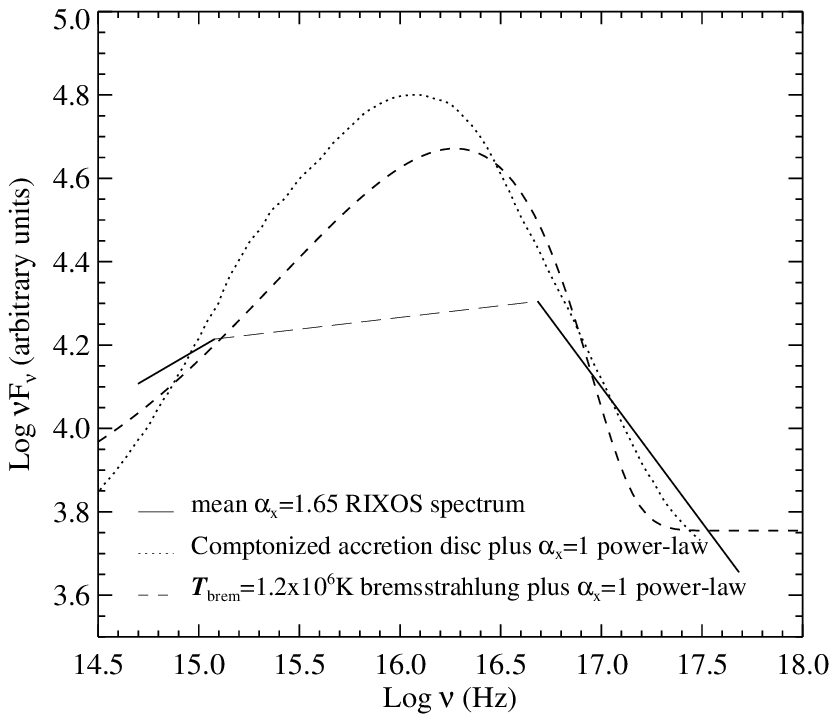,height=3in,width=3.3in,angle=0}
\caption{{\bf Figure 10.} A comparison of the mean convex RIXOS spectrum (\ie
the  mean \ax=1.65 spectrum from Fig 9d; plotted here as a solid/long-dashed 
line )  with two models for the BBB and soft X-ray excess; {\sl 1.} the Czerny
\& Elvis (1987) model of an accretion disc surrounded by  a hot ($T=10^8$~K)
corona (dotted line),  and {\sl 2.} a \tbrem=1.2$\times10^6$~K simple
bremsstrahlung spectrum (dashed line). An \ax=1.0 power-law has also  been
added to both models.}
\endfigure

A simple thermal bremsstrahlung spectrum with a temperature,
\tbrem=1.2$\times10^6$~K also compares well with the convex spectrum; 
bremsstrahlung models have previously been used to fit the big bumps of other
AGN with soft excesses, \eg PG1211+143 was fitted with a \tbrem=8$\times10^5$~K
spectrum (Barvainis 1993) and E1346+266 compared well with a bremsstrahlung
model with \tbrem=7$\times10^6$~K (Puchnarewicz \etal 1994).

The AD-plus-corona and bremsstrahlung models are plotted in Fig 10 with the
mean \ax=1.65 `observed' spectrum (an $\alpha$=1.0 power-law has been added to
the model spectra). In the subsequent modelling, for simplicity, we have used a
thermal bremsstrahlung spectrum for the big bump, although the AD-plus-corona
provided an equally acceptable alternative. 

\subsection{Finding the best fit}

In this section, we assess the ability of the model to reproduce the observed
correlations between \ax\ and \aox\ and between \ax\ and \aopt. While we have
searched for the best-fitting models over a wide range of parameter space (\ie
in \tbrem\ and the relative normalization of the bremsstrahlung component), the
model has not been allowed to converge formally, therefore the presence of
other minima in the data are possible. This approach was considered adequate
for the purpose of a first evaluation of the model.

\beginfigure*{11}  
\psfig{figure=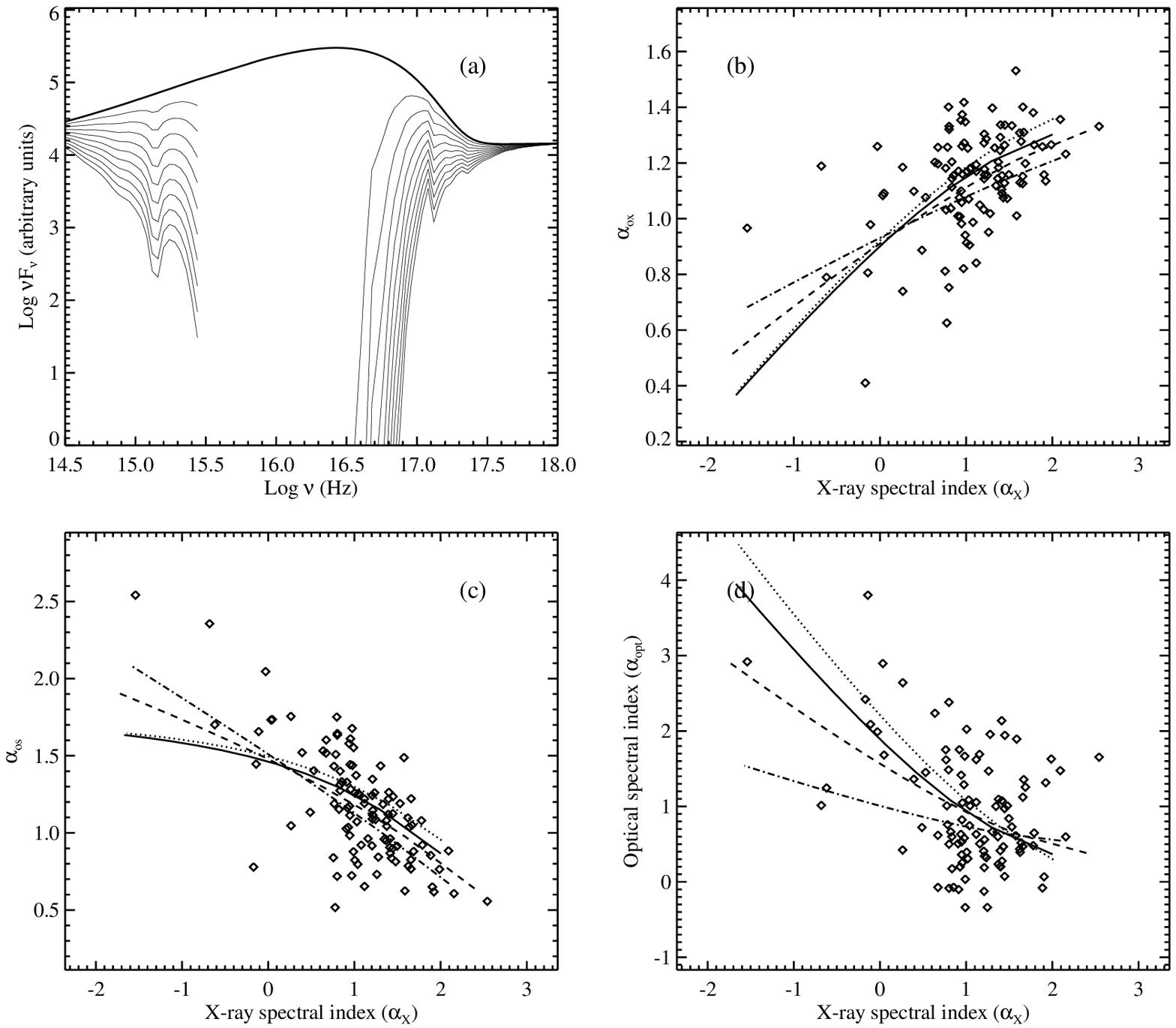,height=6in,width=7.0in,angle=0}
\caption{{\bf Figure 11.} A comparison of the RIXOS AGN data  with the model
which assumes the sum of  a bremsstrahlung spectrum and an  \ax=1.0 power-law
and which is absorbed by cold gas and dust.  The flux in the power-law and the 
bremsstrahlung spectrum remain constant while the absorbing column density is
varied; {\sl (a)} illustrates the range of continua predicted by the $z$=1.0
model: the thick line is the unabsorbed spectrum while each curve plotted as a
thin line represents absorption by an additional 6$\times10^{20}$ cm$^{-2}$. In
(b)-(d), models are plotted for $z$=0.2 (dot-dashed line),  $z$=0.5 (dashed
line), $z$=1.0 (solid line) and  $z$=1.5 (dotted line). Model spectral indices
are compared with the data in {\sl (b)} where \aox\ is plotted against \ax; 
{\sl (c)} \aos\ is plotted against \ax, and {\sl (d)} \aopt\ is plotted against
\ax. In all cases, the observed data are plotted as diamonds.}
\endfigure

In searching for the model which provided the best representation of the data
at each redshift,  we changed two of the parameters, \tbrem\ and the flux in
the bremsstrahlung component relative to the power law. The model was
calculated over a grid of \tbrem\ and big bump-to-power-law flux ratios and at
each point in the grid, \tbrem\ and the flux ratio were fixed while the
intrinsic \nh\ was varied. The \aox\ versus \ax\ and \aopt\ versus \ax\
distributions were fitted simultaneously and the goodness-of-fit was evaluated
using the $\chi^2_\nu$ statistic (taking into account the errors on \aox\ and
\aopt; see Table 1).  The `best-fits'  described throughout the remainder of
the paper represent the points in the grid with the lowest $\chi^2_\nu$. The
\aos\ versus \ax\ distribution is biased by a selection effect and may not be a
real physical dependence (Section 2.1.2), therefore it was not included in the
fits (although the predicted model has been compared with the data in the plots
as a check for consistency).  The models were only fitted to the data
appropriate to the redshift assumed, \ie the models at $z$=0.2, 0.5, 1.0, 1.5
and 2.5 were fitted to the data in the $z$=0-0.25, 0.25-0.75, 0.75-1.25,
1.25-1.75 and 1.75-3.5 ranges respectively.

\begintable{6} 
\caption{{\bf Table 4} Parameters of the best-fitting models to the RIXOS AGN
continua (plotted in Figs. 11 and 12)}
\halign{  \quad\quad\hfil#\hfil\quad\quad          
     &\quad\hfil#\hfil\quad\quad    
     &\quad\hfil#\hfil\quad\quad    
     &\quad\hfil#\hfil\quad\quad \cr
\noalign{\medskip} 
   $z$  &       \tbrem\       & \ratio\ & \nh\ \cr
        &       (1)           &   (2)   &  (3) \cr
\noalign{\smallskip} 
  0.2   &   1.7$\times10^6$K  &   10    &  1.6 \cr
  0.5   &   1.7$\times10^6$K  &   20    &  4.0 \cr
  1.0   &   1.3$\times10^6$K  &   20    &  6.0 \cr
  1.5   &   1.2$\times10^6$K  &   30    &  7.0 \cr
  2.5   &   1.0$\times10^6$K  &   40    &  4.0 \cr
} 
\tabletext{(1) Temperature of the bremsstrahlung component. (2) Ratio of flux
in the bremsstrahlung component to flux in the power-law component at 0.2~\keV.
(3) Maximum column density of the gas used to produce the model in units of
$10^{21}$ cm$^{-2}$.}                                 
\endtable 

The `best-fitting'  models at all redshifts except $z$=2.5 are plotted in Fig
11 and show a good agreement with the data, considering the simplistic
assumptions made. Parameters of the models are listed in Table 4; the
normalization of the bremsstrahlung component relative to the power-law is
expressed as \ratio, the ratio of flux in the bremsstralung spectrum to flux in
the power law at 0.2~\keV.

\beginfigure*{12} 
\psfig{figure=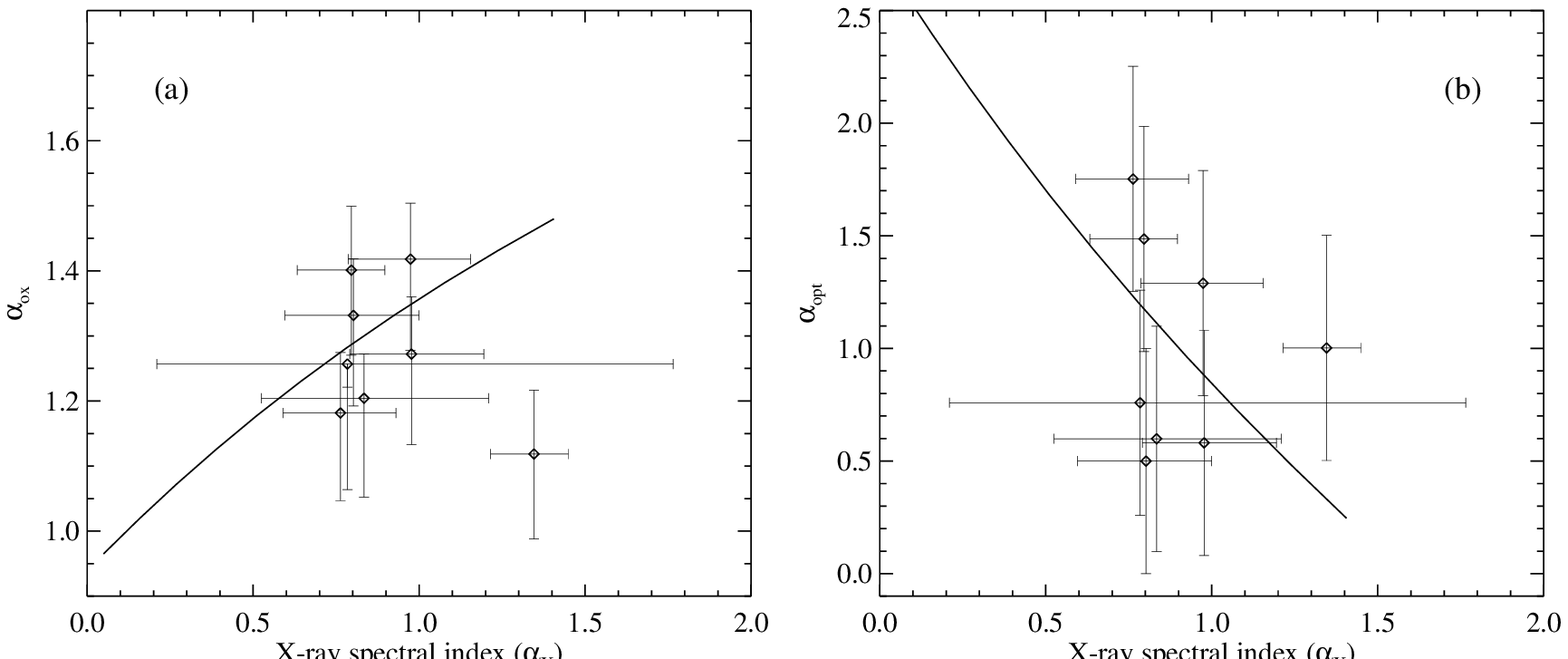,height=3in,width=7.0in,angle=0}
\caption{{\bf Figure 12.} A comparison of the model predictions with the data
for the high redshift ($z>1.75$) AGN; {\sl (a)} \aox\ versus \ax, and {\sl (b)}
\aopt\ versus \ax. The model (plotted as a solid line) has a temperature,
\tbrem=1$\times10^6$~K and \ratio=50, and assumes a redshift of 2.5.}
\endfigure

There are very few differences between the best-fitting \aox\ vs \ax\ models as
the redshift changes, which is consistent with the observations (see Figs 4, 5
and 7, and Section 3.1). The model does predict however that for any given \ax,
\aopt\ should be relatively soft at higher-$z$, whereas the observed values for
\aopt\ tend to be softer at {\sl low}-$z$ (\ie $z<0.5$; although this may be
due to contamination by the host galaxy for which no corrections have been
made). The slopes of the predicted \aos\ versus \ax\ models are harder than the
observed data; this may be a consequence of the selection effect discussed in
Section 2.1.2.

At higher redshifts ($z>1.75$), the range of \ax\ is relatively small and the
relationships between the indices may not be the same (Section
3.1), therefore we discuss these objects separately.  When $z$ is high, the
model predicts that the  effects of absorption in soft X-rays are lessened
(because the strongly absorbed part of the spectrum has been shifted out of the
observed range), while in  the optical, the absorption has a greater effect
(because the heavily-absorbed UV is shifted {\sl into} range). Therefore, the
scatter  in \ax\ at high-$z$ would be reduced while the range in \aopt\ and
\aox\ would remain relatively wide. In Fig 12 we compare the \aox\ versus \ax\
and \aopt\ versus \ax\ distributions with the best-fitting  model at $z$=2.5
(parameters of the model are listed in Table 4) and we find that  the
high-redshift data are consistent with the predictions of the model. Thus the
model can predict the observed data from $z$=0 to $\sim$3.

\subsubsection{Errors on the Galactic \nh}

The measurements of \ax\ used in this analysis have already been corrected for
\nhgal, therefore any absorption implied by the modelling is additional  to
this. In principle, this could be caused by errors on the assumed \nhgal\
rather than  absorption external to our Galaxy.  To test whether the changes in
the continuum slopes of the RIXOS AGN could be due to errors on  \nhgal, we
have compared the \nh\ distribution implied by the model with the typical
uncertainties on the Stark \etal (1992) measurements of \nhgal. By default the
model assumes that  the gas and dust absorption occurs in the rest-frame of the
quasar; to measure the equivalent {\sl Galactic} column, the model was modified 
to assume that the absorption is entirely within the frame of the observer. 

We found that the typical additional observer-frame column density suggested by
the model would be $\sim2\times10^{20}$ cm$^{-2}$ (and this model also
provided a poorer representation of the data, especially at high-$z$). Elvis,
Lockman \& Wilkes (1989) calculated a 1$\sigma$ error on the Stark \etal
measurement of \nhgal\ of 11 per cent for pointings at high Galactic latitude,
implying typical errors on \nhgal\ for the RIXOS AGN of $\sim2\times10^{19}$
cm$^{-2}$; this is significantly lower than the additional `Galactic' columns
implied by the model. Thus we conclude that the variations in continuum shape
are not simply a reflection of the errors on the measurements of \nhgal.

\subsection{Range of big bump temperature and strength}

Although the correlations between \aox\ and \ax\ and between \aopt\ and \ax\
are significant, there is still a large degree of scatter in the data. While
the modelling suggests that the {\sl correlations} may be due to  absorption,
the {\sl scatter} may reflect a range in other parameters such as the
temperature and relative normalization of the big bump.  Therefore, by
calculating models for different values of \tbrem\ and the big
bump-to-power-law ratio and comparing these with the observed distributions, we
can find the approximate ranges of temperature and normalization implied by the
model.

\subsubsection{\tbrem}

To investigate the range of bremsstrahlung temperature, different models were
calculated at low and high \tbrem\ until they fell outside the regions occupied
by the distributions of the spectral indices (\ie \aox\ versus \ax\ and \aopt\
versus \ax). The normalizations and column densities were adjusted where
necessary to provide the best comparison with the data. 

\begintable{7} 
\caption{{\bf Table 5} Range of bremsstrahlung temperature at $z$=1: 
parameters of models shown in Fig. 13}
\halign{   \quad\quad\hfil#\hfil\quad\quad
     &\quad\quad\hfil#\hfil\quad\quad    
     &\quad\quad\hfil#\hfil\quad\quad \cr
\noalign{\medskip} 
       \tbrem\       & \ratio\ & \nh\ \cr
        (1)          &   (2)   &  (3) \cr
\noalign{\smallskip} 
  1$\times10^5$K    &   10$^i$ &  3.0 \cr
  1.3$\times10^6$K  &   20    &  4.0 \cr
  3$\times10^6$K    &   30   &  5.0 \cr
} 
\tabletext{(1) Temperature of the bremsstrahlung component. (2) Ratio of flux
in the bremsstrahlung component to flux in the power-law component at 0.2~\keV.
(3) Maximum column density of the gas used to produce the model in units of
$10^{21}$ cm$^{-2}$. All models are calculated at $z$=1.0. $^i$ Ratio
calculated at 0.02~\keV; the 0.2~\keV\ ratio (\ratio) is 1$\times10^{-7}$.}
\endtable 

The limits on the bremsstrahlung temperature implied by the model at a redshift
of 1.0 are shown in Fig 13 and listed in Table 5, and show that  a wide range
in temperature can be accommodated (from 10$^5$ to 3$\times10^6$K).  At lower
temperatures, the bremsstrahlung spectrum does not reach far enough into soft
X-rays to reproduce the soft excess, while at higher temperatures, the \aox\ is
too low for any given \ax, as the big bump component shifts too far out of the
optical range. Both upper and lower limits are constrained by the \aox\ versus
\ax\ distribution; the scatter in the observed  \aopt\ versus \ax\ data falls
outside these limits (at low and high \aopt) but because of  the relatively
high errors on the measurements of \aopt\ we are unable to determine whether
this is physically significant.

\begintable{8} 
\caption{{\bf Table 6} Range of bremsstrahlung normalization at $z$=1:
parameters of models shown in Fig 14}
\halign{   \quad\quad\hfil#\hfil\quad\quad    
     &\quad\quad\hfil#\quad\quad    
     &\quad\quad\hfil#\quad\quad \cr
\noalign{\medskip} 
       \tbrem\       & \ratio\ & \nh\ \cr
        (1)          &   (2)   &  (3) \cr
\noalign{\smallskip} 
  1.3$\times10^5$K   &   1000   &  10.0 \cr
  1.3$\times10^6$K   &   20     &  5.0 \cr
  1.3$\times10^6$K   &   0      &  4.0 \cr
} 
\tabletext{(1) Temperature of the bremsstrahlung component. (2) Ratio of flux
in the bremsstrahlung component to flux in the power-law component at 0.2~\keV.
(3) Maximum column density of the gas used to produce the model in units of
$10^{21}$ cm$^{-2}$. All models are calculated at $z$=1.0.}                                 
\endtable 

\subsubsection{Ratio of big bump to power-law flux}

A similar procedure was followed to estimate the range of the big
bump-to-power-law flux ratio implied by the model. In this case, \tbrem\ was
fixed while the normalization of the bremsstrahlung spectrum was varied until
the model fell outside the regions defined by the data. Only the column density
was adjusted where necessary to extend the model to hard \ax.

\beginfigure*{13}  
\psfig{figure=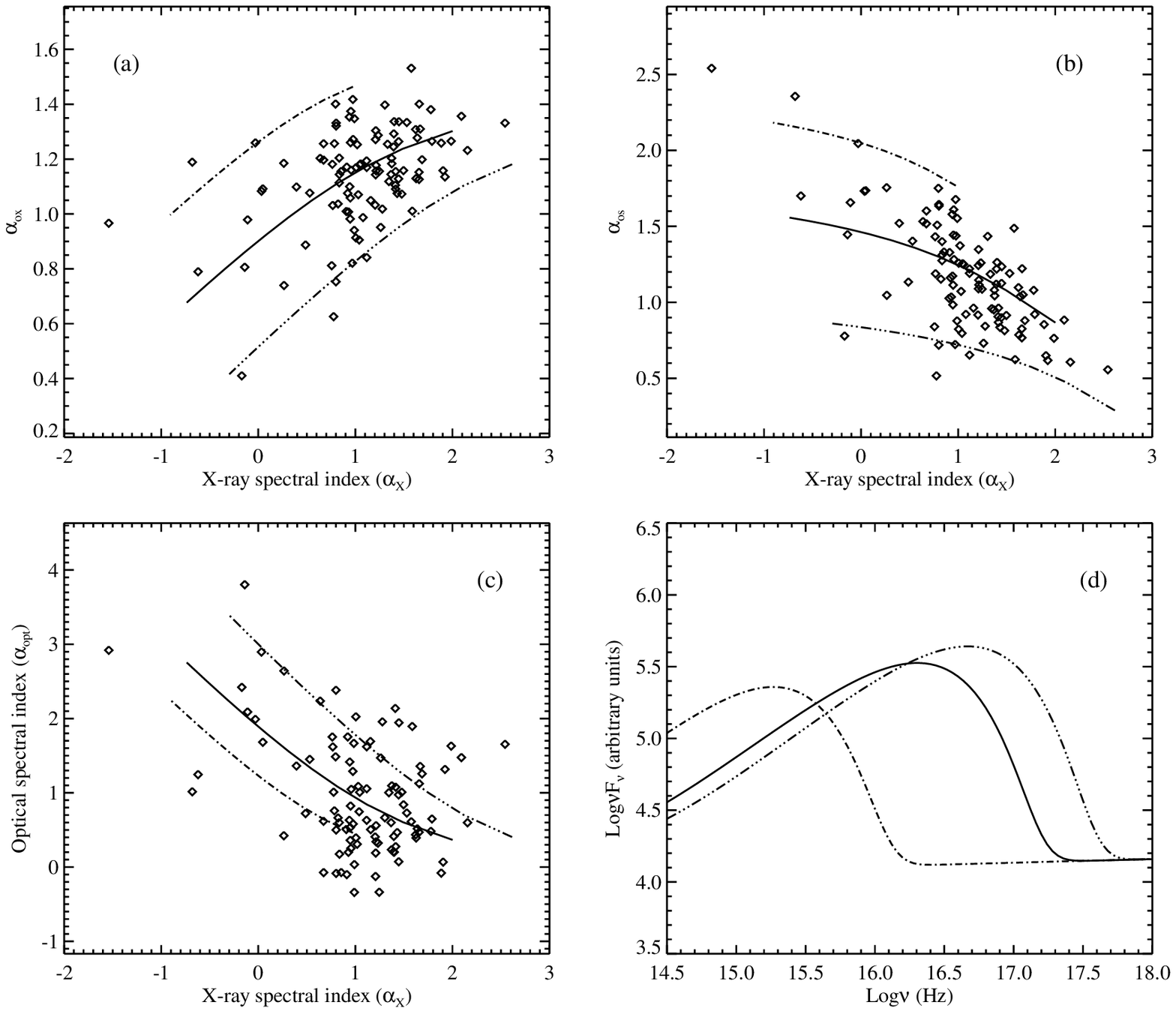,height=6in,width=7.0in,angle=0}
\caption{{\bf Figure 13.} The range of \tbrem\  at a redshift of 1.0 which is
implied by the scatter in the observed data. Three different models are shown:
for the dot-dashed line, \tbrem=10$^5$~K; for the solid line,
\tbrem=1.25$\times10^6$~K; for the dot-dot-dot-dashed line,
\tbrem=3$\times10^6$~K. All three models are compared with {\sl (a)} the
observed \aox\ versus \ax\ distribution; {\sl (b)} the observed \aos\ versus
\ax\ distribution; {\sl (c)} the observed \aopt\ versus \ax\ distribution. 
The three unabsorbed model spectra are plotted in {\sl (d)}.}
\endfigure

The results (at $z=1$) are shown in Fig 14 and are compared with the
best-fitting model (parameters are listed in Table 6). When the bremsstrahlung
normalization is zero, there is no soft X-ray excess or BBB and  only sources
with a relatively hard \ax\ and soft \aopt\ can be predicted by the model. At
the other extreme when \ratio\ is very high, there is little change in the
model for the predicted \aox\ and \aos\ versus \ax\ distributions, but at any
given \aopt, the X-ray slope is much softer. This is because the bremsstrahlung
dominates throughout the optical/UV so that the slope is unchanged by the flux
in the big bump (except for very low normalizations), whereas in soft X-rays,
the bremsstrahlung spectrum only dominates at relatively low X-ray energies
(\ie less than $\sim$0.5~\keV) and therefore its strength relative to the
power-law has a much greater effect on the measured X-ray slope. 

\beginfigure*{14}  
\psfig{figure=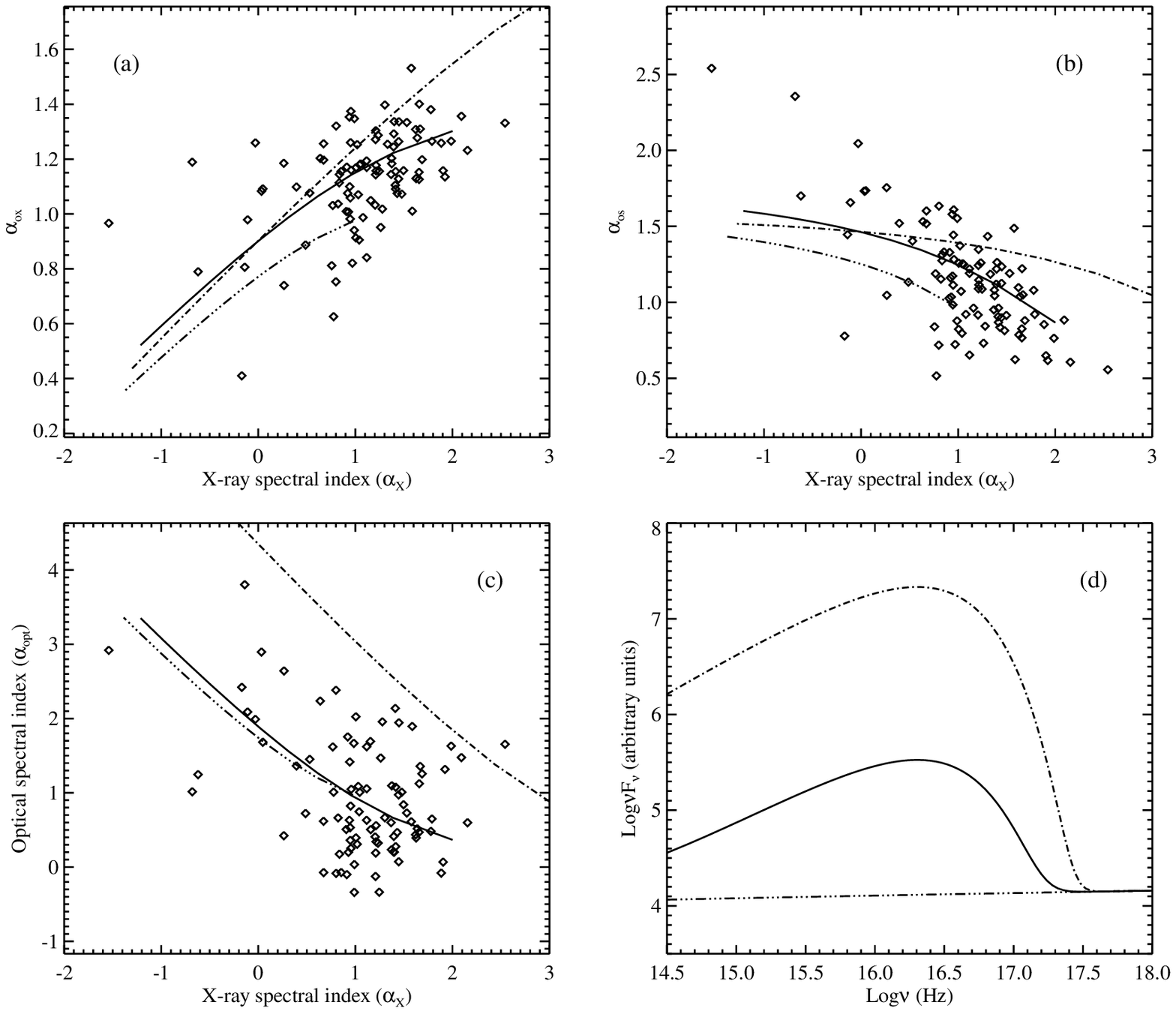,height=6in,width=7.0in,angle=0}
\caption{{\bf Figure 14.} An illustration of the effect of changing the
strength of the bremsstrahlung component relative to the underlying power law.
Three different models (with z=1.0) are shown and compared with the data: for
the dot-dot-dot-dashed line, there is no bremsstrahlung component;  for the solid line,
the \ratio=20;  for the dot-dashed line, \ratio=1000. All three models
have \tbrem=1.25$\times10^6$~K and are compared with {\sl (a)} the observed
\aox\ versus \ax\ distribution; {\sl (b)} the observed \aos\ versus \ax\
distribution; {\sl (c)} the observed \aopt\ versus \ax\ distribution. 
The three unabsorbed model spectra are plotted in {\sl (d)}.}
\endfigure

The model thus suggests that when \tbrem=1.25$\times10^6$K (at $z$=1.0),
\ratio\ can vary from 0 to $\sim$1000. The upper limit is constrained by the
\aopt\ versus \ax\ distribution  (although we note that the absorption required
is correspondingly high to predict any given \ax) but there is no lower limit
(\ie the model can accommodate AGN without big bumps).

\section{Discussion}

We have presented the optical and X-ray continuum parameters  of 108 X-ray
selected AGN from the RIXOS survey (Mason \etal 1996). From the correlation
analysis of this sample,  we find significant evidence for relationships
between \aox\ and \ax\ and between \aopt\ and \ax\ (especially at $z>1.0$).
These suggest that as the 0.2-2~\keV\ spectrum softens,  the
2500~\AA-to-2~\keV\ ratio increases  and the slope of the optical spectrum
hardens. The trends  apply over a wide range of \ax\ for soft {\sl and} hard
X-ray slopes (\ie  from \ax$\sim$0 to 2.5) and  suggest that the overall
6000~\AA-to-2~\keV\ spectrum changes from convex to concave as \ax\ hardens.

\beginfigure*{15} 
\psfig{figure=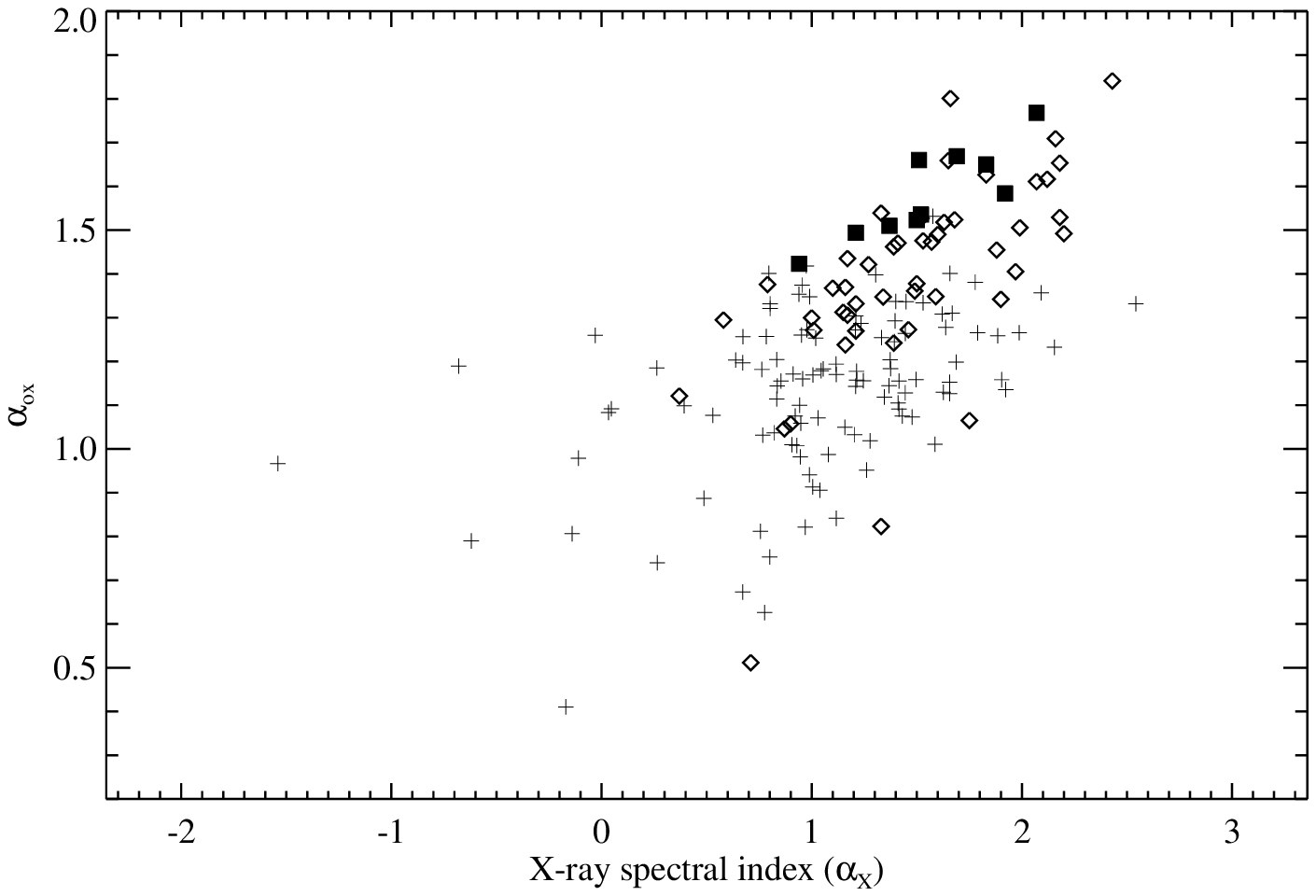,height=4.0in,width=6.0in,angle=0}
\caption{{\bf Figure 15.} The distribution of  \aox\ versus \ax\ for the RIXOS
AGN (crosses) compared with that of the WF (diamonds) and L94 (squares)
samples. The WF and L94 data have been corrected for the differences in the
definitions of \aox\ between all three samples (see Section 3.1.3).}
\endfigure

WF and L94 have also identified \aox\ versus \ax\ correlations in their
relatively soft X-ray samples and WF suggested that it is due to changes in the
strength of a big bump relative to an  underlying continuum. However, this
explanation is not appropriate for the RIXOS sample (which extends to harder
X-ray slopes than WF) because it implies that the underlying continuum from
6000~\AA\ to 2~\keV\ must be concave (see \eg the dot-dashed line in Fig 9d)
and a spectrum of this shape is difficult to reconcile with observations at
longer and shorter wavelengths (Section 4). 

We have illustrated the main differences between the WF, L94 and RIXOS AGN
samples, by plotting \aox\ as a function of \ax\ in Fig 15 (these have been
corrected for the differences in the definition of \aox\ for the three samples,
\ie the WF and L94 \aox s have been increased by 0.02 and 0.04 respectively;
see Section 3.1.3). The figure shows that the WF objects are clustered towards
high \aox\ and high \ax\ relative to the RIXOS AGN; a similar and more extreme
pattern is seen for the L94 objects. This would have led WF and L94 to draw
different conclusions regarding the nature of the optical-to-X-ray continua in
AGN, \eg the WF data do not require the effects of absorption perhaps because
they observed very few highly absorbed sources (with \ax$<$1), while L94
suggested that sources with a soft \ax\ were weak in X-rays (\ie \aox\ was high
when \ax\ was soft), yet Fig 15 shows that in both the WF and RIXOS samples,
there are also many sources with a much lower \aox\ when \ax\ is soft (\ie the
2~\keV\ flux is high relative to the optical and the X-ray spectrum is soft),
implying a true soft X-ray `excess' in these AGN. 

These differences are largely a consequence of the selection methods, \ie the
PG quasars used by L94 were selected by the strength of their UV excess, thus
they tend to be optically-bright and -hard, while the WF objects are bright in
soft X-rays (0.08-2.4~\keV) and therefore X-ray-bright and generally
X-ray-soft. The RIXOS sample is itself restricted by its selection criteria
(\ie by the strength of the 0.4-2.0~\keV\ emission), therefore its properties
can only be said to be typical of objects which are relatively strong in medium
X-rays (\ie at $\sim$2~\keV\ in the quasar rest-frame). Ultra-soft X-ray
objects like E1346+266 and RE~J1034+396, which have very soft {\sl ROSAT}
spectra yet little or no flux above $\sim$1~\keV\ (Puchnarewicz \etal 1994;
Puchnarewicz \etal 1995), would be selected against  in this sample for
instance; as would AGN which are heavily absorbed through cold gas.

As an explanation for the observed trends in the RIXOS AGN, we have proposed
that the correlations  are due to the effects of moderate absorption by cold
gas and dust. It is possible that the results are also reproducible by a dusty
warm absorber (see \eg Mathur \etal 1994; Brandt, Fabian and Pounds 1996), but
the additional parameters required for the fits are not justified by the
quality of these data. This hypothesis has been tested using simple
combinations of power laws and bremsstrahlung spectra which have been modified
for the effects of dust and gas absorption. We find that the model is
consistent with the data up to  $z\sim3$ (this is effectively the limit of the
sample).

However, the model must also be judged by its ability to reproduce the spectra
of AGN which lie outside the RIXOS parameter space, such as ultra-soft X-ray
sources and the PG quasars (see Fig 15), which are relatively quiet in X-rays.
We have fitted our model to the L94 sample of PG quasars, finding a best-fit 
when \tbrem=6$\times10^5$~K and \ratio=10, which reproduces the optical and
X-ray slopes as well as the optical-to-X-ray ratio (see Fig 16). The
temperature is lower than that required for the RIXOS AGN but \ratio\ is the
same (see Table 4); the amount of cold absorption required is very small (a
maximum of 3$\times10^{20}$ cm$^{-2}$). With regard to ultra-soft X-ray
sources, we note that the optical-to-X-ray spectrum of the ultra-soft X-ray
quasar E1346+266 also compares well  with an unabsorbed bremsstrahlung spectrum
(\tbrem=7$\times10^6$~K; Puchnarewicz \etal 1995), in agreement with our model.  
However for the lower-luminosity ultra-soft Seyfert 1 galaxy RE~J1034+396, a
bremsstrahlung plus power-law provided a poor fit to the data (Puchnarewicz
\etal 1995), thus while this model can be extended to some ultra-soft X-ray 
AGN and UV-excess quasars, it may perhaps be inappropriate for lower luminosity
ultra-soft objects like RE~J1034+396. However given the idiosyncrasies of
individual objects, more stringent tests must be made using higher quality data
before any conclusions can be drawn at this stage.

\beginfigure*{16} 
\psfig{figure=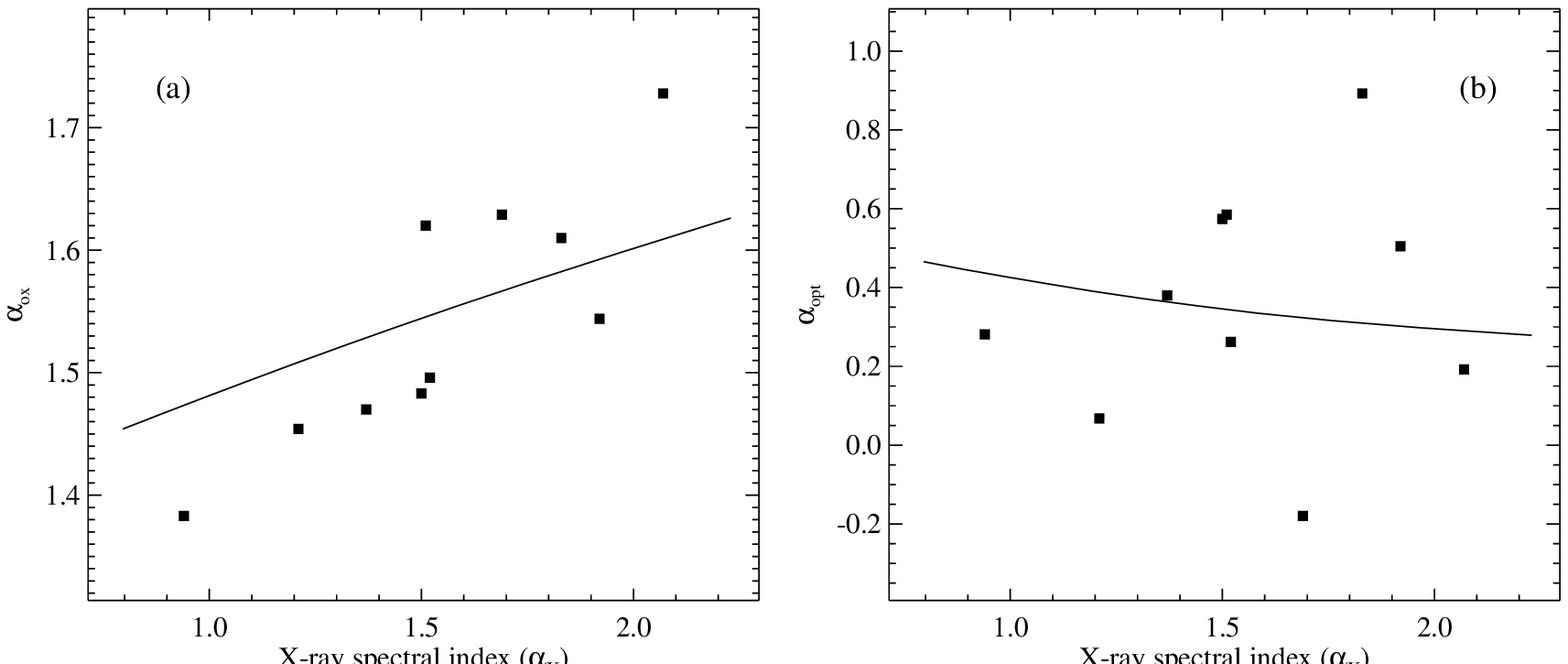,height=3in,width=7.0in,angle=0} 
\caption{{\bf Figure 16.} The L94 data (plotted as squares)  compared with the
best-fitting absorption model;  {\sl (a)} \aox\ versus \ax, and
{\sl (b)} \aopt\ versus \ax. The model (plotted as a solid line) assumes a
redshift of 0.2 (the mean redshift of the L94 sample) and has a temperature,
\tbrem=6$\times10^5$~K, \ratio=10 and uses a range of $N_{\rm H}$ from 0 to
3$\times10^{20}$ cm$^{-2}$.}
\endfigure

We now discuss the implications of the model for an optical-to-soft X-ray big
bump and for the presence and location of dusty, cold gas in AGN.

\subsection{The origin of the BBB}

The correlations in the data suggest that the BBB and the soft X-ray excess are
either part of the same component (the `big bump') or that they are separate
but subject to the same external influences. In the modelling, we used a single
optical/soft X-ray big bump and this was demonstrated to provide a good
representation of the data, therefore throughout this section we have assumed a
big bump as the origin of the BBB and the soft X-ray excess.

\subsubsection{Accretion discs}

The optically-thick AD models are a natural assumption for the origin of the
BBB and  ADs have been  successfully used to reproduce optical-to-X-ray AGN
spectra (\eg Arnaud \etal 1985; Czerny \&\ Elvis 1987; Sun \&\ Malkan 1989),
even in extreme cases such as the EUV-bright Seyfert~1 galaxy RE~J1034+396
(Puchnarewicz \etal 1995). Although some AD models are not appropriate for the
optical/soft X-ray big bump in the RIXOS AGN  (Section 4.1.1), we find that 
the Czerny \& Elvis (1987) model of an AD with a surrounding hot ($T=10^8$~K)
corona does compare well with the most convex (\ie the mean \ax=1.65 spectrum
indicated by the dashed line in Fig 9d) of the mean spectra. It is
similar in shape to the bremsstrahlung spectrum which was successfully used in
the modelling.

The AD corona is optically thin and cooled by inverse Compton scattering so
that the high energy tail of the disc spectrum hardens and extends to higher
energies. The geometry of the corona is not well constrained, for example it
may form a thin `skin' around the disc or be large and spherical. If the
optical depth of a large corona is high ($\tau\sim1$), then the model's
prediction that soft X-ray excesses are only observed from ADs observed
face-on, is relaxed (Czerny \& Elvis 1987).

The Czerny \& Elvis model shown in Fig~10 has a black hole mass of
10$^8$~M$_\odot$. With the naive assumption that the black hole mass scales
inversely with the square root of the temperature (this applies in the simple
thin disc approximation, \eg Pringle 1981), then the range of \tbrem\ implied
by the model (\ie $10^5-3\times10^6$~K at $z$=1) suggests black hole masses of 
$\sim10^6-10^9M_\odot$ for an equivalent AD model.  Furthermore, as the
redshift of the AGN increases, the temperature of the big bump tends to
decrease (see Table 4). In the accretion disc models, the temperature of the
disc is determined by the mass of the black hole and is lower when the black
hole's mass is high. Thus if the big bump spectrum is produced by an accretion
disc, then our  model favours more massive black holes at higher redshifts. A
similar pattern (\ie low-temperature ADs at high-$z$) would be expected in
order for the AD model to be consistent with the restrictions imposed by the
Eddington limit. 

\subsubsection{Optically-thin gas}

The bremsstrahlung model for the big bump has many attractive features which
make it a viable alternative to the more popular, optically-thick AD (Barvainis
1993).  For example, it can predict observed quasar continuum slopes in the
optical (\aopt$\sim$0.2-0.3, \eg Neugebauer \etal 1987) without the need for an
extension of the X-ray power law into the optical to match the observed
spectra; an additional power-law component is often required for AD models.
Also, if the geometry of the gas is toroidal, it can produce linearly polarized
light similar to that which is often observed (see Barvainis 1993 and
references therein). 

A bremsstrahlung spectrum has been assumed for the big bump in the modelling
and we find that this successfully reproduces the observed spectral indices
(\ie \aox, \aopt\ and \ax). Although our model assumes that the X-ray power-law
component continues down to $\sim$10000~\AA, the bremsstrahlung spectrum 
dominates over the power law in the optical so that, in most cases, the removal
of the power law longward of the EUV would make no significant difference to
the model fits. The optical extension to the power-law {\sl is} required
however when the total flux in the bremsstrahlung component is weak relative to
the X-ray power-law (see, for example, the extreme case in Fig 14 where there
is no bremsstrahlung component).

The scatter in the correlations observed in the RIXOS AGN may be satisfied by
differences in the intrinsic \tbrem\ and in the relative normalization of the big
bump to the underlying continuum. The best-fitting temperature of the gas
(\tbrem$\sim10^6$~K) would be high enough to make the Lyman edge very weak and
therefore difficult to detect (Barvainis 1993), which is consistent with
observations (\eg Antonucci, Kinney \& Ford 1989; Koratkar, Kinney \& Bohlin
1992). 

One problem with an optically-thin origin of the big bump is its ability to
reproduce the very short-timescale variability observed in some AGN, although
Barvainis (1993) demonstrated that many small, free-free emitting cloudlets may
have a source size as small as 1 lt-hr across.

\subsubsection{Reprocessing in cool clouds}

In the cool clouds model, the primary quasar spectrum is reprocessed by cool,
dense clouds which lie close to, and perhaps even within,  the central
continuum emitting source (Guilbert \& Rees 1988; Ferland and Rees 1988). The
clouds absorb soft and medium X-rays and re-emit them in the optical/UV. This
model may, for certain values of the cloud density and filling factor, predict
observed AGN features such as the BBB and the soft X-ray excess, and has
already been invoked to explain the medium-to-hard-X-ray spectra observed in
Seyfert galaxies (Nandra \& George 1994).

Barvainis (1993) has suggested that in principle, it is possible for the
reprocessed optical/UV continuum to be optically-thin, thus the reprocessing
(cool) clouds themselves may produce the big bump spectrum observed. The
temperature of the free-free spectrum in this case would be $\sim10^5-10^6$~K,
which is similar to that implied by our modelling ($\sim10^5-3\times10^6$~K),
and the clouds must have relatively low densities ($N<10^{15}$ cm$^{-3}$;
Barvainis 1993), although not so low that the free-free emitting source becomes
very large.

\subsection{The absorbing medium}

The model for the RIXOS AGN optical-to-X-ray continua described in Section 4.1,
suggests that cold gas with a roughly Galactic gas-to-dust ratio and moderate
(\nh$\le5\times10^{21}$ cm$^{-2}$) column density lies  along the line of sight
to the quasar nucleus.  Similar levels of  intrinsic absorption have  been
observed in other AGN, \eg in the quasar 3C109,  Allen \& Fabian (1992) found
evidence of cold gas with a column density of  5$\times10^{21}$ cm$^{-2}$ and a
Galactic dust-to-gas ratio, while Mathur \etal (1994) have shown that the
X-ray and UV spectra in 3C~351 are absorbed by a highly ionized outflow
with a column density of 1-2$\times10^{22}$ cm$^{-2}$.

Based upon the parameters of the gas suggested  by the modelling presented in
Section 4, we now explore the nature of a possible cold absorbing medium in the
RIXOS AGN and whether the results of the modelling can  be consistent with the
emerging picture of quasar and Seyfert nuclei.

\subsubsection{Geometry of the gas}

A `first-order' indication of the covering factor of this dusty gas may be
derived from the distribution of the intrinsic \nh\ which is implied by the
modelling. 

For any given model column density (\nhmod), the absorbed
bremsstrahlung-plus-power-law model makes a prediction of the slope of the
X-ray spectrum (\axmod); \axmod\ is plotted as a function of \nhmod\  in Fig
17a for the best-fitting models at redshifts from 0.2 to 2.5. Therefore, by
using the {\sl observed} \ax\ and interpolating between the points of the
function (appropriate to the redshift of that AGN), the amount
of absorbing material intrinsic to the AGN (\nhint) may be derived.  The data
were thus divided into redshift bins (\ie $z$=0-0.25, 0.25-0.75, 0.75-1.25,
1.25-1.75 and 1.75-3.5) and \nhint\ was calculated for each AGN by 
interpolating between the corresponding \axmod\ vs \nhmod\ functions (\ie the
$z$=0.2, 0.5, 1, 1.5 and 2.5 curves respectively).

The resulting distribution of \nhint\ is shown in Fig 17b; these represent  the
column densities of the absorbing gas and dust intrinsic to the AGN and are
{\sl in addition} to \nhgal. It shows a peak at \nh$<10^{21}$ cm$^{-2}$ and
implies  that approximately one-third of the RIXOS AGN lie behind columns
$\ge10^{21}$ cm$^{-2}$. 

\beginfigure{17} 
\psfig{figure=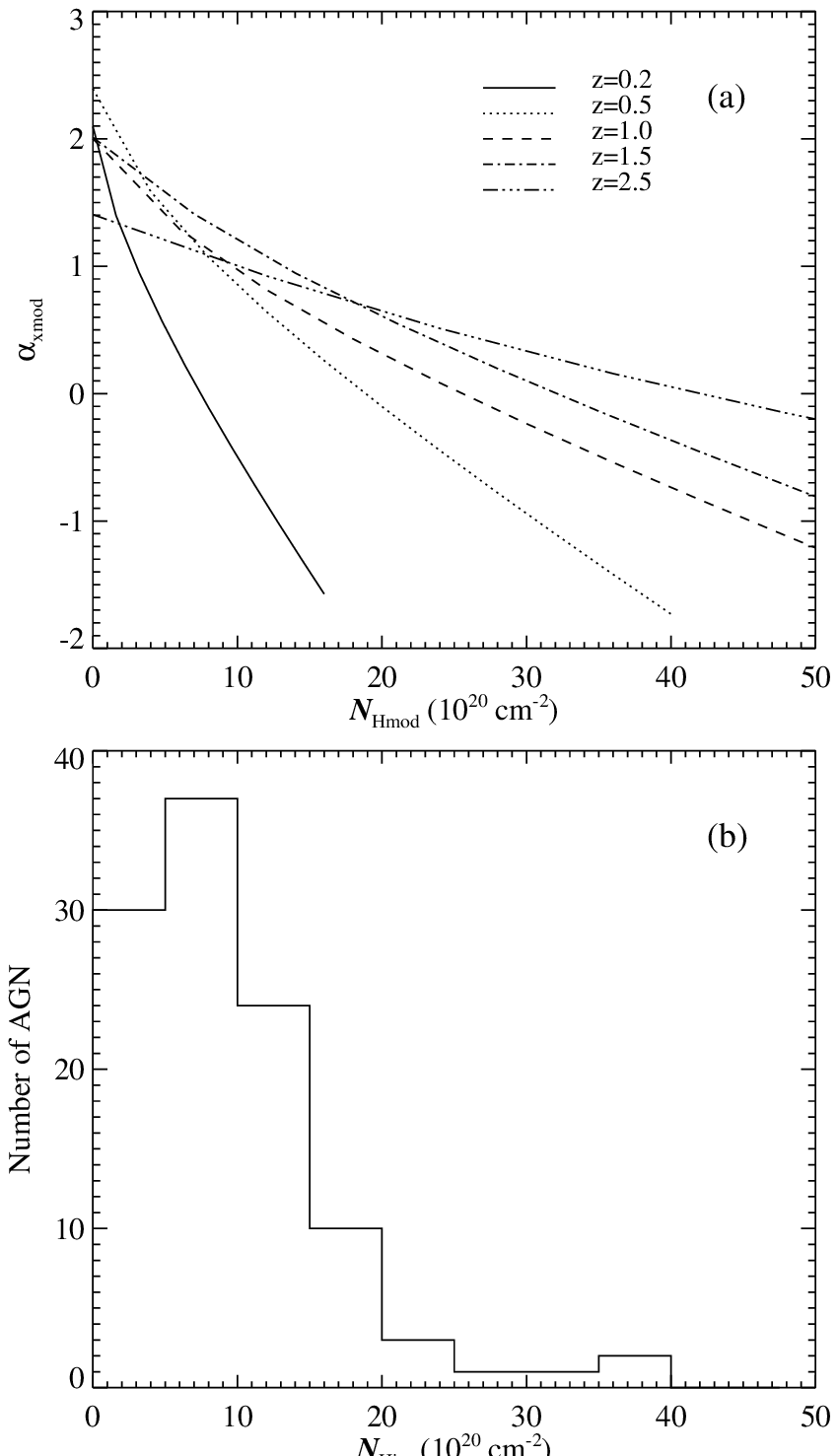,height=6in,width=3.3in,angle=0}
\caption{{\bf Figure 17.} {\sl (a)} The model \ax\ at $z$=0.2, 0.5, 1.0, 1.5 and
2.5 (solid, dotted, dashed, dot-dashed and dot-dot-dot-dashed respectively)
calculated from the best-fitting absorbed bremsstrahlung plus power-law models
at those redshifts  (see Section 4.1), and plotted as a function of the assumed
input column densities.  {\sl (b)} The distribution of inferred intrinsic column
densities for the RIXOS AGN (\nhint; solid line) calculated using the models
plotted in (a) (\ie using bins of $z$=0.0-0.25, 0.25-0.75, 0.75-1.25, 1.25-1.75
and 1.75-3.5 respectively). These represent the column densities which would be
intrinsic to the AGN themselves and are in addition to any Galactic
absorption.}
\endfigure

When considering the distribution of \nhint, a number of points should be borne
in mind. Firstly, the sample used in this paper is not fully complete, although
the \ax\ distribution of these sources is not significantly different from that
of the complete RIXOS survey (Mittaz \etal 1996). Secondly, the defining
criteria of the sample  select against very soft sources which have
little or no flux above 0.4~\keV. However, based on the number  of AGN in the
Ultra-Soft Survey  (Thompson \& C\'ordova 1994) which have no significant
emission above 0.5~\keV, and comparing this with the AGN from  the {\sl
Einstein} Extended Medium Sensitivity Survey (Gioia \etal 1990; Stocke \etal
1991), we estimate that only $\sim$10 very soft sources have been omitted. 
Finally, the RIXOS sample will not include heavily absorbed sources which have
insufficient 0.4-2.0~\keV\ flux to be detected, therefore at relatively high
values of \nhint\ (\eg higher than $\sim10^{21}$ cm$^{-2}$), the distribution
in Fig 17b should be considered a lower limit. 

Taking into account these restrictions on the sample, we can estimate  the
proportion of sources which are relatively unabsorbed (\ie with
\nhint$<10^{21}$ cm$^{-2}$). Assuming (naively)  that this distribution is
typical of the AGN population in general, we find that 67 out of 108 sources
are `unabsorbed' (from Fig 17), which gives a covering factor for the absorber
of $\sim$0.3. Allowing for a possible 10 missing ultra-soft X-ray sources makes
no significant difference to the covering factor, but as we have made {\sl no}
allowance for AGN which are too strongly absorbed to be detected, this  must be
considered a very conservative lower limit. Thus geometries of the gas where
the covering factor is relatively high, \eg  spherical and toroidal
distributions, are favoured while models with a small covering factor (less
than $\sim$0.1), \eg thin discs, are unlikely. 

\paragraph{A thick torus}

If the gas is contained within a geometrically-thick torus, then the range of
absorbing columns implied by the model may be due to orientation effects if the
amount of gas along the line of sight to the nucleus changes with the `viewing
angle' (the angle between the line of  sight and the axis of the torus). For
example, if the density of the gas decreases with increasing distance from the
plane of the torus (whether it is due to fewer clouds,  less dense clouds or a
less dense, smoothly distributed gas), then the amount of absorption observed
would increase as the viewing angle increases. Otherwise, if the density
remains constant, then  the azimuthal profile of the torus may be such that the
path length through the torus changes with the viewing angle. In this
orientation-dependent model then, when the torus is observed relatively
face-on, there would be very little absorption and \ax\ is soft. As the viewing
angle increases, the amount of cold gas along the line-of-sight increases and
\ax\ hardens. 

Alternatively, the range of \nhint\ may simply indicate that  absorbing
tori in AGN have a wide range of optical depths. Note that if the torus is made
up of individual clouds, then the clouds must be able to have low column
densities ($<10^{21}$ cm$^{-2}$) in order to produce the small columns
observed. 

\paragraph{Spherical}

If the gas is spherically distributed, then the observed  range in \ax\
reflects a range in the column density of the absorbing material. The
distribution in Fig 17 would then imply that most RIXOS AGN  are surrounded
by gas with columns of $\sim10^{21}$ cm$^{-2}$.

\subsubsection{Location of the gas}

The presence of dust places limits on the location of the gas, since  the {\sl
minimum} radius at which dust grains sublime ($r_{\rm  min}$) is given by
$r_{\rm min}\simeq 0.20L_{46}^{0.5}$~pc, where $L_{46}$ is the
bolometric luminosity in units of $10^{46}$ erg s$^{-1}$  (Laor \& Draine
1993). For the RIXOS AGN then, the smallest radius at which dust can survive is
the order of a parsec, which places the dust beyond the broad line region
(BLR). Thus the most likely locations of the dust are {\sl (1).} the narrow
line region (NLR), {\sl (2).} the dusty molecular torus, which lies between the
BLR and NLR, {\sl (3).} the host galaxy and {\sl (4).} intervening galaxies
along the line of sight.

\paragraph{NLR}

It has been suggested that dust lies within the NLR itself (\eg Laor \& Draine
1993). The covering factor of the NLR is very small however  ($\sim$0.01) and
since the distribution of \nhint\ implies that a large fraction of the RIXOS
AGN are absorbed, we conclude that the NLR is an unlikely source of the
absorbing medium which produces the continuum changes. 

\paragraph{Molecular torus}

It has been proposed that a dusty, molecular torus lies  between the BLR and
NLR of many Seyfert galaxies  (see \eg Antonucci 1993 for a review). This torus
is probably geometrically-thick [the ratio of the height ($h$) to the inner
radius ($r$), $h/r$, is $\sim$0.7, and $r\sim$1~pc] and has a high column
density (\nh$\ge10^{24}$ cm$^{-2}$; Krolik \& Begelman 1988). It is made up of 
clouds which individually have a large column density (\ie \nh$\ge10^{24}$
cm$^{-2}$) and the torus has a cloud covering factor $\sim$1.

A torus of this kind is not possible as the source of absorbing gas in the
RIXOS AGN, because the column density of individual clouds is much higher than
that implied by the modelling, so that observed AGN continua would be
completely unobscured or very strongly absorbed; in most cases we observe
moderate absorption (\ie \nhint$\sim10^{21}$ cm$^{-2}$). However, if the
Eddington luminosity of an AGN is high,  the torus might instead  be optically
thin or geometrically thin (Krolik \& Begelman 1988).   We have ruled out a
geometrically-thin absorber on the basis of the \nhint\ distribution (Fig
17 and Section 5.2.1); a geometrically-thick torus, subject to the limitations
discussed in Section 5.2.1, is possible however  if the clouds have a low \nh\
($<10^{21}$ cm$^{-2}$). 

\paragraph{Host galaxy}

The column densities implied by the modelling are typical of the interstellar
medium in the disc of our own Galaxy ($\sim10^{21}$~cm$^{-2}$ in a direction
perpendicular to the plane of the disc; Stark \etal 1992), however absorption
by the galactic disc local to the AGN is unlikely as the covering factor is
small. Cold gas and dust in the bulge of spiral galaxies (which are the typical
hosts for Seyfert 1 nuclei) and in ellipticals may produce the absorption
implied; this would suggest a  spherical distribution of the gas.

\paragraph{Intervening galaxies}

It is also possible that the quasar spectra are absorbed through relatively 
face-on galaxies along the line of sight which are too faint to be seen. Spiral
galaxies would be the most likely absorbers as they contain a higher {\sc HI}
mass than ellipticals or lenticulars (Rao \& Briggs 1993).  Any absorption by
external galaxies close to high-redshift quasars would be difficult to measure
in X-rays because the strongest  effects of the absorption are shifted out of
the observed range; the absorption we measure would be dominated by relatively
nearby (\ie low-$z$) galaxies. Nonetheless, we would expect to see a trend
towards more heavily absorbed AGN as the redshift increases, yet there is  no
hardening of \ax\ with $z$ for the RIXOS AGN (Fig 4, this paper; Mittaz \etal
1996). 

Although it is difficult to measure this kind of absorption in X-rays due to
the redshift effect, it should be easier in the optical because  the shorter
wavelength regions, where the dust absorption is higher, are moved into the
observer-frame, making absorption easier to detect. In this case, the optical
slope would soften with increasing $z$, yet again, we find no significant
changes of \aopt\ with $z$ (for $z>0.25$). Thus we conclude that absorption by
intervening spiral galaxies is unlikely.

\subsection{The unified AGN model}

In the `hidden Seyfert 1' picture of a Seyfert 2 galaxy, some (if not all)
Seyfert 2s are actually Seyfert 1 nuclei viewed through the dusty molecular
torus (\eg Miller \& Goodrich 1987); this is a part of the `unified AGN model'
which essentially aims to explain the wide range of AGN properties by
orientation effects (see \eg Antonucci 1993 and references therein). We now
speculate on an extension of this model to the quasar population in the light
of the RIXOS results. 

If indeed the RIXOS AGN are absorbed as we propose, then what we might be
seeing is a mixture of type 1 and type 2 quasars (analogous to type 1 and type
2 Seyferts) whose X-ray and optical spectra are modified by a dusty torus. The
torus has a low column density, (perhaps as a consequence of the relatively
high luminosity of the quasars), low enough to modify the optical-to-X-ray
continuum but still allowing much of the broad permitted line emission to
escape. Thus most quasars would be classified as type 1s based on their optical
spectra, even though their nuclear emission is viewed through the torus (as is
assumed for Seyfert 2s). This would explain the rarity of `type 2' quasars (\ie
quasars with narrow permitted and forbidden lines) since the BLR is not always
completely absorbed by the torus.

\section{Summary}

We have used the RIXOS AGN to probe the nature of the optical-to-X-ray
continuum in a large sample of X-ray selected Seyfert 1s and quasars.  The
sample provides a wide range in parameter space, \ie in luminosity, X-ray
spectral slope and optical-to-X-ray ratio, over which to search for
relationships between the optical and X-ray continua.  We confirm the \aox\ vs
\ax\ correlation previously reported by WF and L94, and identify
anti-correlations between \aopt\ and \aox\ and between \aopt\ and \ax\
(especially at $z>1.0$), suggesting that the optical and X-ray slopes fall and
rise towards the EUV together. 

These relationships imply overall changes in spectral shape from convex (when
\ax\ is soft) to concave (when \ax\ is hard; see Fig 9d). They also suggest
that the BBB and the soft X-ray excess are strongly dependent and they may be
part of the same optical-to-soft X-ray `big bump'. The changes in the mean
optical-to-X-ray continua are not consistent with a component which shifts
between the optical and X-ray regions, \eg ADs or bremsstrahlung spectra with a
range of temperatures. Although these changes might be produced by a big bump
with varying strengths superposed on a relatively invariant underlying
continuum, the underlying continuum must be either (1)  concave from 6000~\AA\
to 2~\keV\ (like the dot-dashed continuum in Fig 9d) yet have $\alpha\sim$1 at
lower and higher frequencies, or (2) absorbed by cold gas and dust while the
BBB and soft X-ray excess are completely unabsorbed: we conclude that both are
unlikely.

We have proposed instead that the changes are due to absorption of the entire
intrinsic quasar spectrum by cold gas and dust and have demonstrated the
viability of this idea from $z$=0 to $\sim$3  by creating models of the
absorbed continua. The model assumes that the intrinsic spectrum is the sum of
an $\alpha$=1 power law and a bremsstrahlung spectrum. The temperature of the
bremsstrahlung component favoured by the modelling is $\sim10^6$~K and the
scatter in the data implies a range in \tbrem\ of $\sim10^5$~K to
3$\times10^6$~K. A wide range in the strength of the big bump is also permitted
by the model, \eg at $z$=1 and for \tbrem=1.25$\times10^6$~K, the ratio of flux
in the bremsstrahlung spectrum to that in the power-law at 0.2~\keV\ can vary
from 0 to $\sim$1000.  Thus the model does not rule out the possibility of
significant differences in the strength and temperature of the big bumps from
source to source. It does suggest however that the correlations observed are
not due to changing big bump parameters (\ie neither the strength nor
temperature) but that they are consistent with the effects of absorption by
dusty cold gas.

The big bump may be produced by an AD with a surrounding hot corona, an
optically-thin gas or perhaps reprocessing in cool clouds.   The absorber is
cold gas with an approximately Galactic dust abundance and its column density
varies from 0 to $\sim4\times10^{21}$ cm$^{-2}$.  The distribution of the
intrinsic \nh\ in the RIXOS AGN implied by the modelling suggests a high
covering factor, \eg a toroidal or spherical geometry of the gas. The presence
of dust places it beyond the BLR and, if it is toroidal, then the different
types of observed continua may be due to an orientation effect with the column
density increasing with the viewing angle. 

If the obscuring torus which lies between the BLR and the NLR is responsible
for the absorption, then the column density of the clouds which make up the
torus must be significantly lower than that observed in Seyfert 2s. We
speculate that the rarity of `type 2' quasars is due to obscuring tori with low
column densities which do not always fully absorb the BLR; this may be a
physical consequence of the high luminosity in quasars compared to Seyferts. An
alternative source of the absorbing gas may be the bulge of a spiral host
galaxy or an elliptical host galaxy.

\section*{Acknowledgments}

We thank all in the RIXOS team for their work in obtaining and reducing the
data. The RIXOS project has been made possible by the award of International
Time on the La Palma telescopes by the Comit\'e Cient\'ifico Internacional.
This research has made use of data obtained from the UK {\sl ROSAT} Data
Archive Centre at the Department of Physics and Astronomy, University of
Leicester (LEDAS) and we would especially like to thank Mike Watson and Steve
Sembay for their kind assistance. We thank the Royal Society for a grant to
purchase equipment essential to the RIXOS project. We are grateful to  Richard
Barvainis, Andy Fabian, Niel Brandt and Harry Lehto for their advice and
comments. The referee, Belinda Wilkes, provided a very thorough report with
valuable comments and suggestions, for which we are also very grateful. KOM
acknowledges the Royal Society for support.

\bye